\documentclass{aa}  

\usepackage{graphicx}
\usepackage{txfonts}
\usepackage{xcolor}
\begin{document}

   \title{Lithium in the lower red giant branch of \\ 5 Galactic globular clusters\thanks{Based on observations collected at the European Southern Observatory under ESO programme 095.D-0735(A).},\thanks{Table A.1 is available in electronic form at the CDS via anonymous ftp to cdsarc.u-strasbg.fr (130.79.128.5) or via http://cdsweb.u-strasbg.fr/cgi-bin/qcat?J/A+A/}}

   \author{C. Aguilera-G\'omez
          \inst{1,2}
          \and
          L. Monaco
          \inst{1}
          \and
          A. Mucciarelli \inst{3,4}
          \and
          M. Salaris \inst{5}
          \and
          S. Villanova \inst{6}
          \and
          E. Pancino \inst{7}}

   \institute{Departamento de Ciencias Fisicas, Universidad Andres Bello, Fernandez Concha 700, Las Condes, Santiago, Chile\\
   \email{craguile@uc.cl}
         \and
        N\'ucleo de Astronom\'ia, Universidad Diego Portales, Ej\'ercito 441, Santiago, Chile
         \and
        Dipartimento di Fisica e Astronomia, Universita degli Studi di Bologna, Via Gobetti 93/2, I-40129 Bologna, Italy
        \and
        INAF - Osservatorio di Astrofisica e Scienza dello Spazio di Bologna, Via Gobetti 93/3, I-40129 Bologna, Italy
        \and
        Astrophysics Research Institute, Liverpool John Moores University, IC2, Liverpool Science Park, 146 Brownlow Hill, Liverpool, L3 5RF, UK
        \and
        Departamento de Astronom\'ia, Casilla 160-C, Universidad de Concepci\'on, Chile
        \and
        INAF - Osservatorio Astrofisico di Arcetri, Largo Enrico Fermi 5, I-50125 Firenze, Italy
        }

   \date{Received ; accepted }

  \abstract
   {Lithium is one of the few elements produced during the Big Bang Nucleosynthesis in the early universe. Moreover, its fragility makes it useful as a proxy for stellar environmental conditions. As such, the lithium abundance in old systems is at the core of different astrophysical problems.}
   {Stars in the lower red giant branch allow studying globular clusters where main sequence stars are too faint to be observed. We use these stars to analyze the initial Li content of the clusters and compare it to cosmological predictions, to measure spreads in Li between different stellar populations, and to study signs of extra depletion in these giants.}
   {We use GIRAFFE spectra to measure the lithium and sodium abundances of lower red giant branch stars in 5 globular clusters. These cover an extensive range in metallicity, from [Fe/H]$\sim-0.7$ to [Fe/H]$\sim-2.3$ dex.}
   {We find that the lithium abundance in these lower red giant branch stars forms a plateau, with values from $\mathrm{A(Li)_{NLTE}}=0.84$ to $1.03$ dex, showing no clear correlation with metallicity. When using stellar evolutionary models to calculate the primordial abundance of these clusters, we recover values $\mathrm{A(Li)_{NLTE}}=2.1-2.3$ dex, consistent with the constant value observed in warm metal-poor halo stars, the Spite plateau. Additionally, we find no difference in the lithium abundance of first and second population stars in each cluster. We also report the discovery of a Li-rich giant in the cluster NGC3201, with $\mathrm{A(Li)_{NLTE}}=1.63\pm0.18$ dex, where the enrichment mechanism is probably pollution from external sources.}
   {}

   \titlerunning{Li in the LRGB of 5 Galactic globular clusters}
   \maketitle
%

\section{Introduction}

Lithium (Li) is one of the few elements that are produced minutes after the Big Bang during the Big Bang Nucleosynthesis phase \citep[BBN,][]{Coc2014}. Predictions of the production of elements in this theory are only dependent on the baryon-to-photon ratio, a number that has been measured from the cosmic microwave background by WMAP \citep{WMAP} and Planck \citep{Planck2014}. Then, the predicted amount of lithium formed in the early universe is A(Li)\footnote{$A(Li)=\log(n_{Li}/n_{H})+12$}$=2.69\pm0.03$ \citep{Coc2014}.
The primitive lithium abundance, to be compared with predictions from the BBN, is measured from old, metal-poor halo stars. This value is expected to be indicative of the primordial abundance, since stars this metal-poor do not have enough time at formation to be enriched with material from the interstellar medium or galactic sources producing Li \citep[e.g.][]{Prantzos2012}. \citet{Spitespite1982b, Spitespite1982a} found that halo dwarf stars with [Fe/H] between $-2.4$ and $-1.4$ dex and effective temperatures between $5700$ and $6300$ K share a similar abundance of A(Li)=2.2 dex, the so-called {\it Spite Plateau}, a result that has been confirmed in the halo \citep[e.g.][]{CharbonnelPrimas2005,Melendez2010} and in other environments, such as globular clusters \citep[e.g.][]{Bonifacio2002}. The discrepancy between the predicted A(Li) from the BBN and the measurements, over 3 times lower, is referred to as the {\it Cosmic Lithium Problem}. Notice that Lithium is the only measured element produced during the BBN that experiences such a discrepancy \citep{Coc2014}.

Moreover, to further complicate the picture, there is a decrease in the mean Li abundance and an increase in scatter at the lower-end metallicity of the Spite plateau, for metallicities [Fe/H]$<-2.8$, known as the {\it meltdown} \citep[][and references therein]{Sbordone2010}, although some puzzling very metal-poor stars have been found with higher Li abundances, closer to the plateau \citep[e.g.][]{Bonifacio2018,Aguado2019}.

Solutions for the discrepancy between Li measurements and predictions from BBN range from modifications to the BBN theory to processes affecting the stellar interiors and changing the Li abundance in old stars \citep[see][for a review]{Fields2011}.

Lithium is an element often used as an indicator of chemical processes affecting the interior of stars, such as mixing, since it burns at temperatures ($2.5\times10^6$ K) and densities found in main-sequence stars. Thus, it is possible that a process such as diffusion \citep{Fu2015} or additional turbulent mixing \citep{Richard2005} is depleting the abundance in the stellar atmospheres of old stars which would not be indicative of the BBN lithium.

Metal-poor globular clusters are among the oldest objects in the Galaxy \citep[e.g.][]{DeAngeli2005}. As such, their lithium should resemble closely the abundance produced during BBN, making these systems important probes and tools to study the Cosmic Lithium Problem. 
However, the measurement of Li abundance requires high quality spectra not easily obtained for the majority of cluster main sequence stars that are too faint. Thus, the Li abundance is known for dwarfs only for a handful of galactic clusters: M4 \citep{Mucciarelli2011, Monaco2012}, NGC6397 \citep{Lind2009, GH2009},  NGC 7099 \citep{Gruyters2016}, NGC6752 \citep{Pasquini2005,Shen2010}, 47 Tuc \citep{DOrazi2010,Dobrovolskas2014}, Omega Centauri \citep{Monaco2010}, and M92 \citep{Bonifacio2002}. In most of these clusters, the Li measured closely resembles that of the Spite Plateau, with the exception of 47 Tuc.

Additionally, globular clusters, once thought to be defined chemically by a single population of stars, with no dispersion in chemical abundances are now known to harbor populations with different light element abundances \citep{BastianLardo18}. The second population of stars, which according to current scenarios would be born from the processed material from the first population, have high [Na/Fe] and low [O/Fe] producing the observed Sodium-Oxygen anticorrelation in globular clusters \citep{Carretta2009a,Carretta2009b}. Given that the thermonuclear reactions that produce this pattern occur at higher temperatures than that required to burn Li, it is expected that a second population of stars should have a lower Li than the first population. However, only two clusters, with Li measured in their main sequence, show a hint of a Li-O correlation, 47 Tuc \citep{Dobrovolskas2014}, with no Li-Na anticorrelation, and NGC6752 \citep{Shen2010}. In M4 there is a weak but statistically significant Li-Na anti-correlation \citep{Monaco2012}, while other clusters are shown to have similar Li in first and second population stars. The lack of a Li anticorrelation could be produced if the polluter of the second population has a significant Li production or if the material from the polluter is mixed with unprocessed material that preserved its initial lithium. Thus, studying the Li in globular clusters can aid to understand their formation.

The lack of more Li measurements in main sequence stars of globular clusters due to their faintness encourages the use of a complementary method, proposed by \citet{Mucciarelli2012}, which uses lower red giant branch stars (LRGB). 

Red giant stars undergo a series of structural changes that produce alterations to their surface chemical abundances. The first of these processes is the {\it first dredge-up} (FDU), where the surface convective envelope of the stars deepens in mass, mixing material from the surface with the chemically processed interior. This translates into a decrease in the carbon and lithium abundances and an increase in the nitrogen abundance.

Standard stellar evolutionary models predict no other surface abundance changes in the red giant branch (RGB) after the end of the FDU. However, observations provide evidence of modified Li, C, N, O abundances and C isotopic ratio after the RGB bump \citep{Gratton2000}. At this moment in stellar evolution, the advancing hydrogen-burning shell encounters and erases the discontinuity left in the chemical profile of the star by the deepest penetration of the convective envelope \citep{DenissenkovVandenberg2003}, allowing extra-mixing to proceed  \citep[or do so more efficiently, e.g.][]{Chaname2005} bringing material from the stellar interior to the surface. The details of how this mechanism acts and how it affects the stellar interiors are, however, not well understood.

LRGBs are located between the end of the first dredge-up and the luminosity function bump. The dilution of lithium during the FDU at the beginning of the red giant phase is mass and metallicity dependent, but it is well characterized by stellar evolutionary models. This is why a complementary way to study Li in old stars is to measure its abundance in LRGBs, where the A(Li) is constant at a given metallicity, mirroring the Spite-plateau but at a lower value of A(Li)$\sim0.9-1.0$ dex that considers its depletion in the FDU.
Moreover, the FDU mitigates the effects of diffusion, one of the main uncertainties for the interpretation of the Li abundance in dwarfs \citep{Mucciarelli2011}.

\citet{Mucciarelli2014} used this technique to study the primordial Li abundance of the globular cluster M54 located in the Sagittarius dwarf Spheroidal galaxy, providing evidence that the primordial Li is the same there as in the Milky Way and thus, that the {\it Cosmic lithium problem} is Universal and not local. More evidence for this can be found in $\omega$ Cen, usually considered to be the remnant core of an accreted galaxy \citep[e.g.][]{Lee1999, Pancino2000}, that also shows a consistent Li abundance with the Spite plateau \citep{Monaco2010}. The discovery of Gaia-Enceladus, a disrupted dwarf galaxy that was once accreted by the Milky Way, and is now forming part of the galactic halo, allows a new way to study the primordial Li content outside our Galaxy, confirming once again, the universality of the cosmic Li problem \citep{Molaro2020, Simpson2020}.

Confirming that the LRGB stars can also be used to study the formation of globular clusters, \citet{Mucciarelli2018} measure Li in LRGB stars of $\omega$ Cen, finding an extended Na-Li anticorrelation. However, this distribution seems to be rather complex, with the most metal-rich stars in the cluster always showing low Li abundances, but the metal-poor stars in the cluster can either show low sodium and normal Li, or high sodium with normal or depleted Li abundances.

Thus, as demonstrated by these works, the study of LRGBs allows to characterize the Li abundance pattern of clusters and the primordial Li in systems where dwarfs are too faint. Following this complementary approach, in this work we study the Li abundance of lower red giant branch stars of 5 Galactic globular clusters, providing new insight in the dependence of the RGB Li plateau with metallicity and its use to calculate the primordial Li abundance in these systems.

Moreover, one of these clusters, NGC6838 is a metal-rich globular cluster. The low Li abundance of dwarfs in the relatively metal-rich cluster 47 Tuc ([Fe/H]=-0.8 dex and A(Li)=1.4-2.2 dex, \citealt{Dobrovolskas2014}) when compared to the Spite plateau suggest that there is a depletion mechanism acting in the main sequence at higher metallicities, that is not found at lower metallicities, given that M4, with [Fe/H]=-1.1 dex shows Li consistent with the Spite plateau \citep{Monaco2012}. The study of NGC6838 will allow us to test if this is a peculiar pattern for 47 Tuc or if all metal-rich globular clusters are Li depleted.

Also notice that NGC3201 is significantly younger ($\sim 2$ Gyr) than the rest of the studied clusters \citep{MarinFranch2009}.

In Section \ref{sec:obs} we report the observations and evaluate membership of our targets to the globular clusters. We measure atmospheric parameters (Section \ref{Sec:Atm}) and lithium and sodium abundances (Section \ref{sec:abund}) of these stars. We report results on the LRGB Li plateau, on the lack of a Li-Na correlation in these clusters, and on the discovery of a new Li-rich giant in NGC3201 in Section \ref{sec:results}. Our summary can be found in Section \ref{sec:summary}.

\section{Observations and membership} \label{sec:obs}
We selected five clusters in order to cover the entire metallicity range of Halo globular clusters, and observe a large number of stars. We end up with NGC4590 (M 68), NGC6809 (M 55), NGC6656 (M 22), NGC3201, and NGC6838 (M 71). For each of these 5 clusters we selected our targets in the LRGB phase.
The spectroscopic observations correspond to the ESO program 095.D-0735 (PI. A. Mucciarelli) and were carried out using the FLAMES multi-object spectrograph \citep{FLAMES} at the Very Large Telescope (VLT). 
The GIRAFFE fibers provide mid-resolution spectra with R$\sim 18000$.
The observations were performed in the setups HR15N, sampling the lithium resonance doublet at $\lambda\sim 6708$\ \AA, and HR12, sampling the sodium D doublet at $\lambda\sim 5890-5896$\ \AA\ for GIRAFFE.

A total of five exposures, 45 minutes each, for NGC6838, NGC6809, NGC6656, and NGC3201, and, 10 exposures for NGC4590 were taken in the HR15N setup. Only one exposure for each star was needed in the HR12 setup, as the large equivalent width of the Na doublet requires smaller signal-to-noise ratios to be measured.

The spectra were bias-subtracted, flat-fielded and wavelength-calibrated using the standard ESO pipelines\footnote{http://www.eso.org/sci/software/pipelines/}. In each exposure, some fibers were dedicated to measure spectra of the sky. These are median-combined to create a master sky, then subtracted to each of our science spectra.

Radial velocities for each individual spectra in the HR15N setup are measured using the IRAF\footnote{IRAF is distributed by the National Optical Astronomy Observatories, which are operated by the Association of Universities for Research in Astronomy, Inc., under cooperative agreement with the National Science Foundation.} task {\it fxcor}. This task uses the cross correlation method, where we use as template a synthetic spectrum from \citet{Coelho2005}, typical of a metal-poor red giant, with a resolution reduced to be similar to our spectra. The typical radial velocity precision is $\sim2-3\ \mathrm{km\ s^{-1}}$ for each HR15N spectra of each star. This value is the formal fxcor error, related to the fitted function used to calculate the velocity \citep{TonryDavis1979}. After shifting every spectrum to their rest-frame, we median-combine all spectra that correspond to a particular target to obtain an individual spectrum for each star that is later used in the analysis. By combining all the exposures we obtain an additional error in the radial velocity, corresponding to the standard deviation of different measurements for the same target. These are typically from $\sim0.6-2.2\ \mathrm{km\ s^{-1}}$.
Signal-to-noise ratios per pixel (S/N) in the HR15N setup are typically around $\sim70-300$, while the single exposure for the HR12 allows to obtain spectra with S/N$\sim20-70$. Given the lower S/N of these spectra and the fewer number of lines clearly visible, we do not measure a radial velocity from this setup, but instead assume the average radial velocity measured from the HR15N spectra.

After obtaining a unique radial velocity for each target star, we use these values to construct radial velocity distributions in each cluster. These are fitted with a Gaussian profile, where a mean and a standard deviation are calculated and used as criteria for membership.

\subsection{Cluster membership}

For each cluster, we exclude stars with radial velocities significantly different with respect to the mean radial velocity of the sample stars (a difference higher than $3\sigma$). Additionally, we use the membership probability reported for each star in these clusters by \citet{VasilievBaumgardt2021} which makes use of the Early Data Release 3 for the Gaia mission \citep{GaiaDR2}, considering that stars are members if they have a membership probability $\mathrm{P_{mem}}>0.9$,

The cluster NGC6838 is in a particularly contaminated field. With a gaussian distribution we get the mean radial velocity of the cluster and remove all the stars outside $2\sigma$ as field contaminants. Only 35 out of 117 observed stars are within that radial velocity range and have astrometric parameters consistent with the cluster.

In NGC6809, from the originally observed 110 stars, 95 remain after excluding stars by their membership probability or radial velocity. In NGC6656, 101 of the 112 observed stars are consistent with being cluster members. In NGC4590, we kept 50 of 69 observed stars, and in NGC3201, 98 out of the 117 observed stars are consistent with the cluster membership.

\begin{table}
\caption{Mean radial velocity (RV) for each cluster.}              
\label{table:RVs}
\centering                                      
\begin{tabular}{c c c c c}          
\hline\hline                        
Cluster & Mean RV & SD & Harris RV & Harris SD\\    
   & ($\mathrm{km\ s^{-1}}$) & ($\mathrm{km\ s^{-1}}$) & ($\mathrm{km\ s^{-1}}$) &  ($\mathrm{km\ s^{-1}}$)\\
\hline
    NGC4590 & $-94.2$ & $3.2$ & $-94.7$ & $2.5$\\
    NGC6809 & $174.9$ & $4.6$ & $174.7$ & $4.0$\\
    NGC6656 & $-146.5$ & $7.8$ & $-146.3$ & $7.8$\\
    NGC3201 & $495.0$ & $3.8$ & $494.0$ & $5.0$\\
    NGC6838 & $-22.9$ & $3.5$ & $-22.8$ & $2.3$\\
\hline                                             
\end{tabular}
\tablefoot{Mean radial velocity (RV) for each cluster and standard deviation (SD) of each radial velocity distribution and comparisons with \citet[][2010 edition]{Harris1996}.}
\end{table}

All of our measured mean radial velocities are reported in Table \ref{table:RVs} and are consistent with those in the catalog by \citet[][2010 edition]{Harris1996}.

\section{Atmospheric Parameters} \label{Sec:Atm}

\begin{figure*}[!hbt]
\begin{center}
\includegraphics[width=0.8\textwidth]{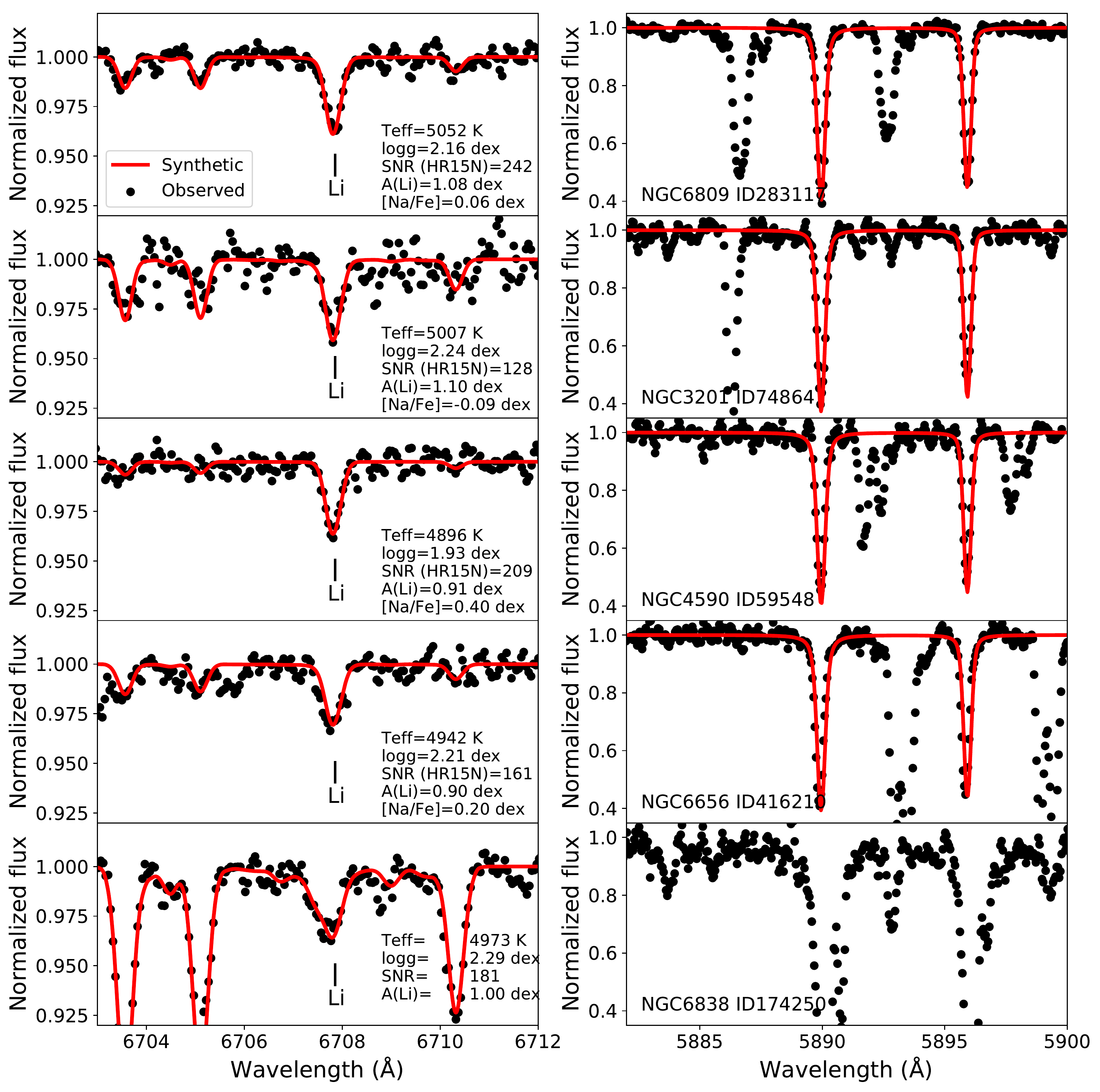}

\end{center}
\caption{Spectra from one typical sample star in each of the clusters in the region of the Li line (left) and the Na doublet (right), and the respective synthetic fits to the lines to measure the abundances. Stellar parameters, NLTE abundances, and S/N of the spectra are indicated for each star.}
\label{fits}
\end{figure*}

Effective temperatures for stars are derived photometrically, using the (V-I) colour and the \citet{Alonso1999} relations. 
For all of our clusters, we use the photometry of \citet{Stetson2019} and transform to Johnson (V-I) colors using the relation in \citet{Bessell1983}.

\begin{table}
\caption{Colour excess E(B-V) and distance modulus for each cluster.}              
\label{table:ebv_mM}
\centering                                      
\begin{tabular}{c c c c c}          
\hline\hline                        
Cluster & E(B-V) &  Source\tablefootmark{a} & $\mathrm{(m-M)_0}$ & Source\tablefootmark{b}\\    
 - & - & - & mag & -\\
\hline
    NGC4590 & $0.06$ & SF11 & $15.00$ & K+15\\
    NGC6809 & $0.12$ & SF11 & $13.95$ & VBD18\\
    NGC6656* & $0.33$ & S+98 & $13.60$ & H10\\
    NGC3201* & $0.24$ & B+13 & $14.20$ & H10\\
    NGC6838 & $0.28$ & SF11 & $13.80$ & H10\\
\hline                                             
\end{tabular}
\tablefoot{
Adopted colour excess E(B-V) and distance modulus adopted for each cluster. Clusters marked with * were corrected by differential reddening.\\
\tablefoottext{a}{SF11: \citet{SandF2011}; S+98: \citet{Schlegel1998}; B+13: \citet{Bonatto2013}.\\}
\tablefoottext{b}{VBD18: \citet{VandenBergDenissenkov2018}; K+15: \citet{Kains2015}; H10: \citet[][2010 edition]{Harris1996}.}
}
\end{table}

To calculate dereddened colors, we use extinction coefficients from \citet{McCall2004}.
The adopted colour excess E(B-V) and distance modulus for each cluster can be found in Table \ref{table:ebv_mM}.

Although the color excess of NGC6838 is high, it does not suffer from significant differential reddening \citep[$\langle \delta E(B-V)\rangle=0.035$ mag,][]{Bonatto2013}.
For the clusters NGC6656 and NGC3201 we corrected for differential reddening using the maps of \citet{AlonsoGarcia2012} with zero point E(B-V)$=0.33$ \citep{Schlegel1998} for NGC6656, as suggested by that work, and of Pancino et al., (in preparation), with zero point E(B-V)=$0.24$ for NGC3201 \citep{Bonatto2013}.

The surface gravity was calculated for each star using isochrone fitting and a set of MIST isochrones \citep{MIST0,MIST1} restricting ages to be higher than 12 Gyr. In particular, we placed the star in an effective temperature - absolute magnitude plane and compare its position there with the theoretical isochrones, using the estimated metallicity of the cluster by \citet[][2010 edition]{Harris1996} to prevent degeneracy. We build a probability distribution with all the isochrone points in a $3\sigma$ radius from the input parameters. This method was preferred to measuring $\log g$ spectroscopically given the lack of a relevant number of Fe II lines in our spectra. Additionally, \citet{MucciarelliBonifacio2020} recommend to use photometric temperatures and gravities in low-metallicity stars of globular clusters, since spectroscopic parameters are lower than photometric determinations, and inconsistent with the position of the giants in color-magnitude diagrams.

We compared the calculated $\log g$ with $\log g$ estimated using bolometric luminosity of the giants and find a good agreement between both methods. Uncertainties obtained are $\Delta \log g\simeq 0.2$ dex. As for the error in effective temperature, we adopt typical uncertainties of $\Delta\mathrm{T_{eff}}\simeq125$ K, that correspond to the standard deviation of the colour relation used \citep{Alonso1999}.

Microturbulence velocities are then calculated for each star using their effective temperature, $\log g$, and the relation in \citet{Bruntt2012}. The typical errors when using this relation are reported to be $0.13\,\mathrm{km\ s^{-1}}$ by the authors. However, this value depends on uncertainties in effective temperature and $\log g$, assumed to be $100$ K and $0.1$ dex respectively. Given that our uncertainties are slightly higher, we assume a typical error in microturbulence velocity of $0.15\,\mathrm{km\ s^{-1}}$.

We calculated metallicities for each star individually by measuring the equivalent width of Fe I lines using the code DAOSPEC \citep{DAOSPEC} through the wrapper 4DAO\footnote{http://www.cosmic-lab.eu/4dao/4dao.php} \citep{4DAO}. Then, the Fe abundances were derived with the code GALA \citep{GALA}.
Our Fe I line list was constructed by using the spectra of the coldest and hottest giants in the sample. We inspected visually the spectra to identify lines that were visible in both stars covering the entire effective temperature range. Although we considered initially a large list of lines, a spurious correlation between metallicity and effective temperature was identified in some of the clusters, with an extremely strong correlation in NGC3201. We only selected lines that did not saturate in any cluster, and as such were good indicators of the real metallicity of the star. Based on this test, we ended up selecting 6 lines with a linear correlation between equivalent widths and temperatures to only estimate the metallicity, and used it as an additional criteria of membership. We then calculated the mean metallicity of the sample and adopted that value for all stars of the cluster when we determine chemical abundances.
We estimated the NLTE corrections for our selected Fe I lines, that could be relevant for metal-poor stars \citep{Bergemann2012}. Only two of the used lines have NLTE corrections reported by \citet{Mashonkina2016}; in the temperature and $\log g$ ranges of NGC4590 stars, they amount to $\sim0.072-0.085$ dex. These are the maximum values expected for our sample of stars, because NLTE corrections are larger for lower metallicities. We find no way to apply consistently Fe NLTE corrections, but these seem to be smaller than the reported uncertainty in metallicity, and would not affect significantly the measured abundances.

\begin{figure*}[!hbt]
\begin{center}
\includegraphics[width=0.6\textwidth]{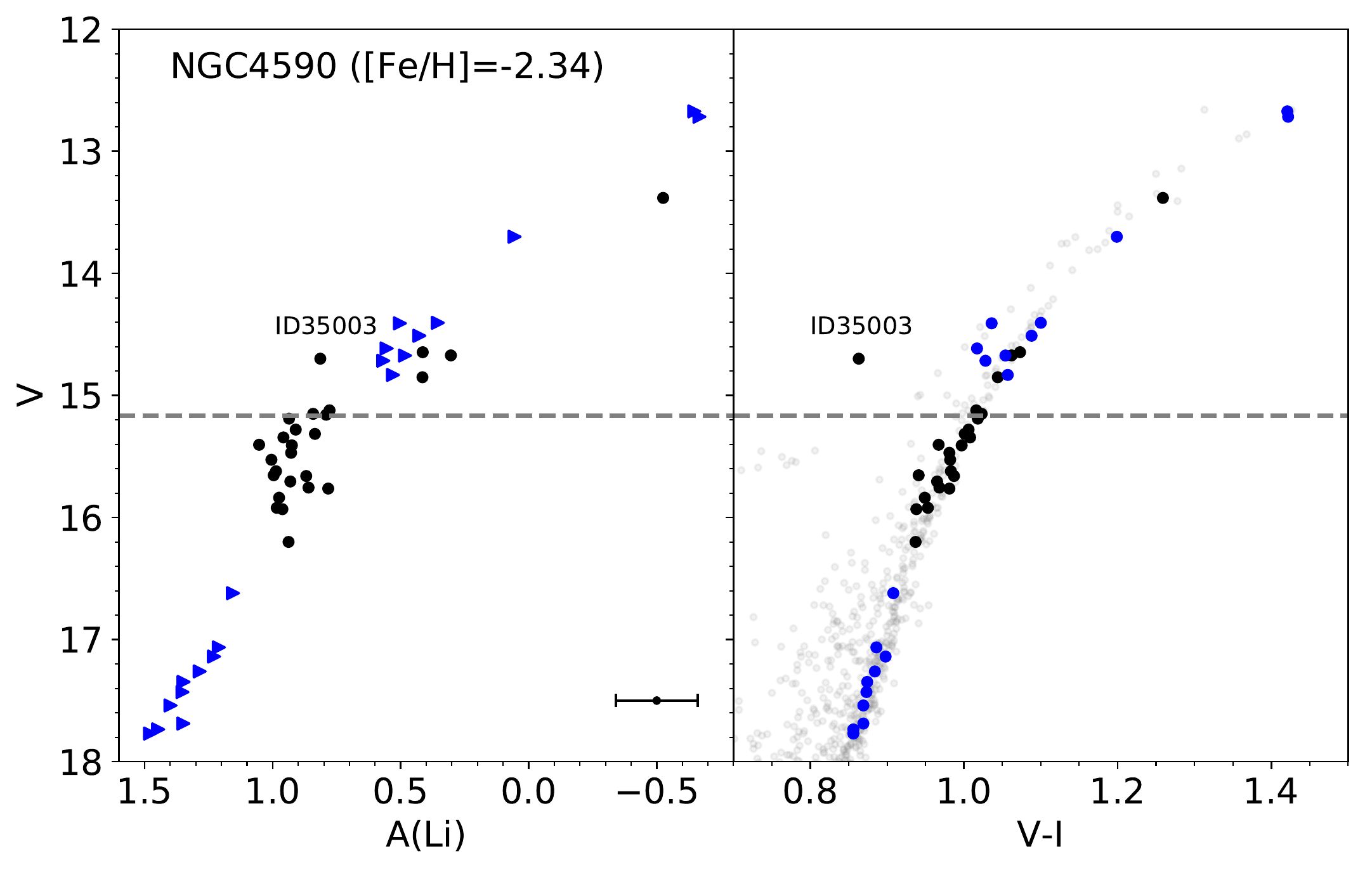}
\includegraphics[width=0.6\textwidth]{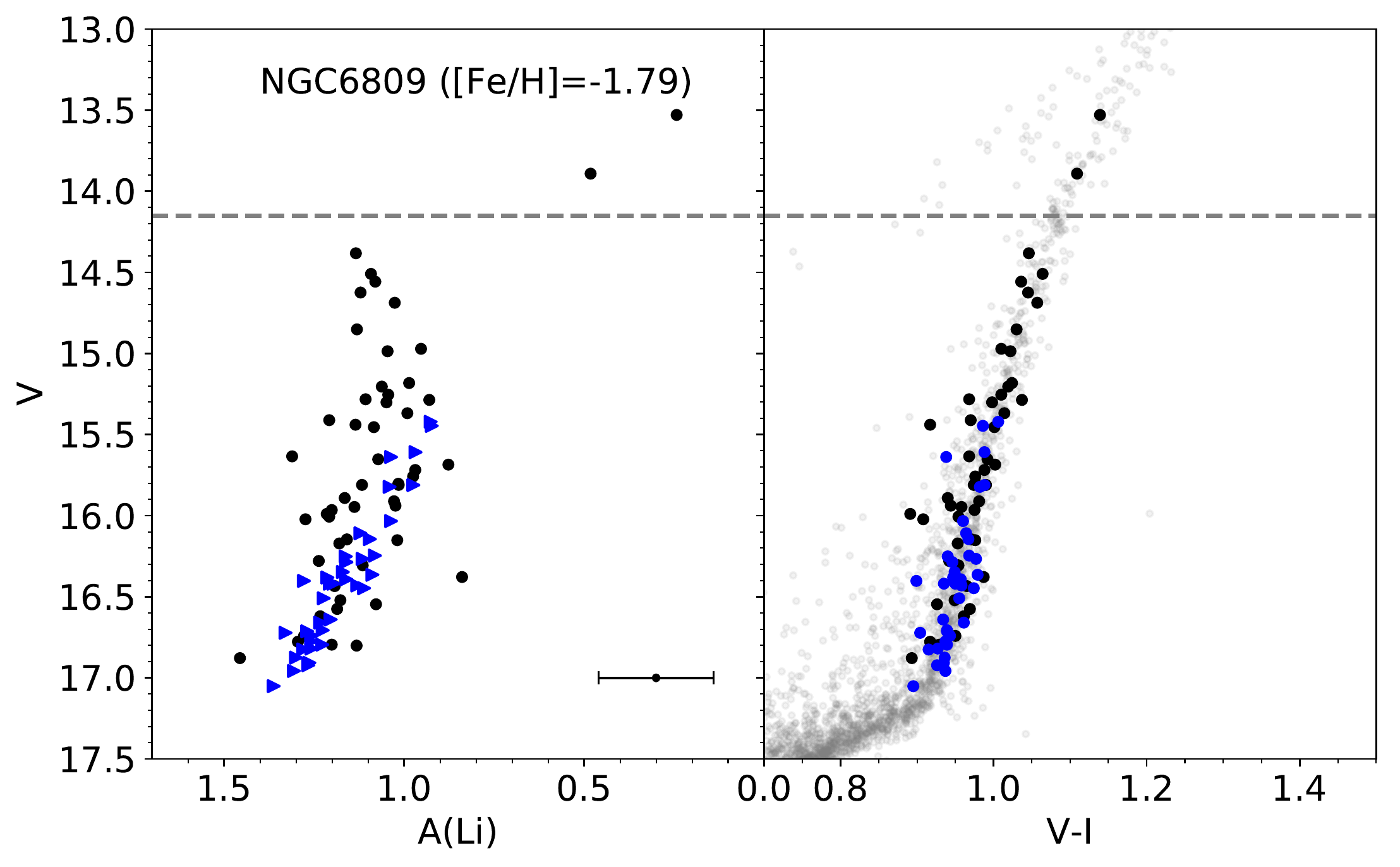}
\includegraphics[width=0.6\textwidth]{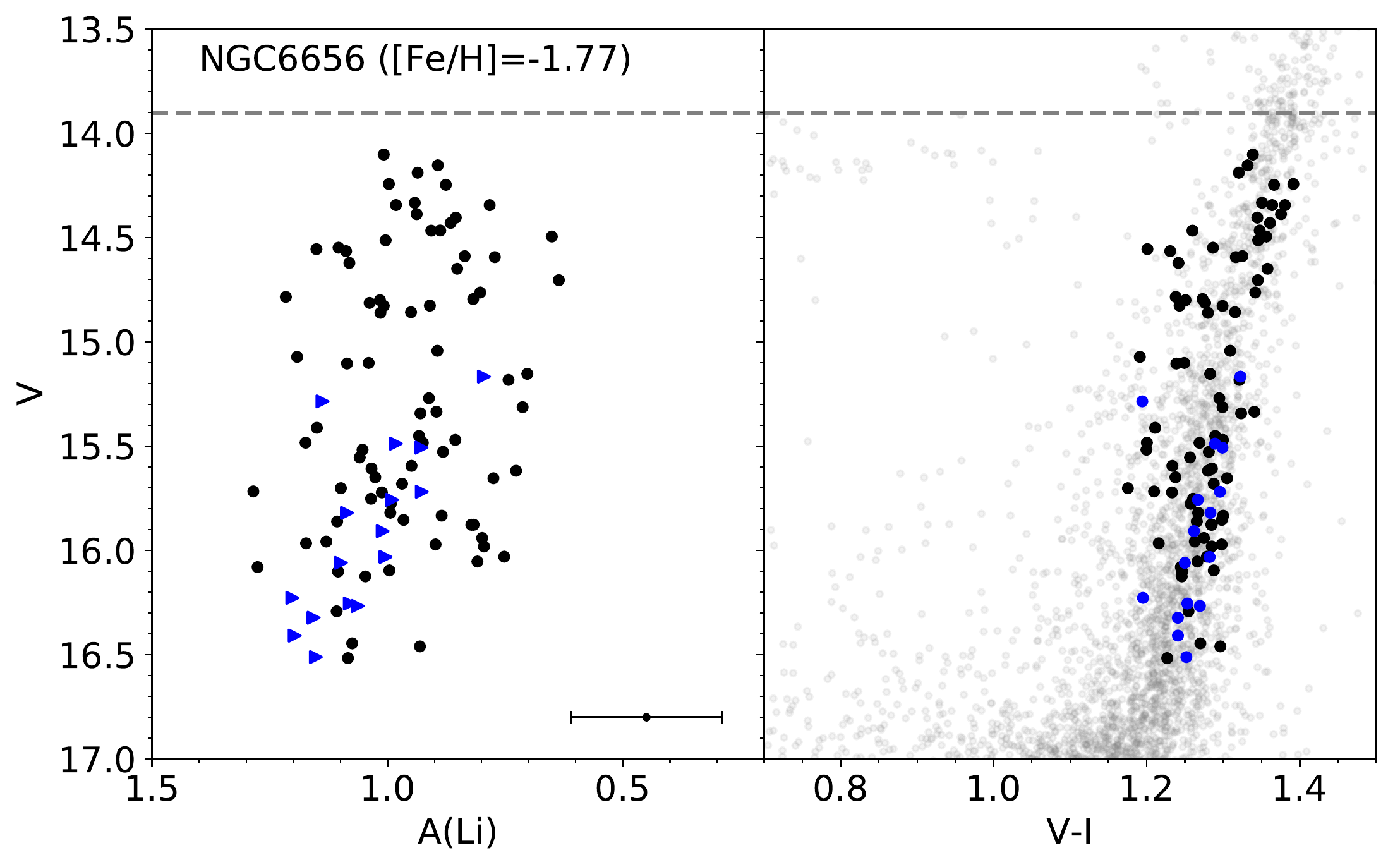}
\end{center}
\caption{Lithium abundances of NGC4590 (top panels), NGC6809 (middle panels), and NGC6656 (bottom panels) as a function of V magnitude (left panels) alongside their respective color-magnitude diagrams (right panels). Blue symbols are Li upper limits. The approximate position of the luminosity function bump in each cluster is marked with a dashed line.}
\label{Li_vs_V_1}
\end{figure*}

\begin{figure*}[!hbt]
\begin{center}
\includegraphics[width=0.6\textwidth]{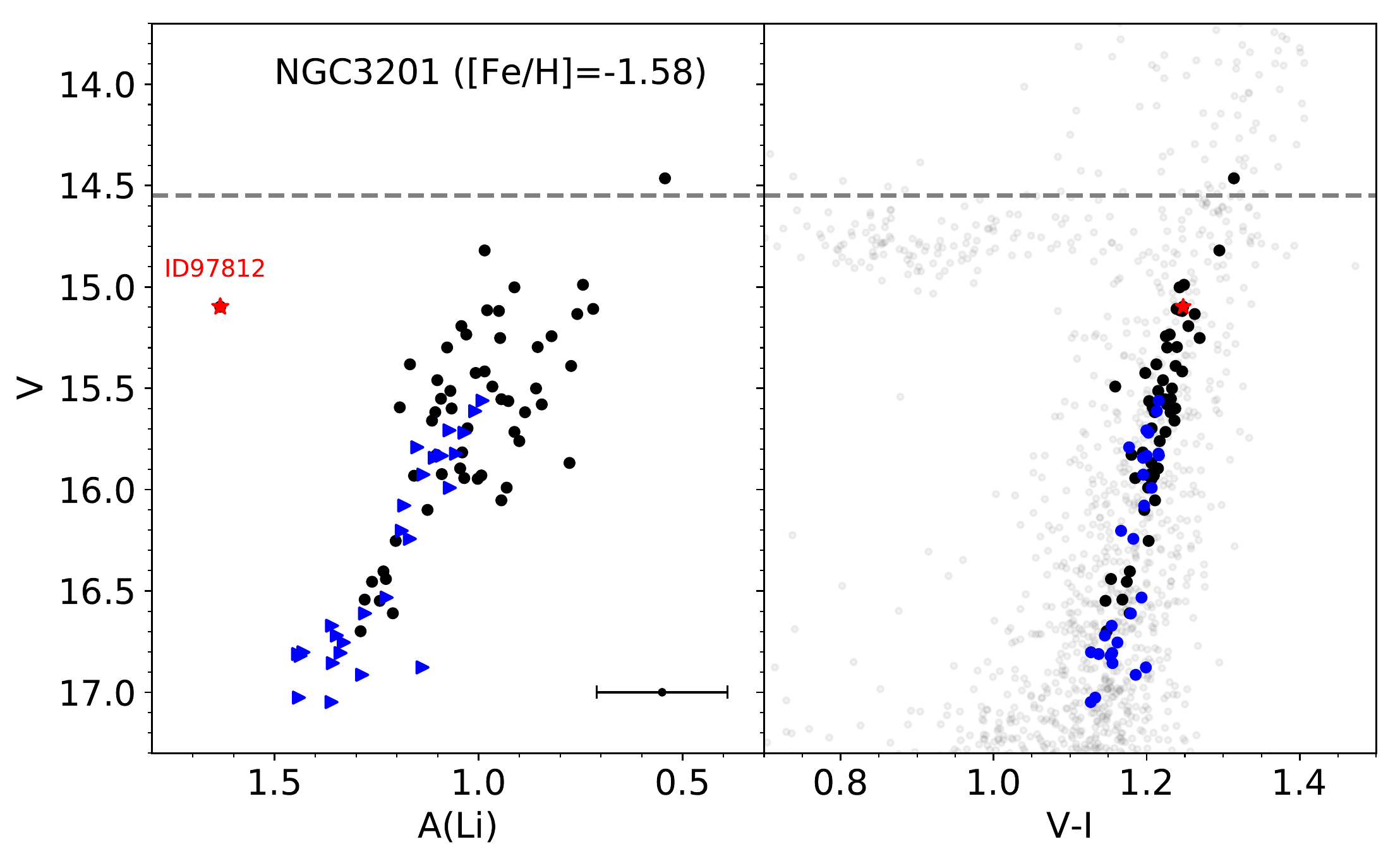}
\includegraphics[width=0.6\textwidth]{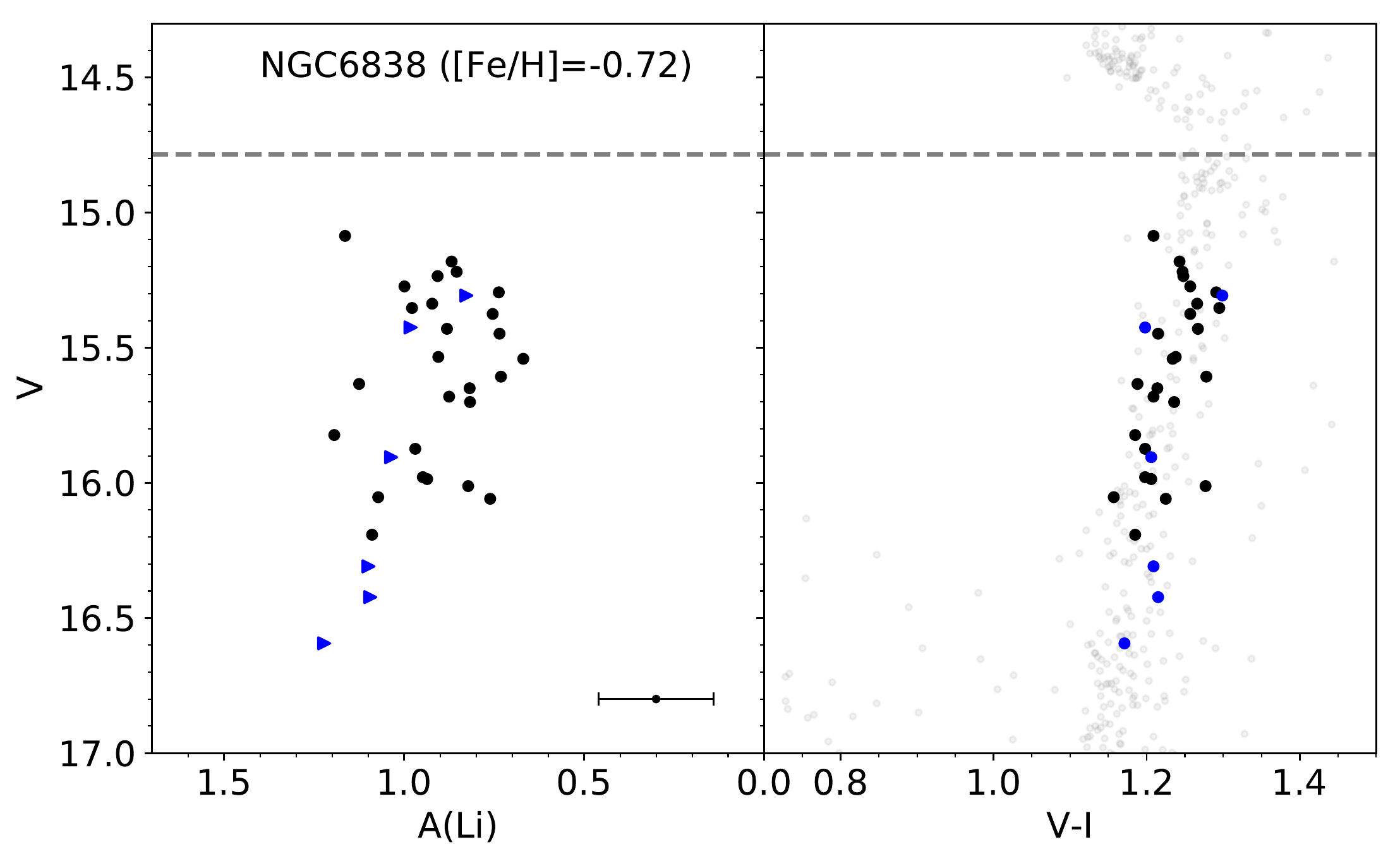}
\end{center}
\caption{Lithium abundances in NGC3201 (top panels) and NGC6838 (bottom panels) as a function of V magnitude (left panels), with their respective color-magnitude diagrams (right panels). Blue symbols are Li upper limits. The Li-rich red giant ID 97812 is marked as a red star. The dashed lines shows the position of the luminosity function bump.}
\label{Li_vs_V_2}
\end{figure*}

\begin{table}
\caption{Mean metallicity and number of members for each cluster.}              
\label{table:Metmem}
\centering                                      
\begin{tabular}{c c c c c}          
\hline\hline                        
Cluster & [Fe/H] & $\mathrm{SD}_{\mathrm{[Fe/H]}}$ & Harris [Fe/H] & Members\\    
 - & dex & dex & dex & -\\
\hline
    NGC4590 & $-2.34$ & $0.10$ & $-2.23$ & 46\\
    NGC6809 & $-1.79$ & $0.10$ & $-1.94$ & 90\\
    NGC6656 & $-1.77$ & $0.12$ & $-1.70$ & 98\\
    NGC3201 & $-1.58$ & $0.06$ & $-1.59$ & 83\\
    NGC6838 & $-0.72$ & $0.07$ & $-0.78$ & 32\\
\hline                                             
\end{tabular}
\tablefoot{Mean metallicity, standard deviation (SD), and number of members for each cluster. Metallicity values reported by \citet[][2010 edition]{Harris1996} are also included.
}
\end{table}

Table \ref{table:Metmem} shows the mean metallicity and final number of members in each cluster.

In NGC3201 we found some outliers in the metallicity distribution, which are likely non members, and were removed with sigma-clipping, using a criteria of $2\sigma$. We aim for the maximum purity of the sample rather than completeness, and thus, although we might exclude some members, this procedure increases the probability of membership by selecting stars with metallicities closer to the mean of the cluster.
Notice that while the mean metallicity of NGC3201 is similar in different literature sources, there is an ongoing debate about the existence of an intrinsic metallicity spread in the cluster \citep[][and references therein]{LA2021}. Given that we are selecting only stars with a metallicity similar to the mean of the cluster, the conclusion about the possible spread in the metallicity distribution should not change our results. The mean value of metallicity we obtain for NGC6809 is [Fe/H]$=-1.79\pm0.10$. The metallicity of this cluster seems to be controversial, with some measurements similar to the value we report \citep[e.g.][]{Kayser2008}, and others closer to the [Fe/H]$=-1.94$ value found in the Harris catalogue \citep[e.g.][]{Carretta2009}. 

The final parameters for member stars in each of the 5 clusters can be found in Table \ref{table:params}, in Appendix \ref{ap:param}.

\section{Chemical abundances} \label{sec:abund}

\subsection{Lithium}

Lithium abundances are calculated using spectral synthesis around the Li doublet at wavelength $\sim6708$ \AA. The observed spectrum was compared to synthetic spectra generated using MOOG \citep[][2018 version]{MOOG}, with ATLAS9 model atmospheres \citep{ATLAS9} and the abundance derived through $\chi^2$ minimization. The continuum level, one of the greatest uncertainties in the determination of Li abundance with this method, was set by using a region of $\sim10\,$ \AA\, around the Li line. For some of the giants where the Li line is not detected, only upper limits are reported. We estimated the detection limits on Li using the relation by \citet{Cayrel1988} for the minimum equivalent width that could be measured in each spectra, and calculate the corresponding lithium abundance for 3 times that limiting equivalent width. 

We calculated non local thermodynamical equilibrium (NLTE) corrections using the grid of \citet{Lind2009NLTE}. In NGC3201, NGC4590, and NGC6656, corrections are usually smaller than $0.1$ dex, while in NGC6809 corrections are even smaller ($<0.06$ dex). In contrast, NGC6838 has larger corrections, from $0.01$ dex to $0.15$ dex. Two stars in NGC4590 are outside the limits of the grid, and thus we use the closest grid point as the values for the Li correction.

The error in lithium is calculated by adding in quadrature the uncertainties associated to the synthetic spectra producing the best fit, something that depends greatly on the positioning of the continuum, and the propagation of errors in stellar parameters, in particular, of effective temperature that produces the largest deviations in A(Li). The typical uncertainties due to the quality of the data are of the order of $\mathrm{\Delta A(Li)}\sim0.05$ dex, which depends on the signal-to-noise ratio of the spectrum. This refers to uncertainties in the fitting procedure, including continuum placement and adjustments in the fit due to small changes in the radial velocity and line broadening.
NGC3201 is the cluster with the overall worst quality spectra, and as such it can show higher errors of up to $\mathrm{\Delta A(Li)}\sim0.08$ dex.
Uncertainties in the Li abundance arising from  the  propagation of errors in the stellar parameters are $0.10-0.17$ dex due to $\mathrm{T_{eff}}$, $0.00-0.02$ dex due to the metallicity, and $0.01-0.03$ dex because of errors in $\log g$. Errors propagated from microturbulence velocity (of $\sim0.01$ dex) are added linearly, given that microturbulence velocity depends on effective temperature and $\log g$. Our typical uncertainties are $\mathrm{\Delta A(Li)}\simeq0.16$ dex. 

As a sanity check, we calculated the Li abundance in the cluster M4 using the same spectra used in \citet{Mucciarelli2011}. The A(Li) we calculated using the same parameters as reported in that work is very similar for the RGB stars. We calculate an average difference $\mathrm{\left<\delta A(Li)\right>=0.09}$. No attempt was made to reproduce the measurements in turn-off stars, as those are out of the scope of this paper. To check that our parameter determination was also consistent with previous literature, we also re-calculated stellar parameters from photometry directly, finding similar values and A(Li), with an average difference of $\mathrm{\left<\delta A(Li)\right>=0.16}$.

\subsection{Sodium}

The sodium abundance was measured from the Na D doublet at $5890 - 5896$ \ \AA, also using spectral synthesis of the region, generating a grid of synthetic spectra using SPECTRUM \citep{SPECTRUM}, with the same ATLAS9 model atmosphere. The choice of using one code over other was only done by convenience of our available wrappers and methods. However, we have tested to see the difference between abundances using the two radiative transfer codes. Abundances of sodium change at most by $0.05$ dex over the range of parameters of our sample. The best fit was selected with $\chi^2$ minimization. The continuum placing is complicated by the low signal to noise of some of our spectra in the region. Accordingly, the errors in Na abundance consider these uncertainties. Uncertainties due to the continuum placement and quality of the fit can be as high as $0.10$ dex. The propagation of errors in the stellar parameters gives typical Na uncertainties of $\sim0.11$ dex due to $\mathrm{T_{eff}}$, $\sim0.14$ dex because of errors in $\log g$, $\sim0.01$ dex uncertainty due to metallicity, and $\sim0.02$ dex due to uncertainties in the microturbulence velocity. Typical uncertainties then are $\mathrm{\Delta A(Na)}\simeq0.18$ dex. 

We applied the Na NLTE corrections computed by \citet{Lind2011}, that can be substantial for sodium measured from the D doublet, even reaching values of $-0.5$ dex within our sample.

It was not possible to measure sodium in NGC6838 due to the presence of contamination by interstellar sodium in that region of the spectra, close to the position of the stellar Na lines.

Typical spectra from our sample in the regions of the Li line and Na doublet is shown in Figure \ref{fits}. These spectra show the quality of our data, and typical fits to the Li and Na lines to measure abundances. In the right panels, we also see the strong interstellar sodium lines.

\section{Results} \label{sec:results}

NLTE lithium and sodium abundances of the LRGB stars in the studied globular clusters are found in Table \ref{table:params}, in Appendix \ref{ap:param}. We also include LTE abundances in the online data. We show our measured Li abundances as a function of V magnitude and the position of member stars in their color magnitude diagrams in Figures \ref{Li_vs_V_1} and \ref{Li_vs_V_2}, left and right panels respectively. The magnitudes are also corrected by differential reddening in these figures. Lithium upper limits are shown as blue arrows. The position of the luminosity function bump \citep{Samus1995,Ferraro1999} is indicated for each cluster as a dashed line, and the background corresponds to the catalog from \citet{Stetson2019}, cleaned using membership probabilities by \citet{VasilievBaumgardt2021}, but without corrections by differential reddening. Lithium upper limits in all five clusters are above our reported measurements, and as such are consistent with the abundances reported.
 
 We first notice here that some of the stars show unusual positions in the color-magnitude diagram. This is because our initial selection of targets was not done using the Stetson photometry. We study each of these stars independently and follow them throughout our analysis to make sure they are not contaminating our sample and confusing our results.
 
 In NGC4590 the star ID35003 is not located in the locus of RGB stars of the cluster. This star has a proper motion, radial velocity, and metallicity consistent with NGC4590. We also check an independent measurement of its $\log g$ using bolometric corrections and find a very similar value between both determinations. However, given its color, this star has a higher temperature and higher lithium abundance (A(Li)$=0.81$ dex) than other stars in the RGB of that magnitude. This star can be clearly identified in the left panel of Figure \ref{Li_vs_V_1} with higher Li abundance. Thus, although consistent given our analysis, we consider that this star may not be part of the cluster due to its unusual position in the color-magnitude diagram and remove it from further analysis.
 
 NGC6656 shows a broad RGB even after corrections by differential reddening. The bluer sequence does not show a particular spatial location, indicating that this is probably not the effect of additional differential reddening and that the broad sequence might be due to the spatial resolution of the used reddening maps. This cluster is suspected to have an intrinsic iron spread \citep{DaCosta2009}, but see also \citet{Mucciarelli2015}.
 
 In some of the clusters we can identify the position of the end of the FDU and the RGB bump in both panels, with abrupt decreases in A(Li). In NGC4590 we can clearly identify stars going through the FDU, diluting their Li at V$\sim17$ mag, then a plateau formed by the LRGB and a second decrease in A(Li) produced by the inclusion of giants located after the luminosity function bump. Most of the upper RGB stars have only Li upper limits. Cluster NGC6809 has two stars after the RGB bump, while NGC3201 has only one, that shows a smaller Li abundance than the rest of RGB stars, at the level of the lower envelope of the Li distribution of LRGBs.
 We are able to identify the end of the FDU in NGC3201, where stars at the bottom of the RGB decrease their Li abundances. The Li dilution due to FDU in NGC6656 or NGC6838 is not clearly visible. In NGC6656, the abundance slowly decreases as we move up in the giant branch, with a large scatter, and the plateau is not as clear as in other clusters. 
 
 We also notice that there is a giant in NGC3201 (namely giant ID 97812, a red star symbol in Figure \ref{Li_vs_V_2}) with unusually high Li abundance $\mathrm{A(Li)_{NLTE}=1.63\pm0.18}$ dex. Given that it is located before the onset of efficient extra-mixing, this Li-rich giant has probably experienced pollution from an external source. We analyze it further in Section \ref{Sec:Lirich}.

\begin{figure*}[!hbt]
\begin{center}
\includegraphics[width=0.35\textwidth]{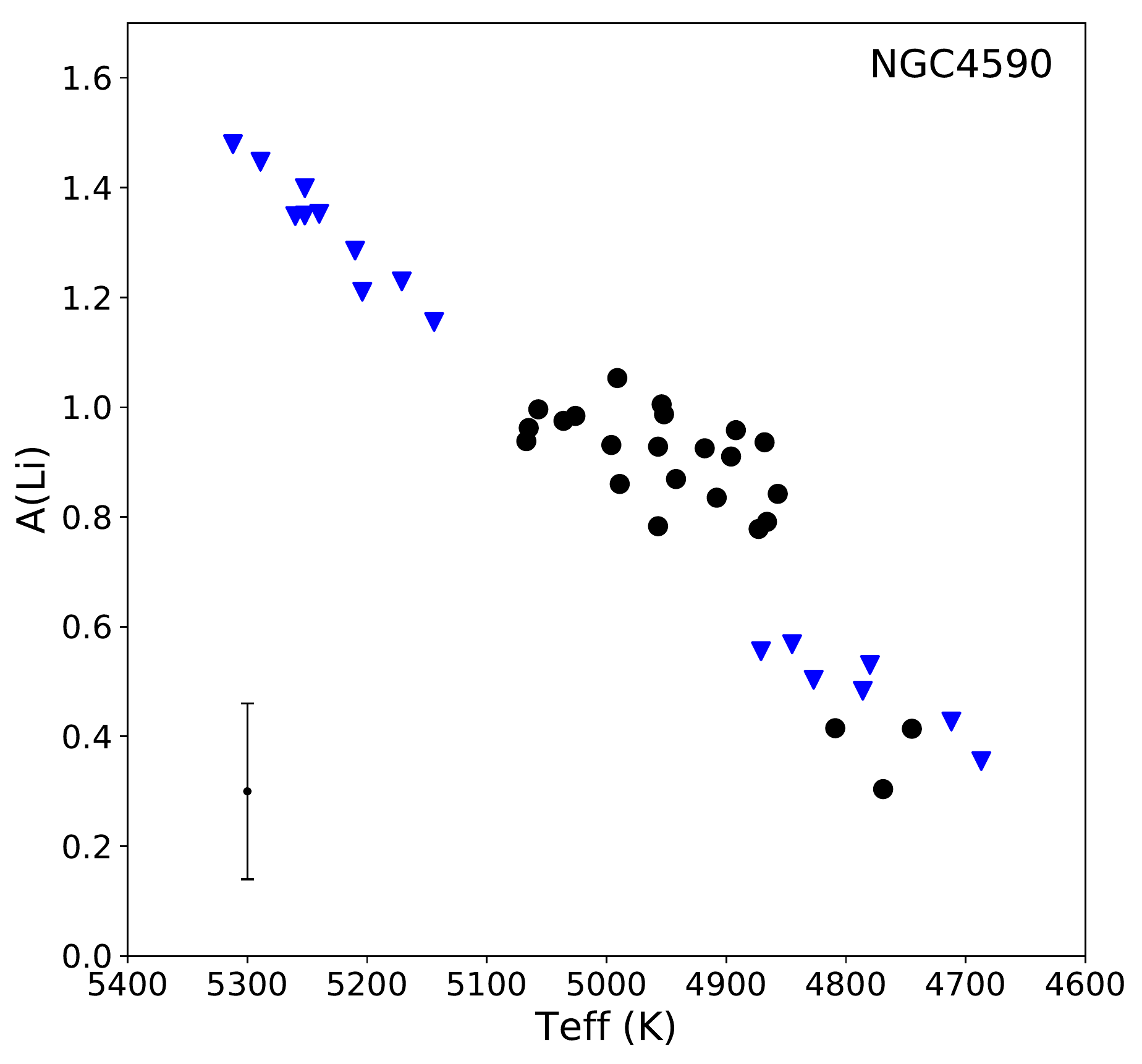}
\includegraphics[width=0.35\textwidth]{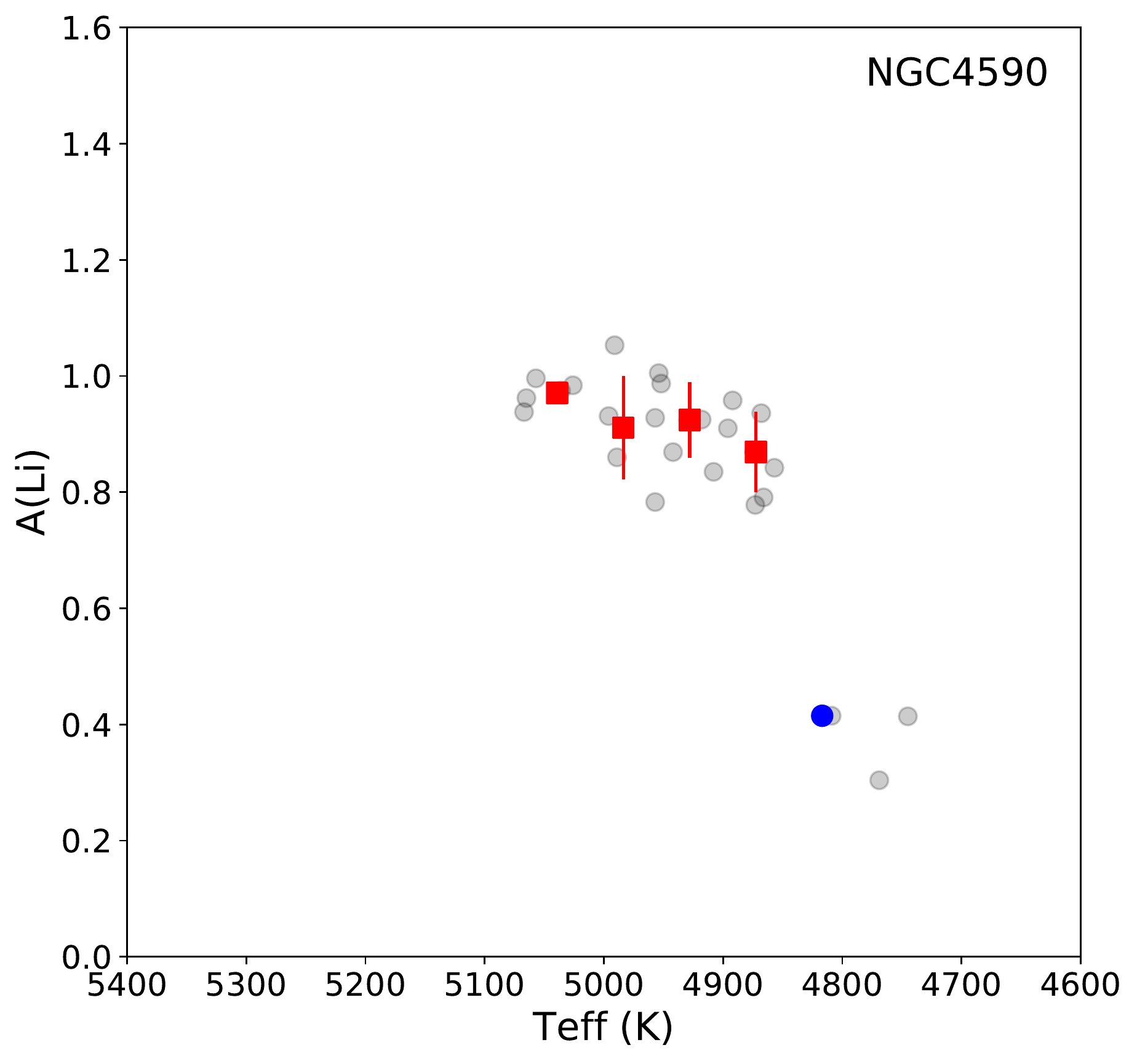}
\includegraphics[width=0.35\textwidth]{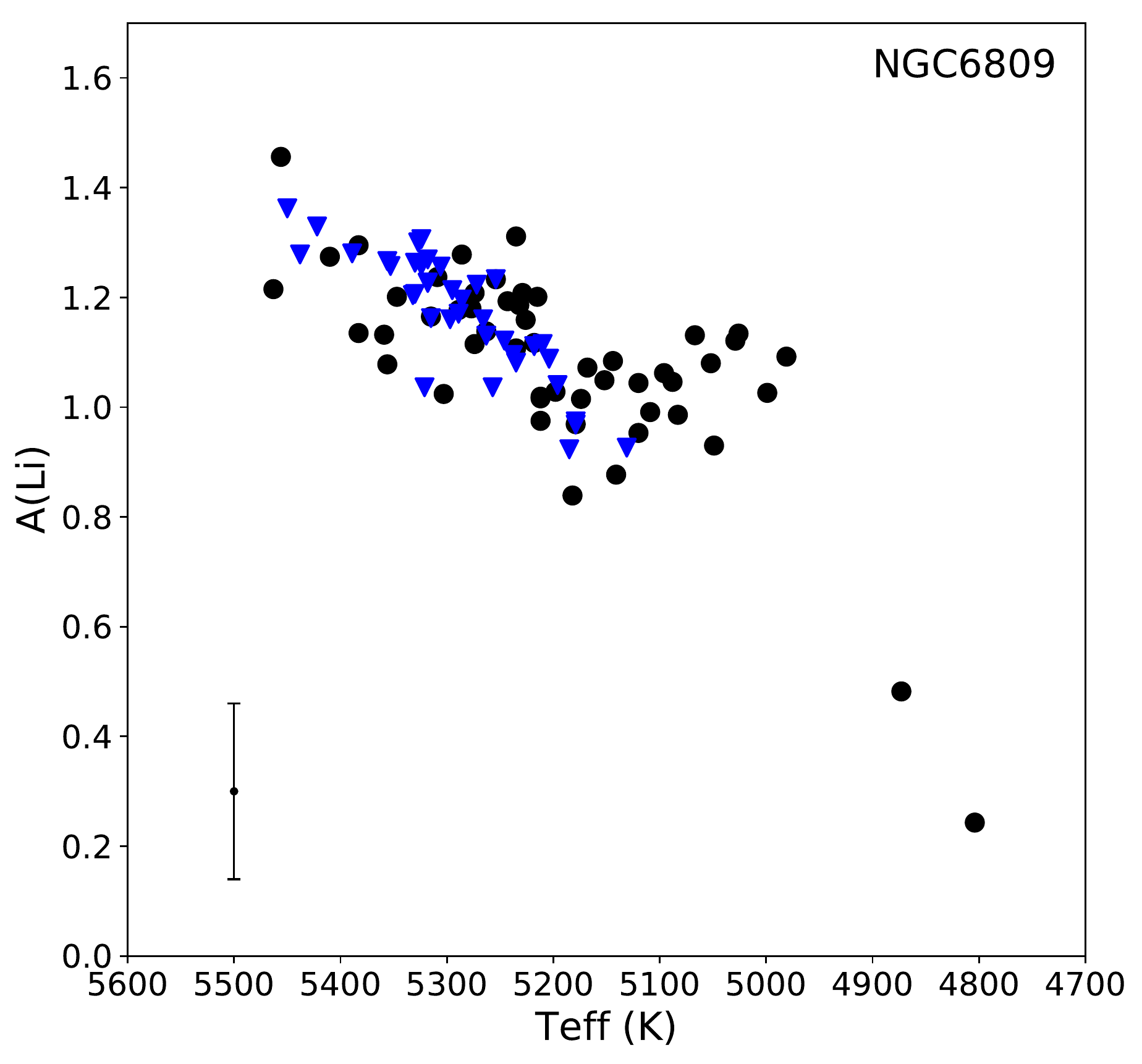}
\includegraphics[width=0.35\textwidth]{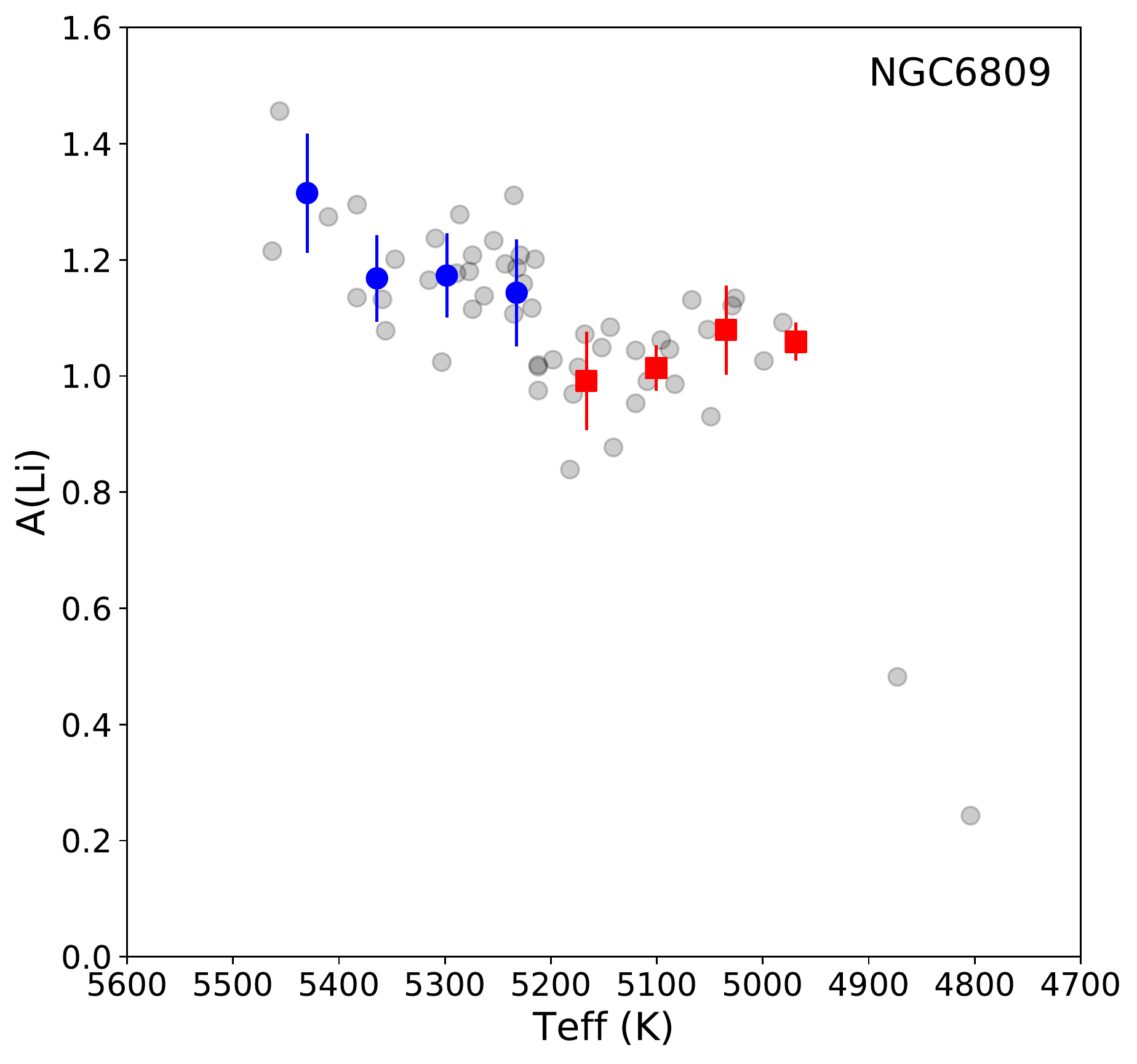}
\includegraphics[width=0.35\textwidth]{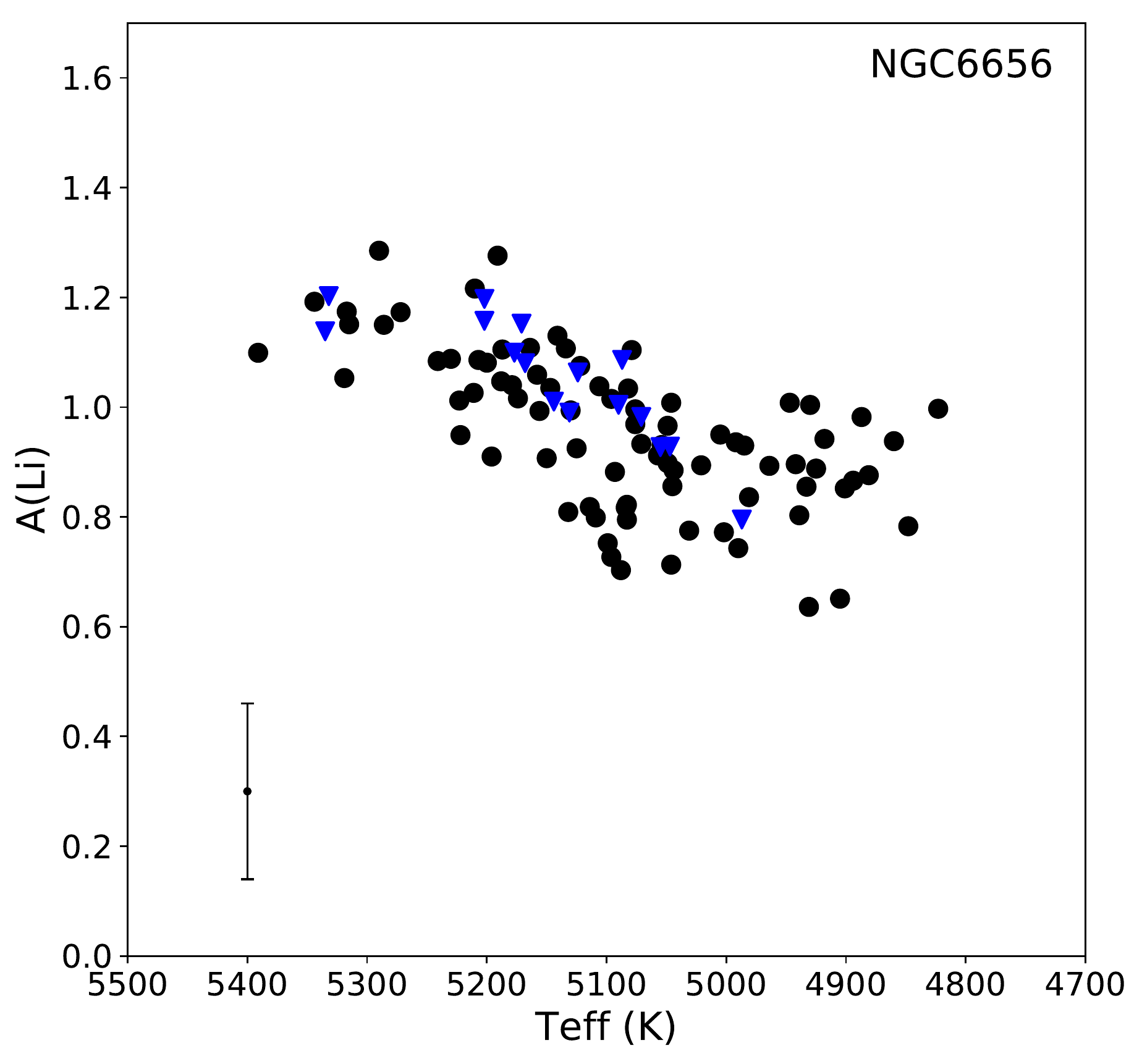}
\includegraphics[width=0.35\textwidth]{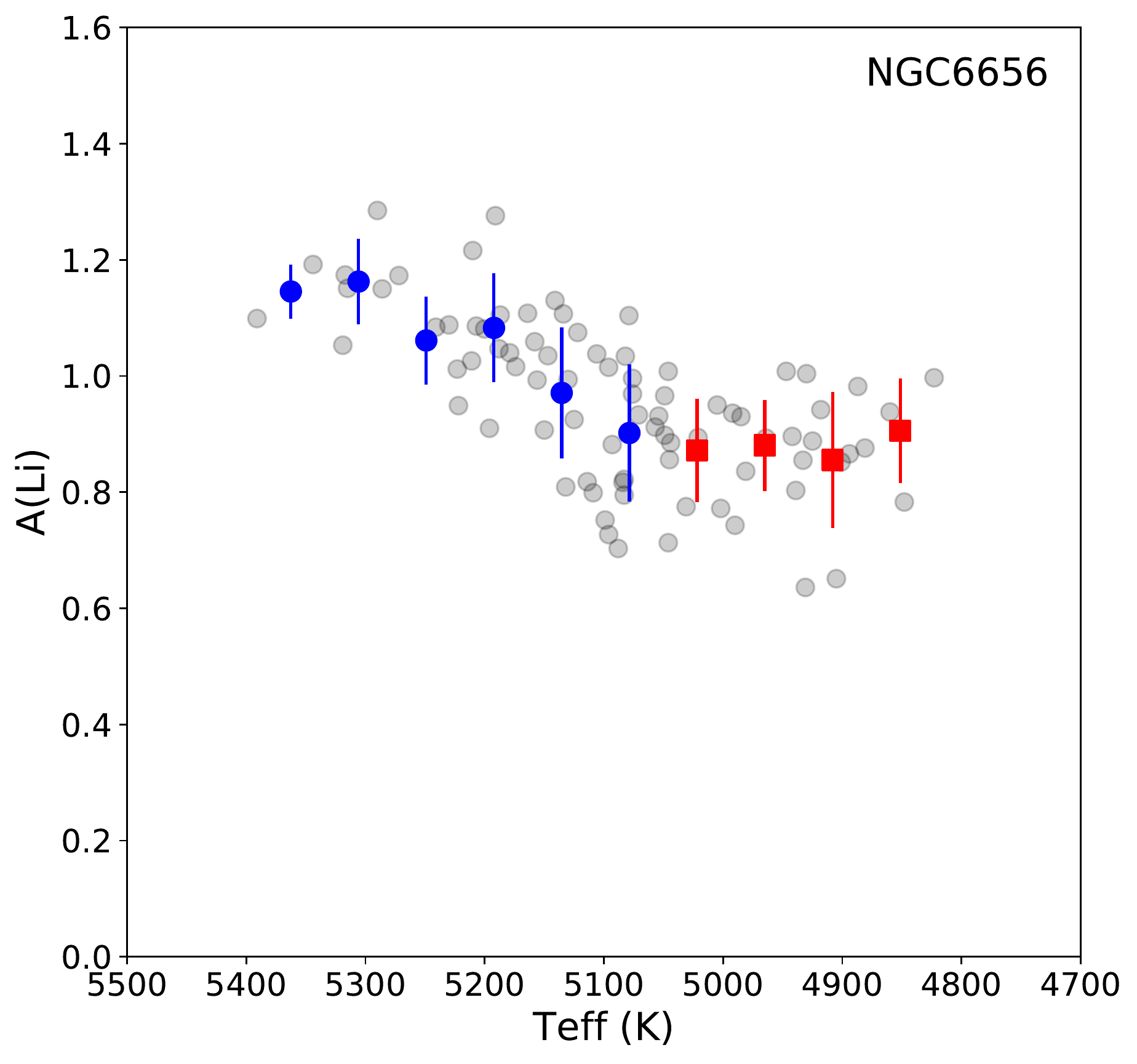}

\end{center}
\caption{Behavior of lithium abundances as a function of effective temperatures for NGC4590, NGC6809, and NGC6656, from top to bottom panels. The left panels show the abundances (black points) and upper limits (blue arrows). The right panels only show the measurements in gray and the binned Li abundances, considering equal-sized bins (blue points). Red squares mark the bins considered to be part of the Li LRGB plateau and used to calculate mean values reported.}
\label{Li_vs_Teff_1}
\end{figure*}

\begin{figure*}[!hbt]
\begin{center}
\includegraphics[width=0.35\textwidth]{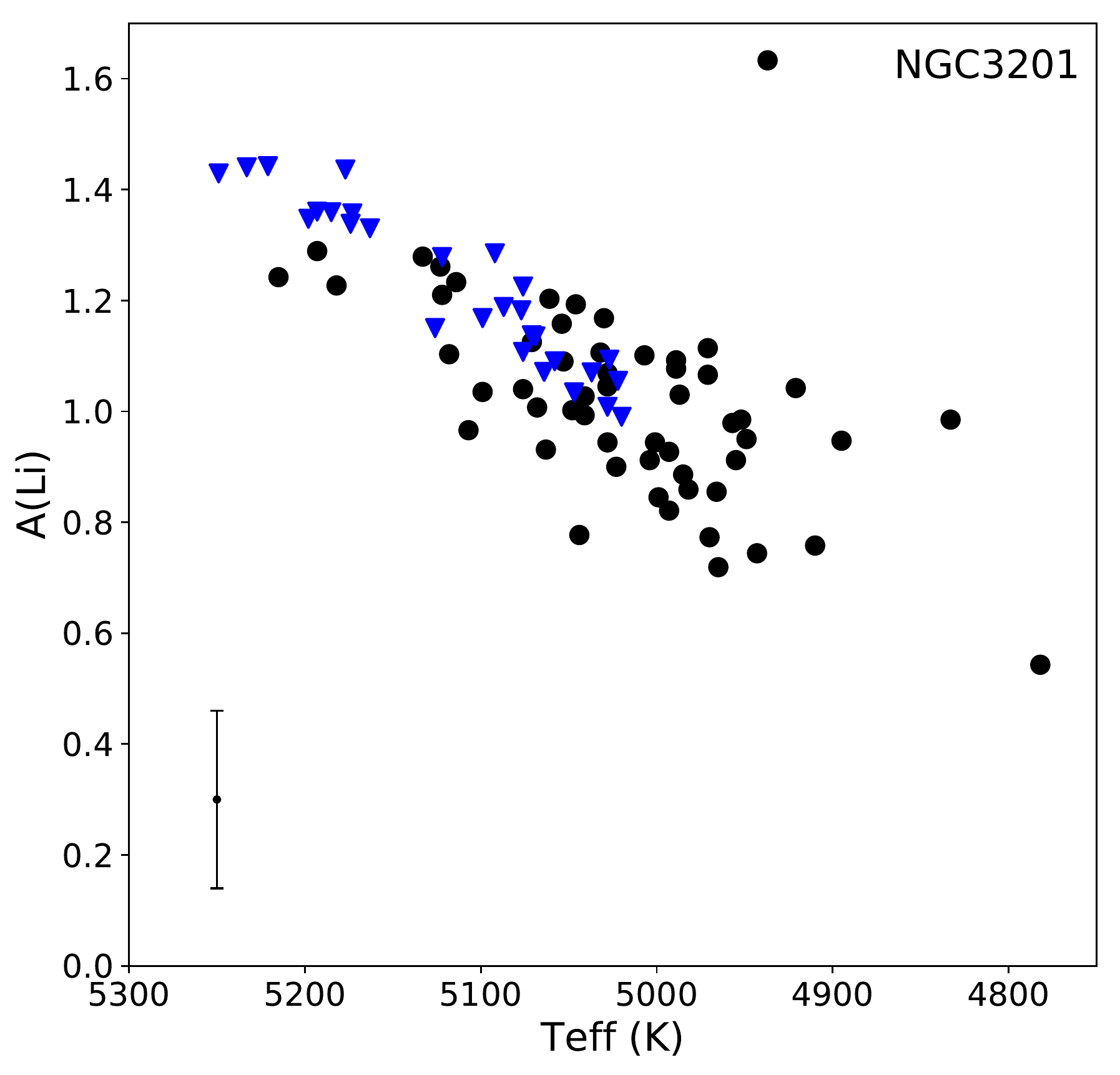}
\includegraphics[width=0.35\textwidth]{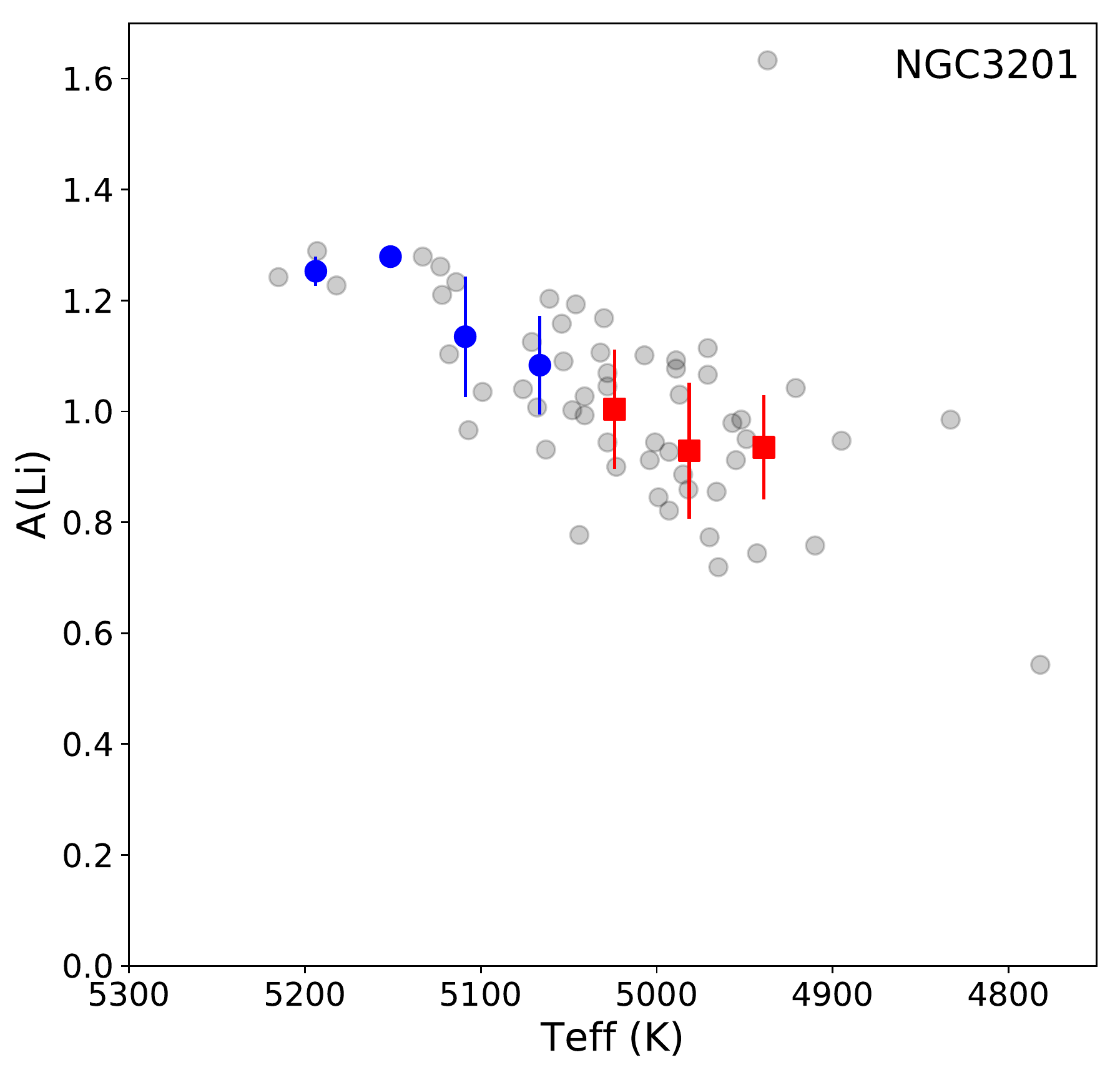}
\includegraphics[width=0.35\textwidth]{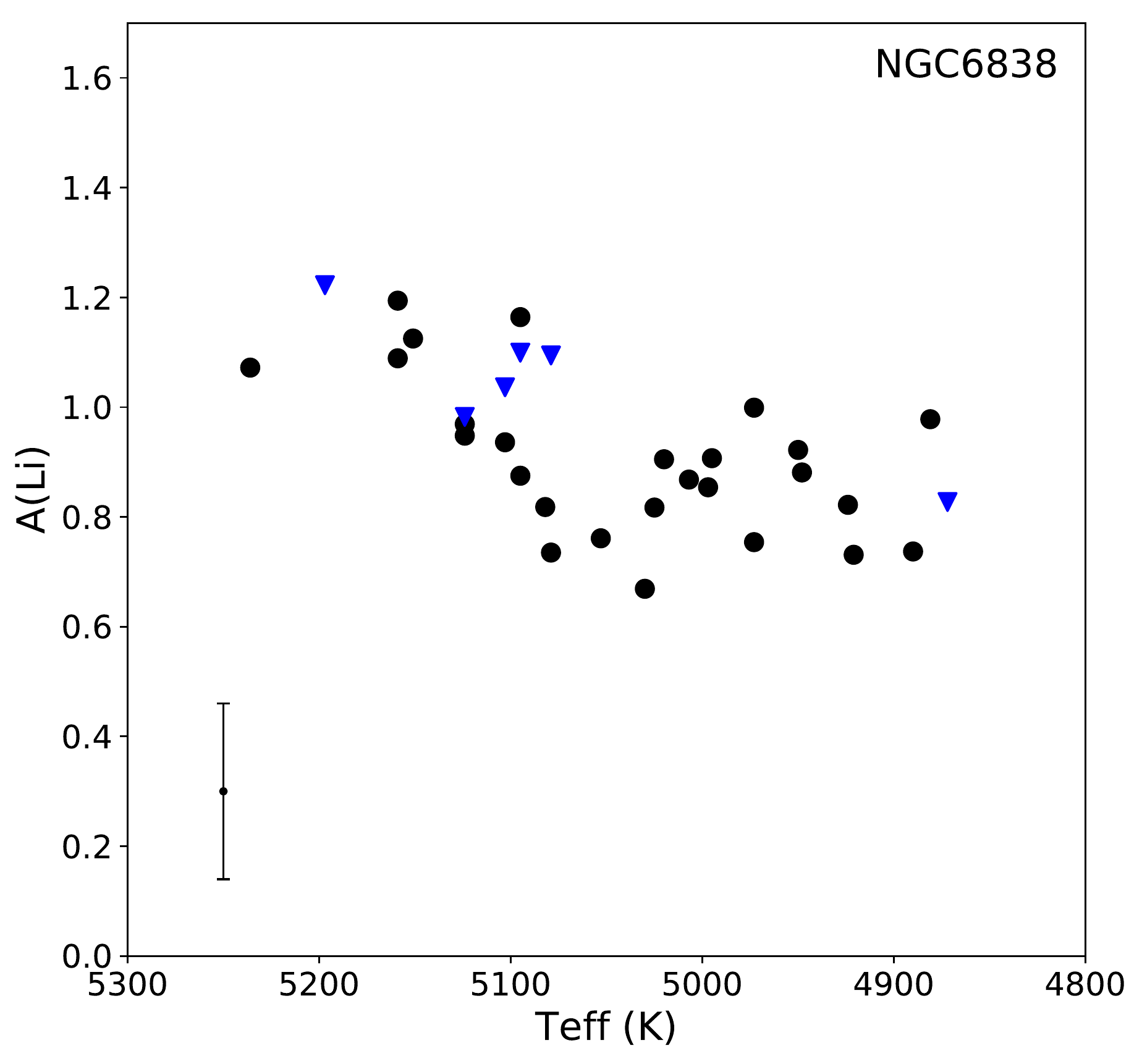}
\includegraphics[width=0.35\textwidth]{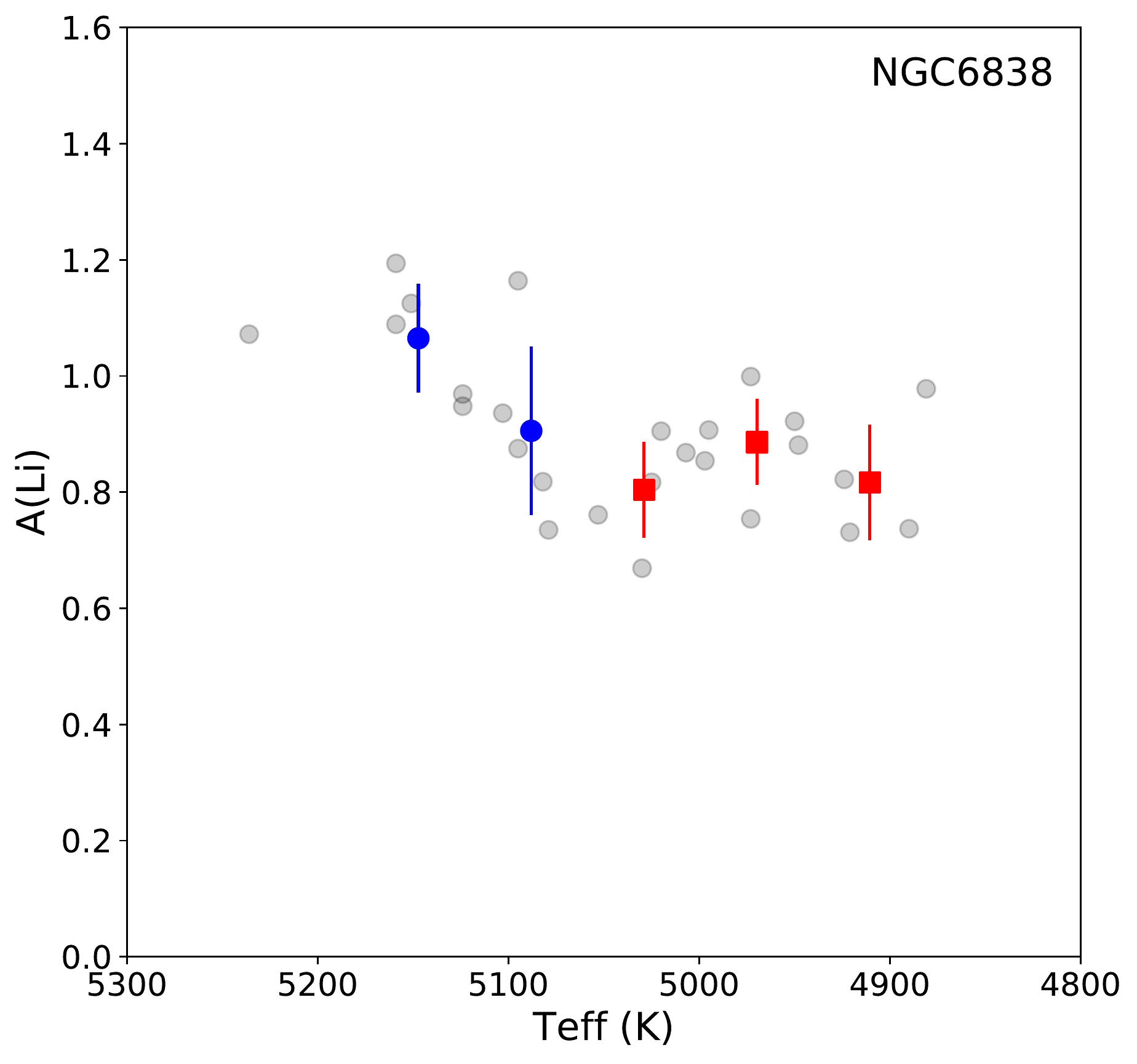}
\end{center}
\caption{Same as Figure \ref{Li_vs_Teff_1} for NGC3201 (top panel) and NGC6838 (bottom panel).}
\label{Li_vs_Teff_2}
\end{figure*}

\subsection{LRGB plateau and the cosmic Li problem}
To better identify a possible LRGB plateau in the 5 globular clusters we have binned the Li abundance as a function of effective temperature in Figures \ref{Li_vs_Teff_1} and \ref{Li_vs_Teff_2}. Left panel shows the Li abundance as a function of effective temperature, where we include upper limits as blue arrows. The right panel shows the binned abundance as blue points but only considering Li measurements (gray points) and no upper limits. To select stars that belong to the plateau, we define the end of the first dredge-up by using the measured Li abundances. We perform a simple bilinear fit to the data in the A(Li)-$\mathrm{T_{eff}}$ diagram, from the highest temperature to the luminosity function bump, which is clearly signposted by the sharp drop of Li abundances with decreasing temperature.
The fit provides the approximate temperature where the abundance plateau starts, whilst the luminosity function bump marks its end. Notice that in NGC4590 there are no stars that have completed the first dredge-up, as suggested by our fit to the data. We identify the bins that belong to the LRGB plateau as red squares, which are the values we use to calculate the mean Li abundance of the plateau in each cluster. The stars are binned in equal-sized temperature ranges within each cluster, to make sure that there is a significant number of stars in each bin. Changing the bin size does not significantly alter our results.

The binned Li abundance allows to better identify the LRGB plateau present in the clusters. In NGC4590, the value of this plateau is A(Li)$=0.90\pm0.08$ dex. The error reported here is the standard deviation of the individual abundances of stars at the plateau. NGC6809 shows a clear decrease in Li at the start of the RGB until reaching a plateau of A(Li)$=1.03\pm0.08$ dex. The effect of the FDU can also be observed in NGC6656, where in this case, the LRGB reach a value of A(Li)=$0.88\pm0.09$ dex. In NGC6838 we also identify a mean Li in LRGB stars of A(Li)=$0.84\pm0.10$ dex. The only cluster where the LRGB do not have a constant value is in NGC3201. Here we observe a decrease in the abundance from higher to lower temperatures. The presence of a plateau is much harder to identify, and its value depends slightly on the position of the bins. If we define the plateau considering the 3 bins between $\sim4900$ K to $\sim 5050$ K where the abundance seems to be constant, we find a mean abundance of A(Li)=$0.97\pm0.10$ dex. Changing the bin size and position of bins, the mean A(Li) varies from $0.93$ to $0.98$ dex, with similar standard deviation. The main difference between this and other clusters in our sample is its age, but there is no clear explanation to the larger scatter near the plateau.

The scatter is fully consistent with the total error, which includes both the uncertainties due to quality of the spectra and uncertainties due to stellar parameters, especially effective temperature.

\begin{figure*}[!hbt]
\begin{center}
\includegraphics[width=0.32\textwidth]{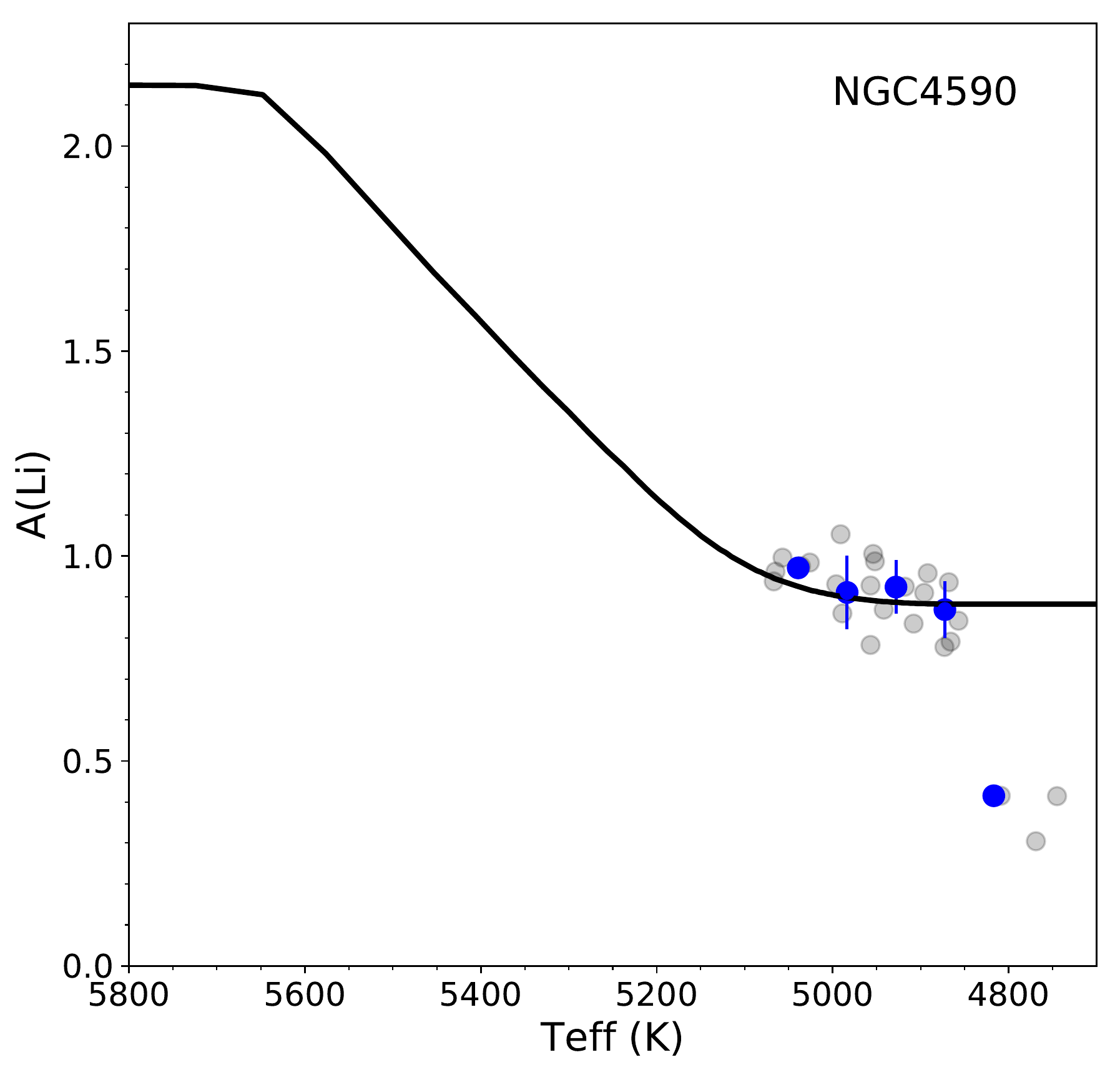}
\includegraphics[width=0.32\textwidth]{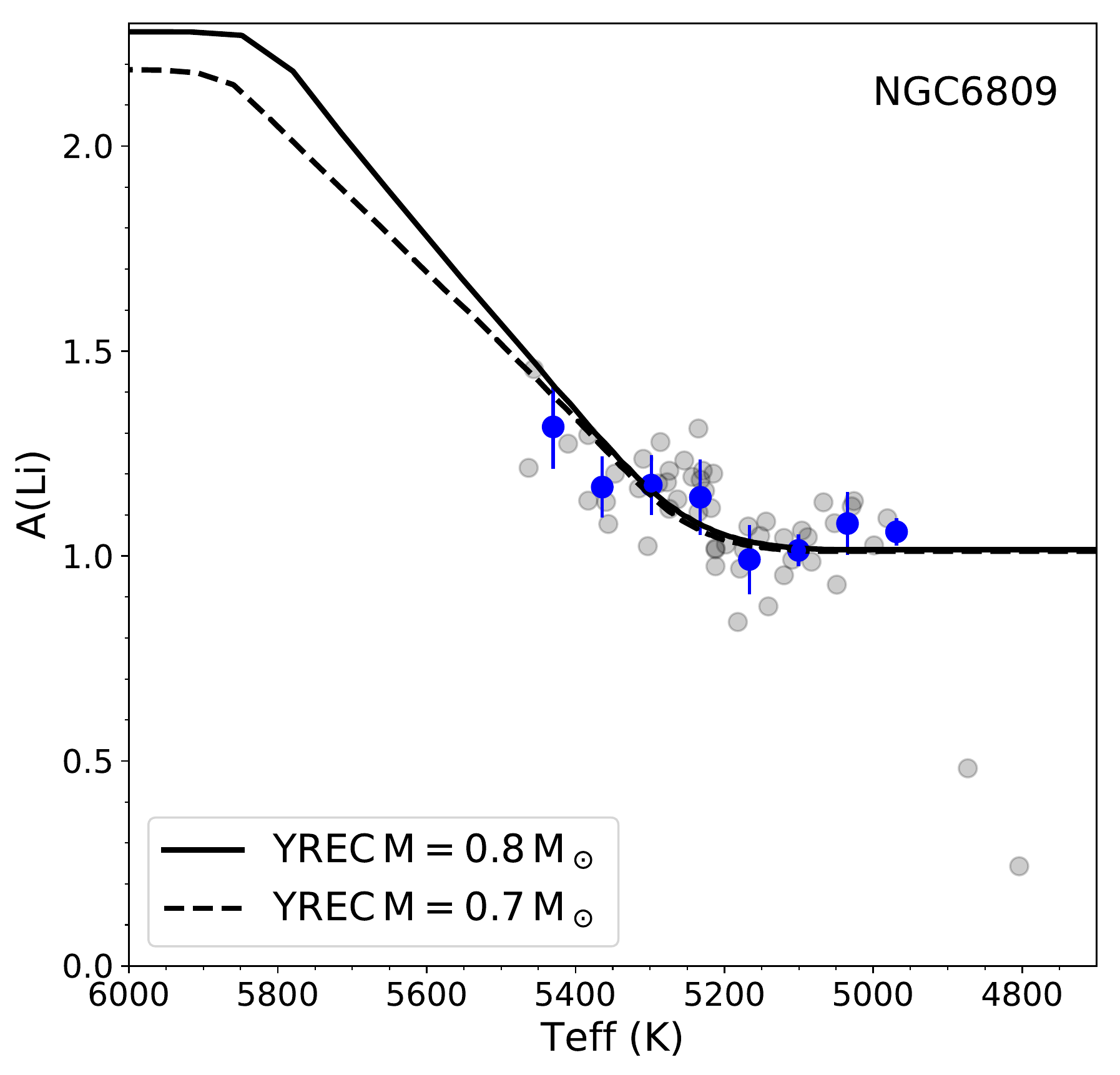}
\includegraphics[width=0.32\textwidth]{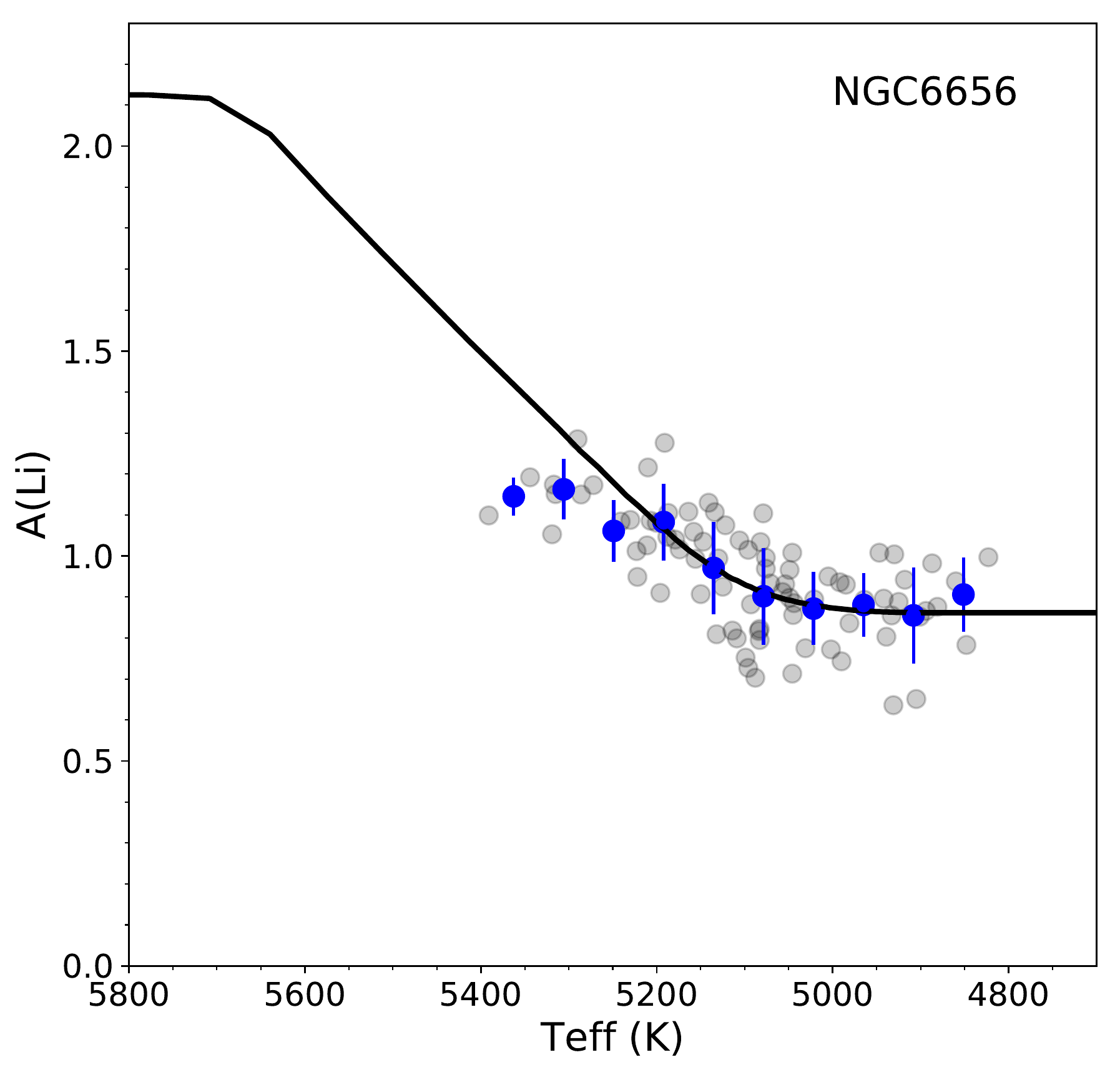}
\hfill
\includegraphics[width=0.32\textwidth]{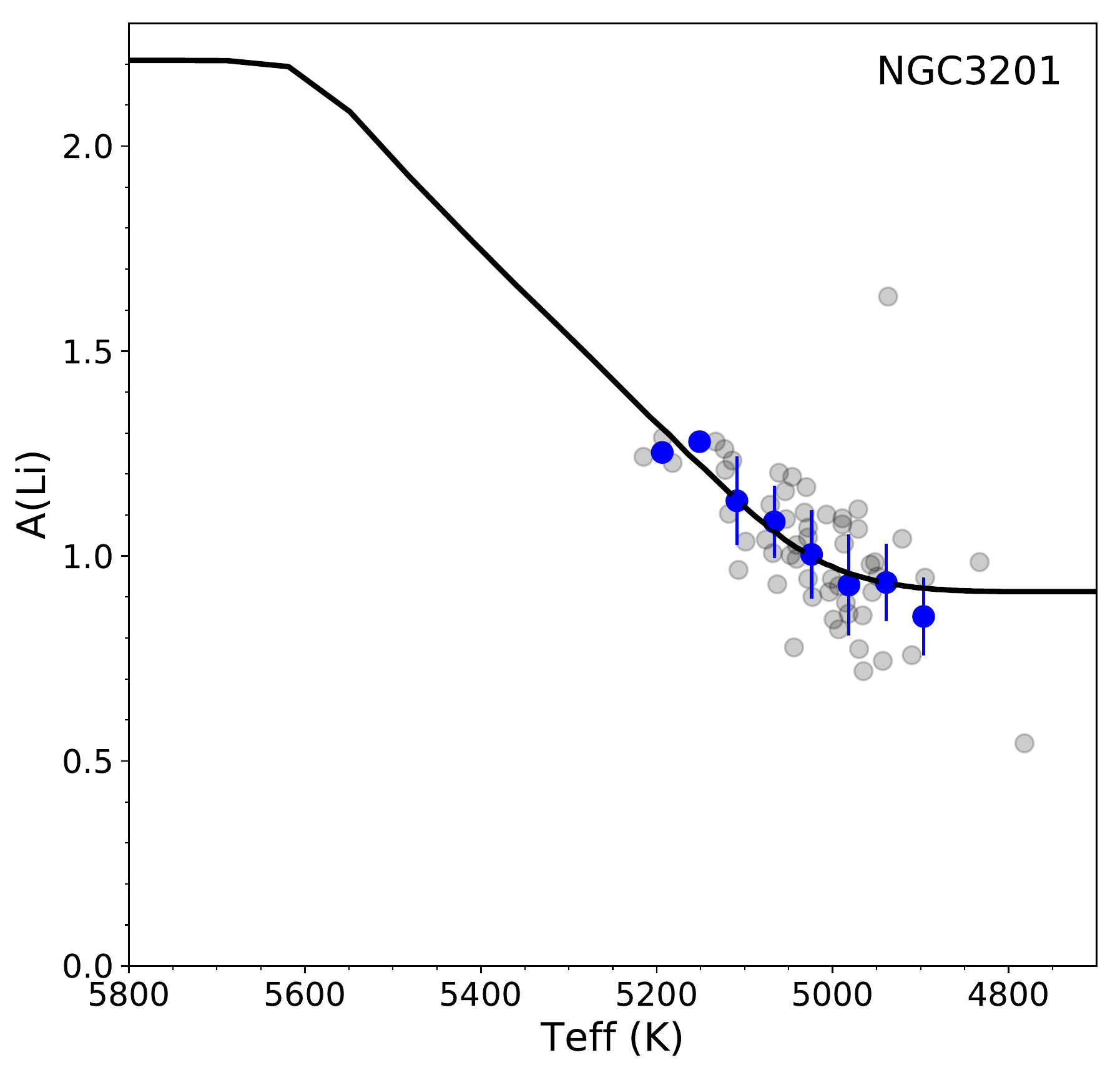}
\includegraphics[width=0.32\textwidth]{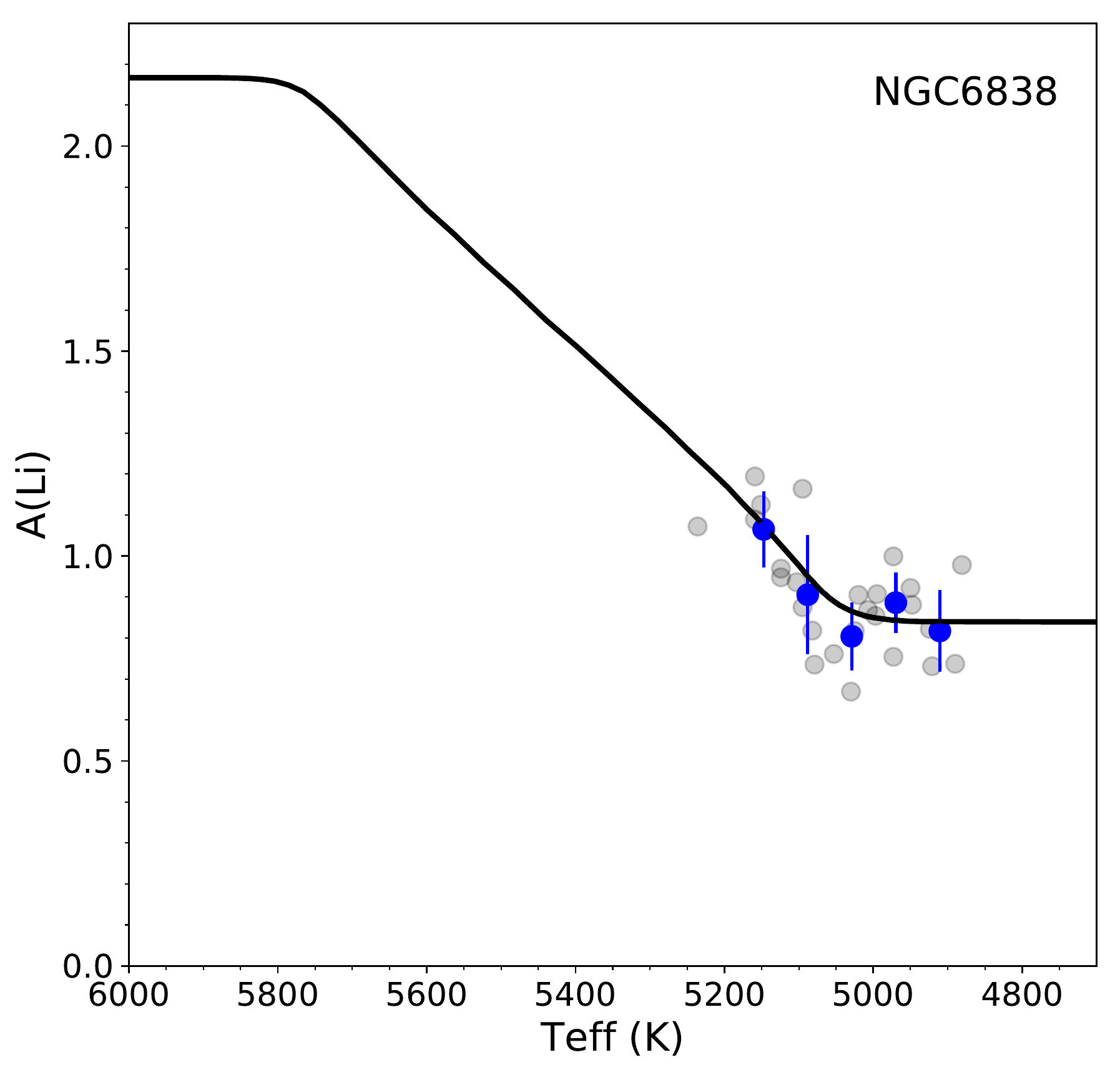}

\end{center}
\caption{Lithium abundance as a function of effective temperature. Blue points are the binned abundance in equal-sized bins. We include theoretical models that fit the LRGB Li plateau. In the cluster NGC6809, we include models with $0.7$ and $0.8\,\mathrm{M_\odot}$.}
\label{Li_models}
\end{figure*}

We use our Li abundances together with theoretical stellar evolutionary models to predict an initial Li value in these clusters. We use the Yale Rotating Evolutionary code \citep[YREC,][]{Pinsonneault1989,Demarque2008}, without diffusion, rotation or overshooting. The models use mixing length theory for convection \citep{CoxGiuli1968}, which acts as the only mixing mechanism inside the star, with no extra-mixing to modify the surface abundances after the RGB bump. Additionally, we use 2006 OPAL equation of state \citep{RogersNayfonov2002}, and cross section for the proton capture by lithium from \citet{Lamia2012}. Other input physics included in these standard models can be found in \citet{AG2016}. For each of the clusters we run models with a mass M$=0.8\,\mathrm{M_\odot}$, considered to be the typical turn-off mass in globular clusters, and a metallicity equal to the median value of the cluster, presented in Section \ref{Sec:Atm}.

The effects of diffusion, that can significantly change the lithium abundance in the main sequence evolutionary phase \citep{Richard2002}, are almost completely erased in the LRGB, given that the diffusion layers are mixed again by the deepening convective envelope. The lithium abundance in the LRGB of standard models is at most $0.07$ dex higher than when models include diffusion \citep{Mucciarelli2012}. 

Models with an initial lithium abundance equal to the standard BBN value of A(Li)=$2.72$ dex produce LRGB Li values much larger than those observed. Instead, we attempt to find this primordial Li abundance. We modify the initial Li abundance of the models to match the Li in LRGB, post FDU dilution. These results can be seen in Figure \ref{Li_models}. The model presented for the cluster NGC4590 has a metallicity of [Fe/H]$\sim-2.2$, which is the lowest metallicity we had available for the models. We tested that at such low metallicity the initial abundance should not change significantly between that metallicity and the metallicity measured for the cluster ([Fe/H]=$-2.34$ dex).

Signatures of diffusion have been found in some globular clusters \citep[e.g.][]{Korn2006,Gruyters2016}. The overall effect of diffusion in main sequence stars is to lower the surface Li abundance when approaching the turn-off, however observations suggest that the efficiency of diffusion is moderated by some competing mixing mechanism of unspecified origin. This makes it difficult to predict theoretically the Li abundance in turn-off stars of our studied clusters using the measured LRGB as starting point. Thus, we do not attempt to predict a turn-off abundance, and instead we recover the initial, primordial Li abundances of these clusters with our models, considering that the effects of diffusion are erased during the FDU.

Predicted initial values in every case are very similar to the Spite plateau and A(Li) found in other globular clusters where the Li abundance can be measured in dwarfs. The inclusion of diffusion could decrease this predicted value by $0.07$ dex at most. 
The primordial Li abundance of NGC6809 is predicted to be $\mathrm{A(Li)_0}=2.28$. We have also included here a model with $0.7\,\mathrm{M_\odot}$, to see the effects that mass can have in the abundance of dwarfs in the cluster. Li burning, including in the pre-main sequence, can be substantially different in this mass range. By changing the mass of the model, and adjusting the Li in the LRGB, the predicted initial lithium changes by $0.08$ dex, with a primordial value of $\mathrm{A(Li)_0}=2.20$, both within values found in halo stars. The initial lithium predictions of the other clusters are $\mathrm{A(Li)_0}=2.16$ in NGC4590, $\mathrm{A(Li)_0}=2.14$ for NGC6656, $\mathrm{A(Li)_0}=2.21$ in NGC3201, and $\mathrm{A(Li)_0}=2.17$ in NGC6838. These predicted primordial A(Li) values calculated for each cluster match the Li abundances of the Spite plateau and other globular clusters in the literature.

\begin{figure}[!hbt]
\begin{center}

\includegraphics[width=0.45\textwidth]{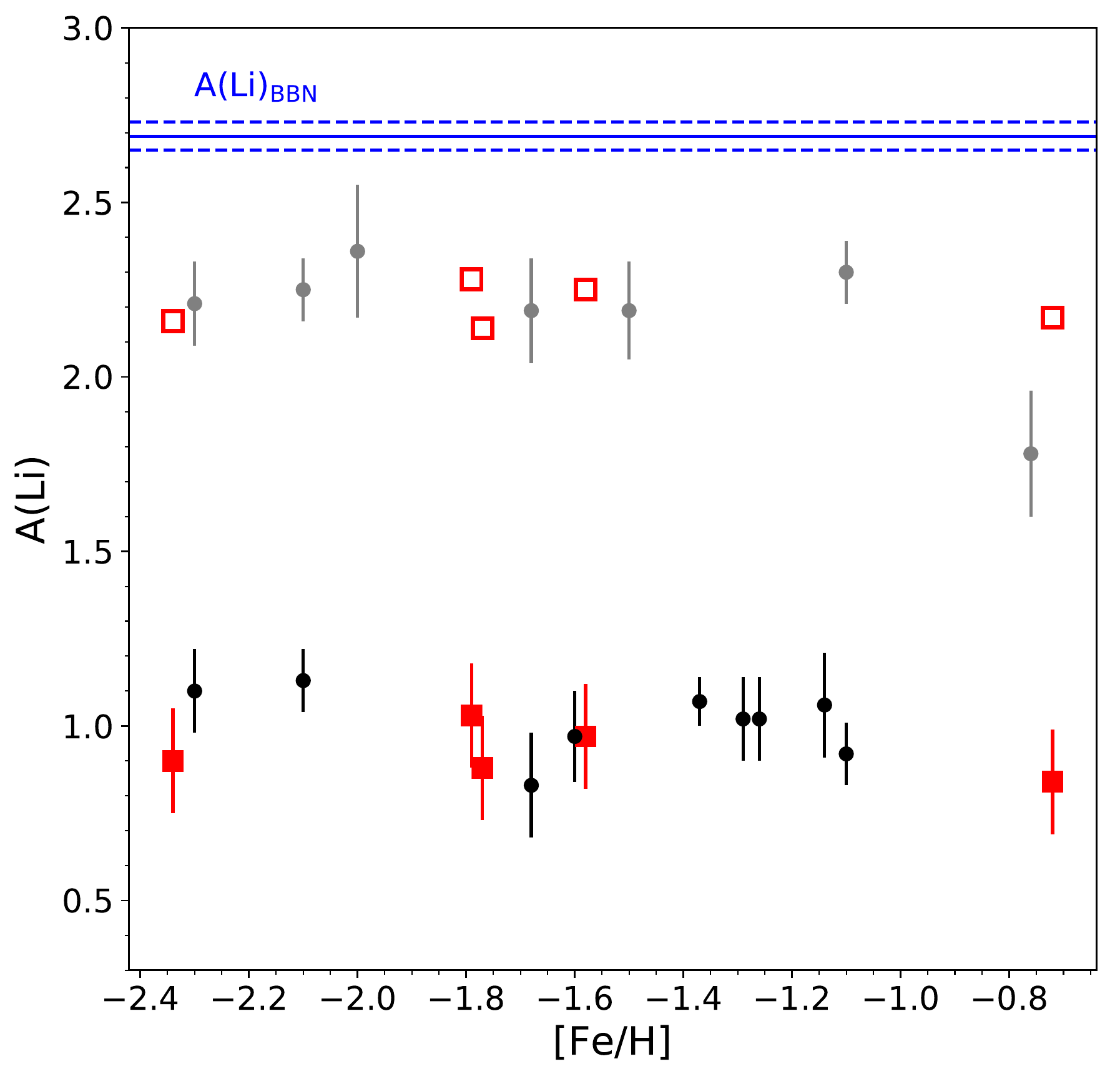}
\hfill

\end{center}
\caption{Li abundance in the LRGB of the 5 studied globular clusters as a function of [Fe/H] (filled red squares). The calculated primordial abundances for these clusters are the empty red squares. These are compared to literature measurements of the lithium abundance in the LRGB (black points at low Li abundances) and in the turn-off stars (grey points at high Li abundances) of globular clusters in the Galaxy. Also the Big Bang Nucleosynthesis prediction from \citet{Coc2014} is included and shown as a blue line, with the dashed blue lines representing its reported uncertainties.}
\label{Li_vs_met_clusters}
\end{figure}

We report in Table \ref{table:limet} LRGB stars abundances of the studied clusters and others in the literature. We also report in this table our predicted initial Li abundances and the Li abundance in the turn-off of clusters where it has been measured. Notice here that measurements are not homogeneous, and temperature scales can change the Li abundance measurements. There does not seem to be any correlations between the Li abundance in the LRGB (or the predicted primordial value) and metallicity for different clusters (Figure \ref{Li_vs_met_clusters}).

The use of a different temperature scale could also change our Li measurements and estimated primordial abundances in the clusters. By using a hotter temperature scale \citep[e.g.,][]{GHBonifacio2009}, our Li abundances should be higher by $\sim0.1$ dex. The use of different stellar evolutionary models, or even different prescriptions in the model used (e.g., including overshooting or changing the efficiency of diffusion) can also modify the predicted estimation of cosmological Li \citep[][]{Mucciarelli2012}, and even make it higher than the Spite Plateau, although differences in the temperature scale of those measurements should also be taken into account \citep[e.g.][]{MelendezRamirez2004}. Thus, our predictions should not be considered as an attempt to precisely obtain the exact primordial lithium of each cluster, but rather an estimation of the possible abundance range.

\begin{table*}
\caption{Lithium abundance in the LRGB plateau and turn-off of galactic globular clusters.}             
\label{table:limet}
\centering
\begin{tabular}{l c c c l}        
\hline\hline                 
Cluster & [Fe/H] & $\mathrm{A(Li)_{LRGB}}$ & $\mathrm{A(Li)_{0}}$ & Reference \\    
- & (dex) & (dex) &  (dex) & -\\
\hline                        
   NGC4590 & $-2.34$ & $0.90\pm0.08$ & $2.16$ & This work \\
   NGC6809 & $-1.79$ & $1.03\pm0.08$ & $2.28$ & This work \\
   NGC6656 & $-1.77$ & $0.88\pm0.09$ & $2.14$ & This work \\
   NGC3201 & $-1.58$ & $0.97\pm0.10$ & $2.21$ & This work \\
   NGC6838 & $-0.72$ & $0.84\pm0.10$ & $2.17$ & This work \\ 
\hline\hline                 
Cluster & [Fe/H] & $\mathrm{A(Li)_{LRGB}}$ & $\mathrm{A(Li)_{TO}}$ & Reference \\ 
\hline
   NGC7099 & $-2.30$ & $1.10\pm0.06$ & $2.21\pm0.12$ & 1\tablefootmark{a}\\
   NGC6397& $-2.10$ &$1.13\pm0.09$ & $2.25\pm0.01\pm0.09$ & 2\tablefootmark{a}\\
   M4 & $-1.10$  & $0.92\pm0.01\pm0.08$ & $2.30\pm0.02\pm0.10$ & 3\tablefootmark{a}\\
   M4 & $-1.31$ & - & $2.13\pm0.09$ & 4\\
   NGC6752 &$-1.68$ &$0.83\pm0.15$ & - 
   & 5\\
   NGC1904 & $-1.60$ & $0.97\pm0.02\pm0.11$ & - & 6 \\
   NGC2808 & $-1.14$ & $1.06\pm0.02\pm0.13$ & - & 6 \\
   NGC362 & $-1.26$ & $1.02\pm0.01\pm0.11$ & - & 6\\
   NGC6218 & $-1.37$ & $1.07\pm0.01\pm0.06$ & - & 7\\
   NGC5904 & $-1.29$ & $1.02\pm0.01\pm0.11$ & - & 7\\
   \hline
   47 Tuc & $-0.76$ & - &$1.78\pm0.18$ & 8 \\
   M92 &$-2.00$ & - &$2.36\pm0.19$ & 9 \\
   $\omega$ Cen & $-1.50$ & - & $2.19\pm0.14$ & 10 \\
   
\hline                                   
\end{tabular}
\tablebib{
(1) \citet{Gruyters2016}; (2) \citet{Lind2009}; (3) \citet{Mucciarelli2011}; (4) \citet{Monaco2012}; (5) \citet{Mucciarelli2012}; (6) \citet{DOrazi2015}; (7) \citet{DOrazi2014}; (8) \citet{Dobrovolskas2014}; (9) \citet{Bonifacio2002}; (10) \citet{Monaco2010}}
\tablefoot{
\tablefoottext{a}{These works present Li abundances in the turn-off and lower red giant branch.}
}

\end{table*}

\subsection{First and second population stars}

Measurements of the Na abundance were made in order to separate populations in the studied globular clusters. This is based on the idea that the more massive stars of the first population, now evolved, had an active nucleosynthesis cycle in its interior able to produce, for instance, fresh Na at the expense of O. Throughout the lifetime of the star, this processed material is carried to the surface of the star, and through mass loss, stellar winds, and the planetary nebula phase, to the interstellar medium. The second population of stars is born from this polluted material, creating different populations of stars coexisting in the same cluster \citep[see e.g.][]{BastianLardo18}. The nature of the polluter is still a matter of open debate, with fast rotating massive stars \citep{Decressin2007} and asymptotic red giant branch stars \citep[AGB, ][]{VenturaDAntona2009} being the main contenders. On the other hand, there could be alternative scenarios to explain this pattern in clusters, not related to nucleosynthesis, or it is possible that the generational scenario is complicated by additional mechanisms acting \citep{Gratton2019}.

\begin{figure*}[!hbt]
\begin{center}
\includegraphics[width=0.42\textwidth]{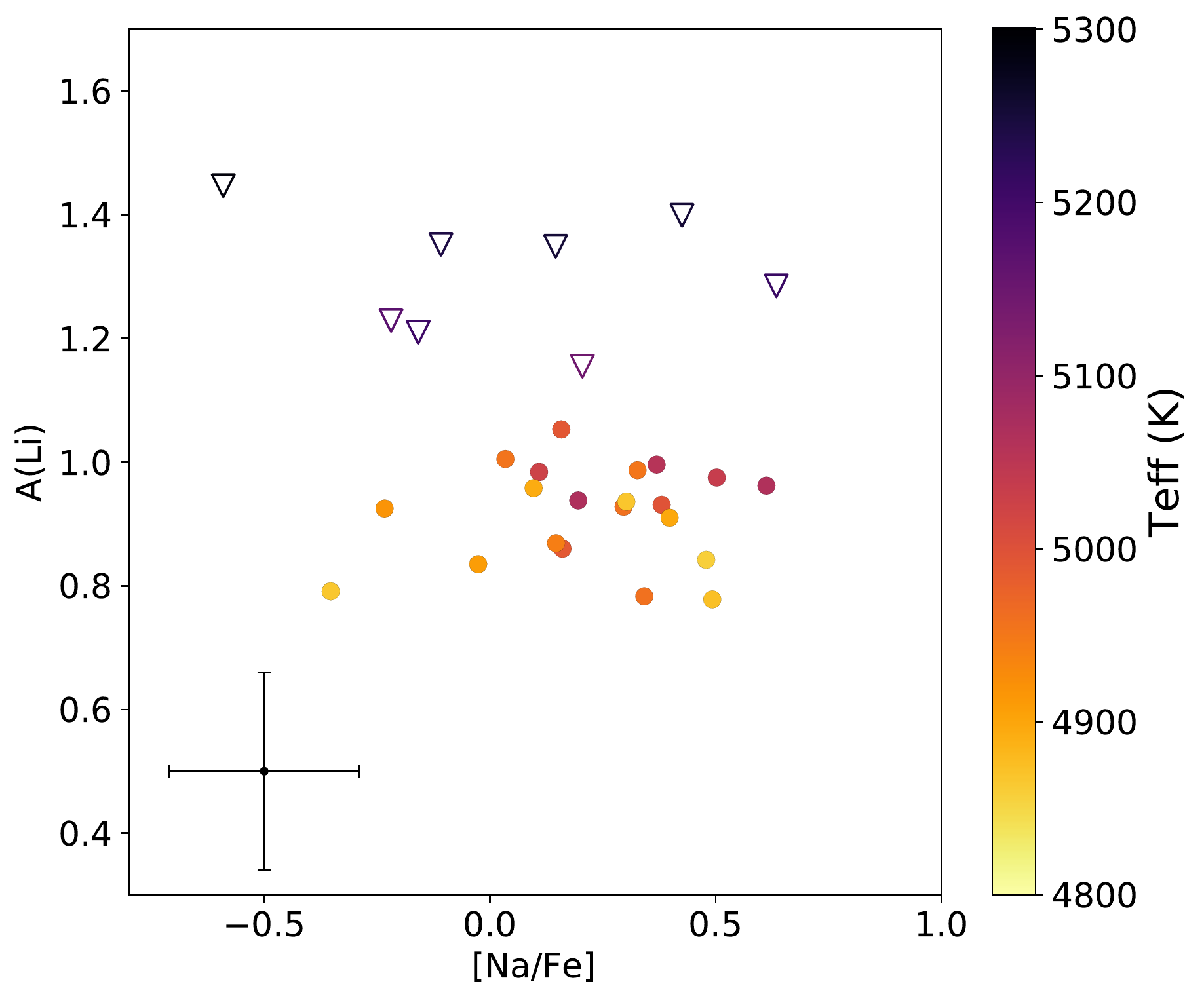}
\includegraphics[width=0.4\textwidth]{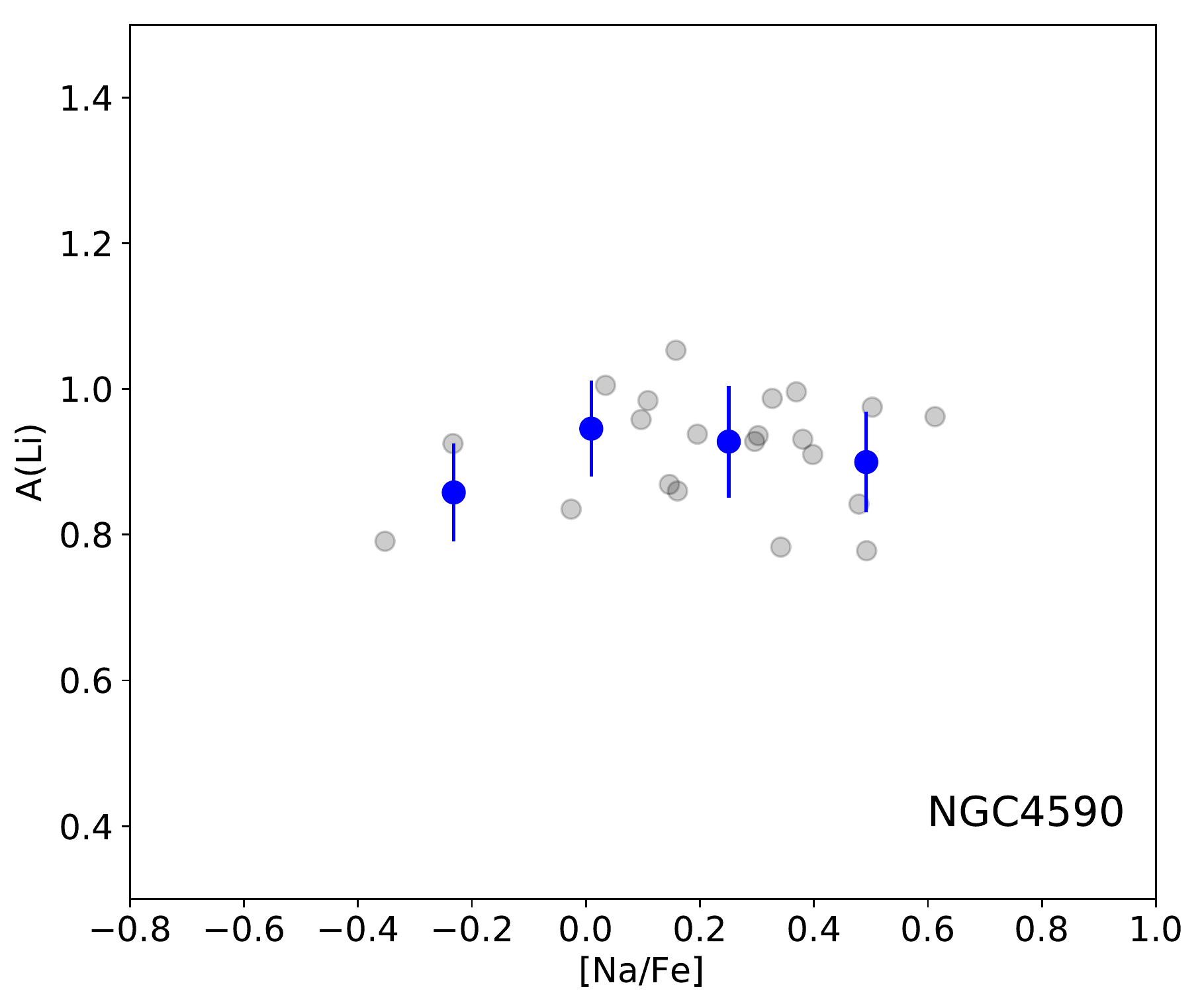}
\hfill
\includegraphics[width=0.42\textwidth]{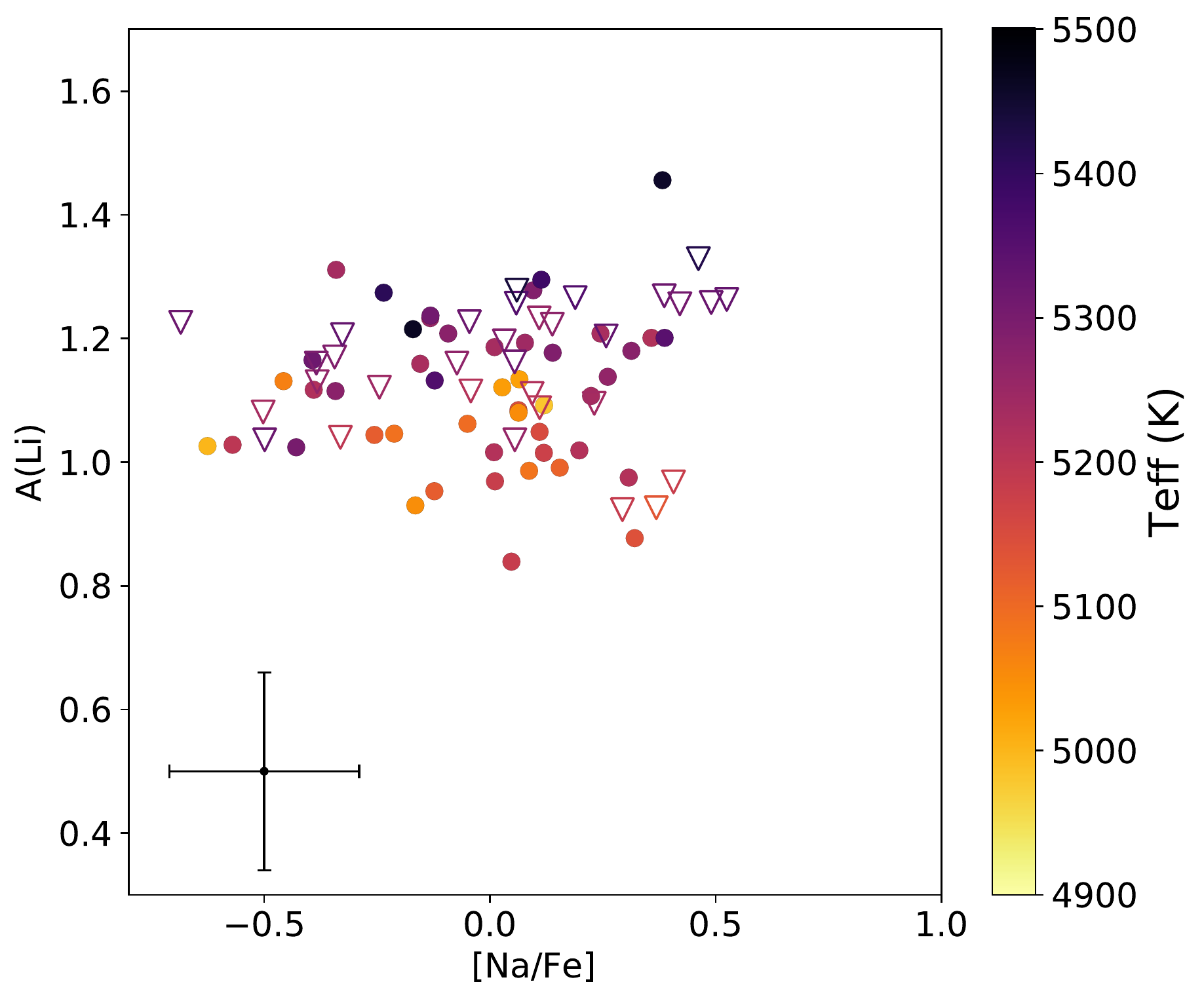}
\includegraphics[width=0.4\textwidth]{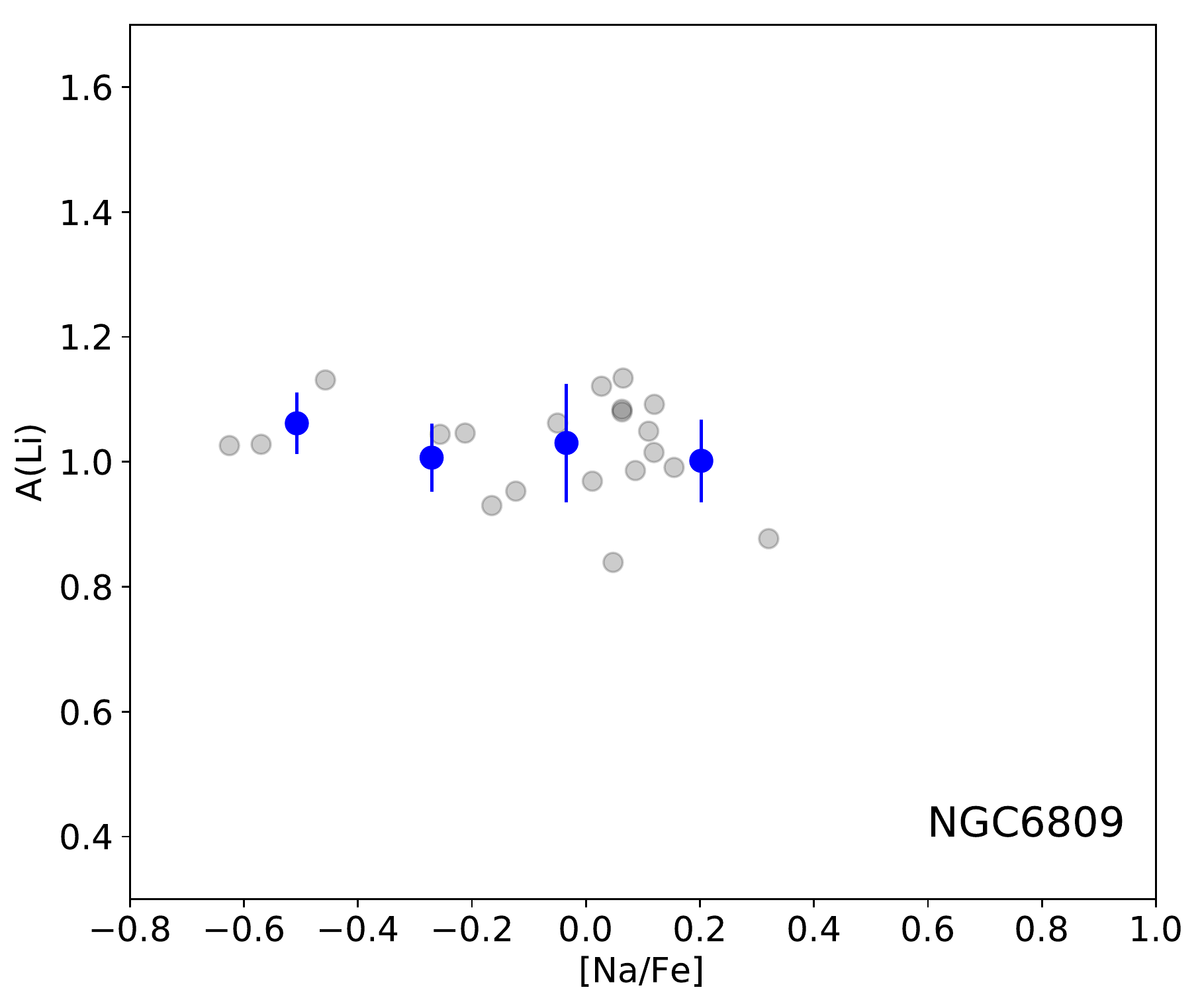}

\end{center}
\caption{Lithium abundance as a function of sodium abundance. The left panel is color-coded by effective temperature, and includes all RGB stars before the luminosity function bump, while the right panel considers only Li measurements (no upper-limits) of plateau LRGB stars, and it is binned (blue points) to show possible trends between both abundances. Top panel is NGC4590, bottom panel shows NGC6809.}
\label{Li_vs_Na_1}
\end{figure*}

\begin{figure*}[!hbt]
\begin{center}

\includegraphics[width=0.42\textwidth]{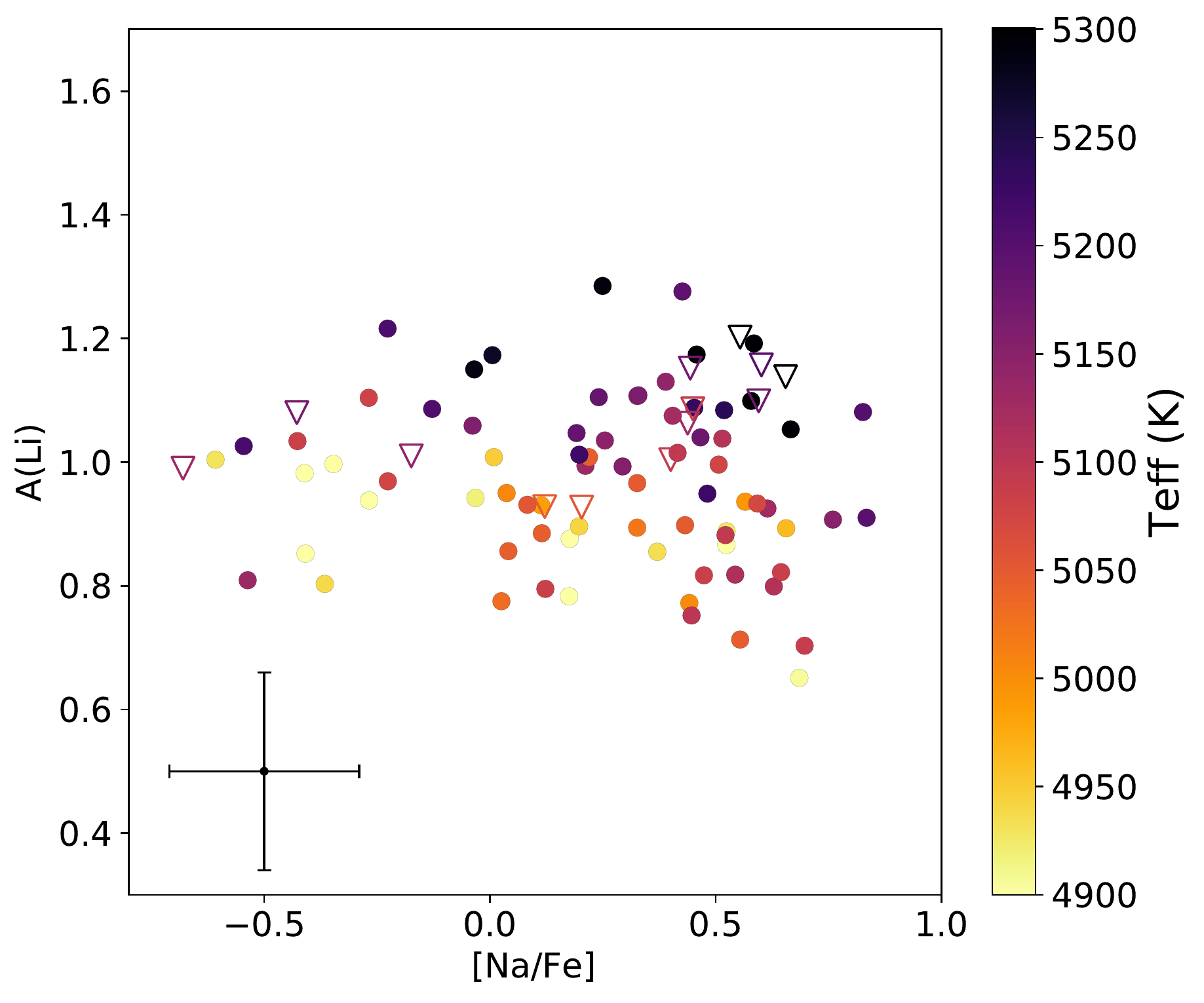}
\includegraphics[width=0.4\textwidth]{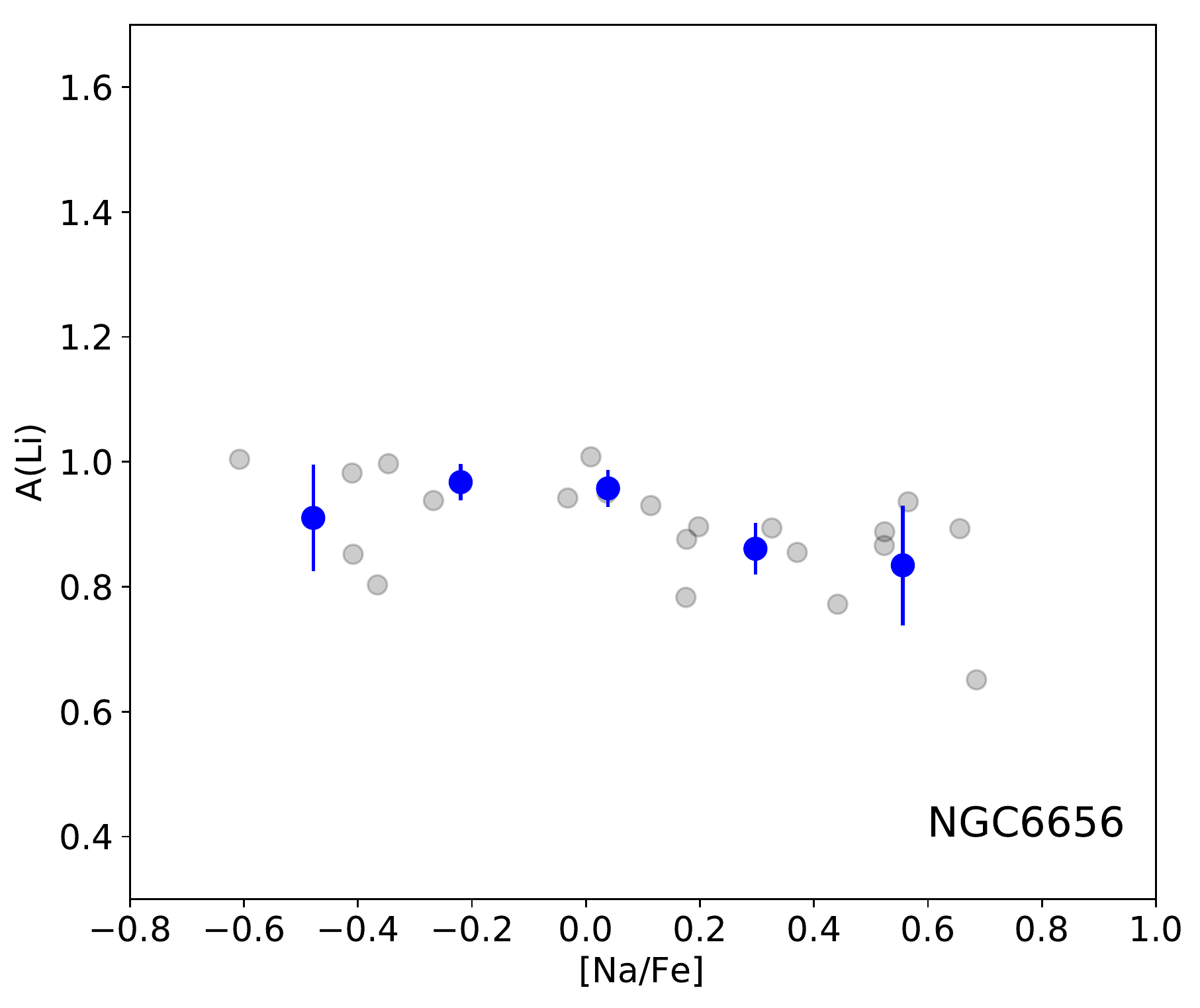}
\hfill
\includegraphics[width=0.42\textwidth]{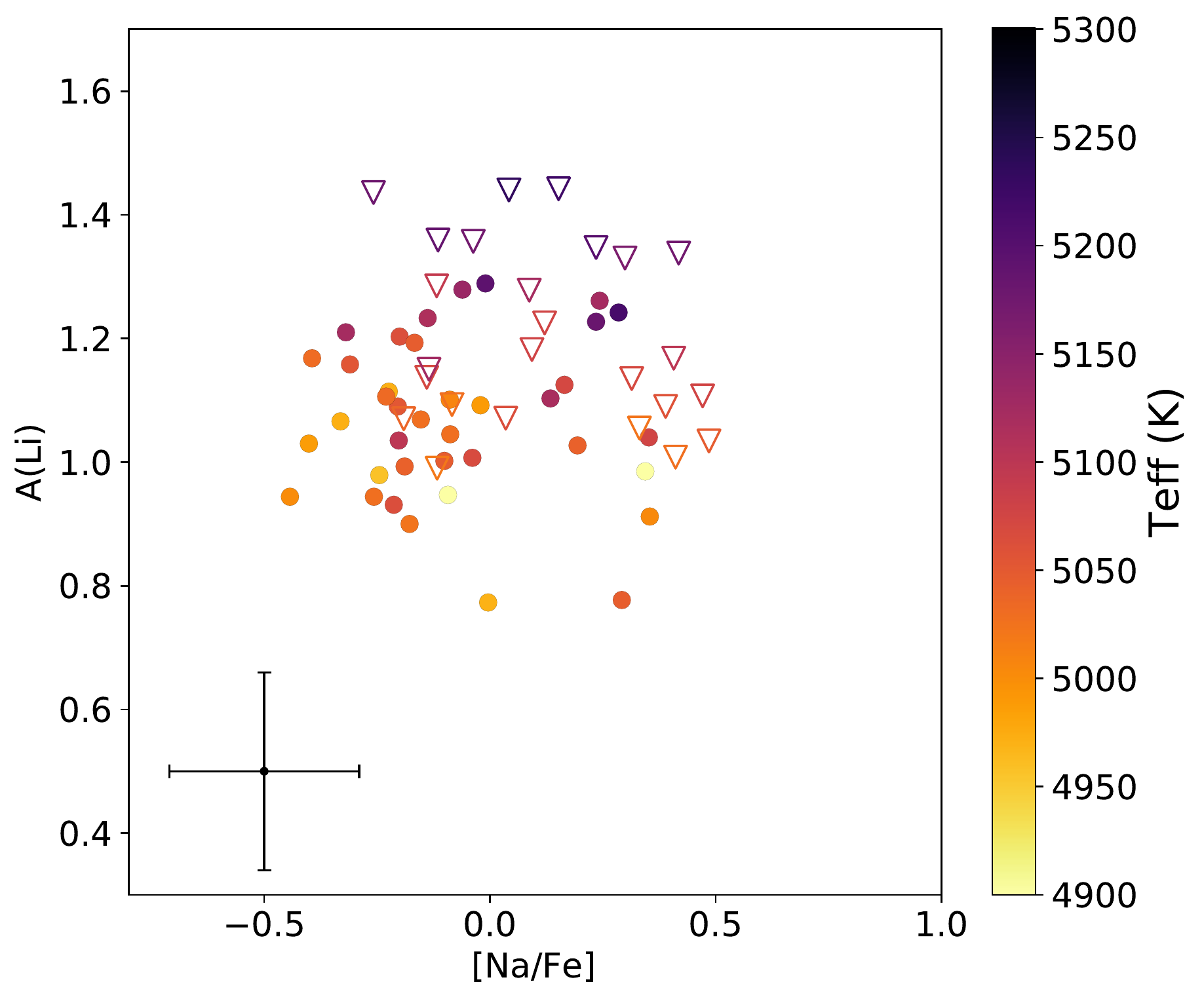}
\includegraphics[width=0.4\textwidth]{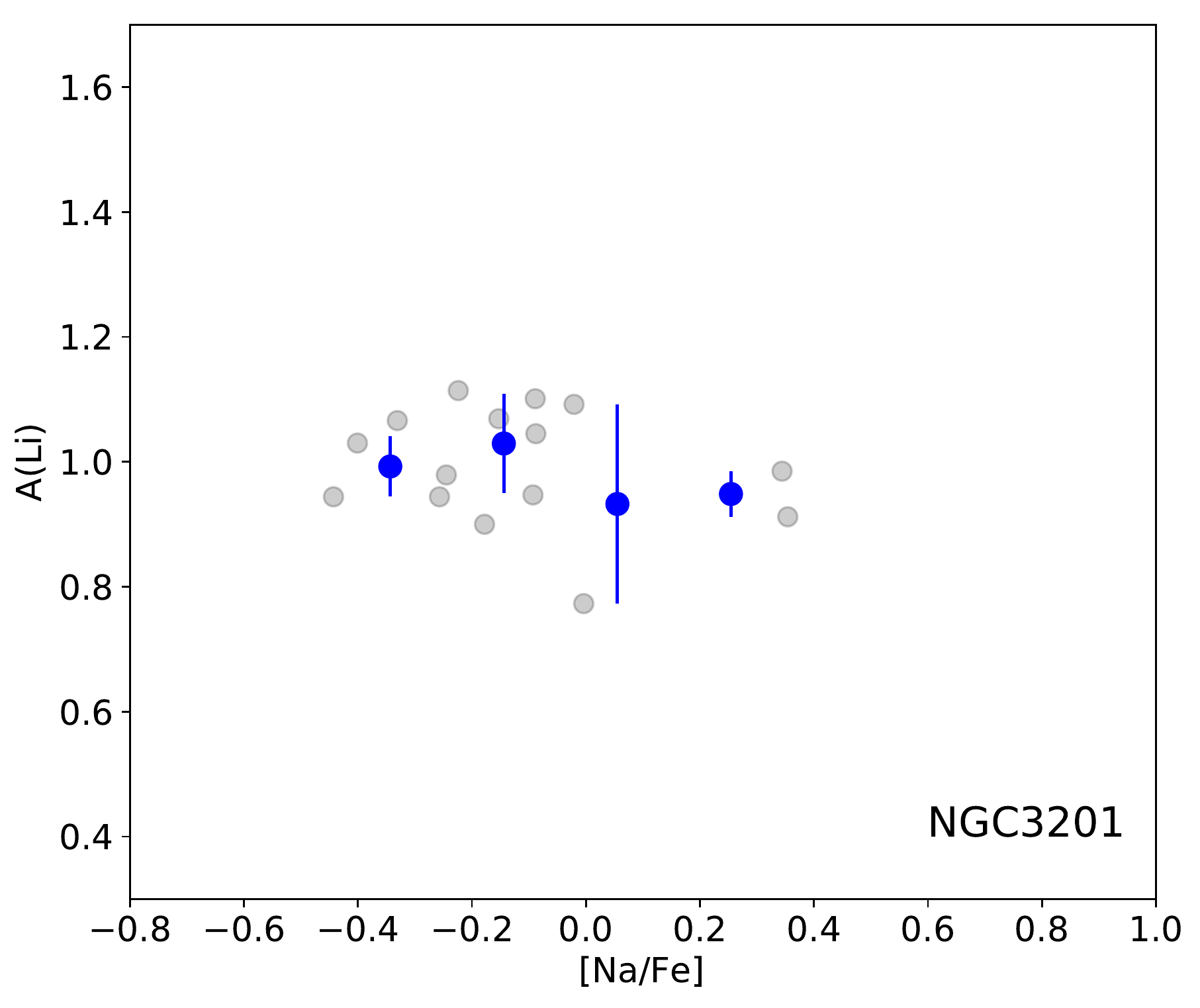}

\end{center}
\caption{Same as Figure \ref{Li_vs_Na_1}. Top panels are NGC6656 and bottom panels NGC3201.}
\label{Li_vs_Na_2}
\end{figure*}

Li, which is destroyed at relatively low temperatures by proton capture, is expected to be depleted in Na-enriched material, such as that from which second population stars are born. An anticorrelation between Na and Li is then expected. However, in certain AGB stars, Li can be created in the interior through the Cameron-Fowler mechanism \citep{CF1971}, and is quickly transported to the surface of the star by convection, where the cooler temperatures prevent it from destruction by proton capture \citep{SackmannBoothroyd1992}. Thus, it may be relevant to compare the Li abundance from the first and second population, with the first population expected to have a cosmological Li content diluted because of the FDU, and the second population may show an abnormally high Li abundance if the polluter is either a Li-enriched AGB star or if the ejecta of polluters is mixed with material that has not burned Li.

Figures \ref{Li_vs_Na_1} and \ref{Li_vs_Na_2} show the behavior of A(Li) as a function of [Na/Fe] only for RGB stars in NGC4590 and NGC6809 (top and bottom panels of Figure \ref{Li_vs_Na_1} respectively), and NGC6656 and NGC3201 (top and bottom panels of Figure \ref{Li_vs_Na_2}). We have removed stars brighter than the RGB bump in these figures. Additionally, we consider only LRGB plateau stars in the right panel of these figures, by removing all stars that have not yet completed their first dredge-up. As mentioned in Section \ref{sec:abund}, we were not able to measure Na in NGC6838, thus we do not present results for that cluster.
Focusing only on LRGB stars, there is no clear correlation between Li and Na in NGC4590, NGC6809, NGC3201, and NGC6656. We do see some star-to-star scatter, but Li does not scale with Na. This, however, does not exclude the possibility to find a trend if more stars are considered in the analysis.

As previously mentioned, there is no statistically significant anticorrelation in NGC6656. By eye, when considering all RGB stars of different effective temperatures in the top left panel of Figure \ref{Li_vs_Na_2}, objects with higher Na would seem to have slightly lower Li. However, when we look for possible correlations by binning both in effective temperature and Na, only the higher temperature bin shows a hint of an anticorrelation, that is however not considered statistically significant.

The lack of a clear Li-Na anticorrelation in our cluster sample needs further confirmation with additional data. In the literature, some clusters do show correlations between Li and other light element abundances. NGC6752 presents a Li-O correlation \citep{Shen2010} and Li-Na anticorrelation \citep{Pasquini2005}. NGC6397 has some stars enriched in Na that are Li poor \citep{Lind2009}. NGC2808 has some stars enriched in Al that are Li depleted \citep{DOrazi2015}. In M4 there is something like a Li-Na anticorrelation \citep{Monaco2012} but no Li-O correlation \citep{Mucciarelli2011}. 47 Tuc shows no sign of a Li-Na anticorrelation \citep{Dobrovolskas2014}.

If it is confirmed that some clusters have a similar Li abundance in both populations, this would point to a higher Li than expected in the second population. This could mean that the birth material of these stars should have been mixed with relatively Li-rich material, pointing to AGB stars as possible polluters. Models have to be fine-tuned to produce such a pattern in globular clusters, given that the Li yields have great uncertainties depending on how physics such as mass loss is introduced \citep{VenturaDAntona2010}. If massive stars were the polluter, this scenario would require mixing the Na-rich material from the ejecta with unprocessed material that has a higher Li abundance. However, confirmation of the lack of a Na-Li anticorrelation is needed before anything can be firmly concluded about the mechanism behind the different populations in clusters.
Additionally, measurements from clusters come from non-homogeneous sources that not only have different parameters scales, spectral qualities, and use different methods, but that also provide abundances of different light elements. An homogeneous determination of properties and abundances could be a major improvement to gain insight about the second generation polluters.

\subsection{Li-rich giant in NGC3201} \label{Sec:Lirich}

Stars in the red giant branch experience abundance changes during the FDU, and then, at the luminosity function bump where the extra-mixing acts. If a solar-like star enters the RGB phase with a meteoritic abundance A(Li)$=3.3$, its predicted Li abundance after the FDU is expected to be A(Li)=$1.5$, only considering FDU dilution. However, values can be much lower, when additional ingredients, such as  a much lower initial Li abundance, Li burning, and main sequence mixing are taken into account. The precise value to classify a giant as enriched is actually mass and metallicity dependent, and standard giants with higher A(Li) than $1.5$ dex can be found, as well as giants with lower abundances that could have experienced a Li-enrichment process \citep{AG2016}. In spite of predictions from canonical models, lithium-rich red giants, with higher Li abundances, even reaching or exceeding the meteoritic value, are known to exist \citep[e.g.][]{WallersteinSneden1982, Monaco2011}.

Globular clusters present an advantage, with all of their giants sharing a similar mass, and possibly, a similar original Li content. Because of this, we can compare the Li abundance of the giants to abundances of other stars with similar parameters and at a similar evolutionary stage, making enriched objects much easier to identify. Although Li-rich giants are unusual in general, they are particularly rare in globular clusters. Only about a dozen giants are known to have a much higher Li abundance than other stars in the same evolutionary stage in a globular cluster. So far, Li enriched RGB stars have been found in  NGC5272 \citep{Kraft1999}, NGC362 \citep{Smith1999, DOrazi2015rich}, NGC4590 \citep{Ruchti2011, Kirby2016}, NGC5053, NGC5897 \citep{Kirby2016}, 2 giants in NGC7099 \citep{Kirby2016}, $\omega$ Cen \citep{Mucciarelli2019}, and only one Li-rich star in NGC1261 \citep{Sanna2020}.

These are located all along the RGB phase, although AGB Li-rich stars have also been found \citep[e.g.][]{Kirby2016}. Some are located after the luminosity function bump of their respective clusters, where extra-mixing is expected to affect the abundance of stars and could be the reason behind the Li-enrichment. Before that point in evolution, other explanations must be invoked that require pollution from an external source or the presence of a binary companion to trigger Li production \citep{Casey2019}. In the case of pollution, the source could be an AGB companion that can produce additional lithium in its interior and could then be transferring mass to the RGB star; it could be a nova, that can produce Li during the thermonuclear runaway \citep{Starrfield1978, Izzo2015}; or a planet or brown dwarf accreted by the star, objects that preserve the Li they have at formation \citep{Alexander1967,SiessLivio1999}.

\begin{figure}[!hbt]
\begin{center}
\includegraphics[width=0.45\textwidth]{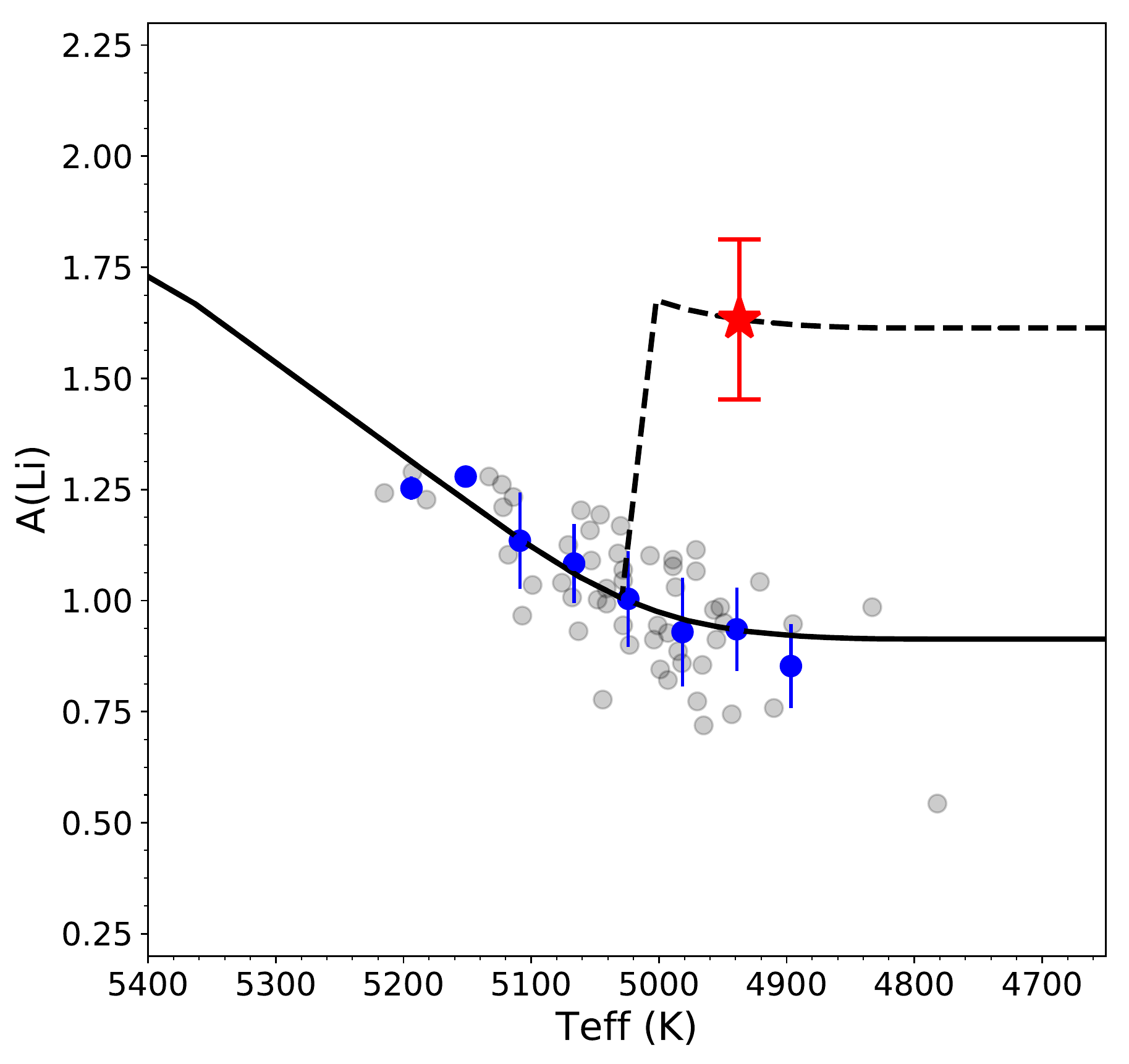}

\end{center}
\caption{Lithium abundance pattern of NGC3201. The Li-rich giant found (red star) could be explained by a model with planet engulfment (dashed line), where the hypothetical planet has a mass of $\mathrm{M_{planet}}=9.3\,\mathrm{M_{Jupiter}}$ (with Jupiter-like composition) or $\mathrm{M_{planet}}=110\,\mathrm{M_{Earth}}$ with Earth-like composition.}
\label{planetengulfment}
\end{figure}

We present the discovery here of one more Li-rich giant in a globular cluster, in this case, NGC3201. The star ID 97812, with $\mathrm{A(Li)_{NLTE}}=1.63\pm0.18$ dex is located before the luminosity function bump, and thus it is not expected to be enriched by the internal production of Li. Instead, pollution, either during the RGB phase or before, is probably the cause of enrichment for this giant, it is still possible that the presence of a binary companion is triggering the Li enhancement. Considering accretion as a possible scenario, we calculate the Li abundance of the star after the engulfment of a planet, using the models and parameters from \citet{AG2016}. Applying as initial conditions the Li abundance of the rest of the cluster, we calculate the engulfed mass of a hypothetical planet needed to explain the high A(Li) of this star. This model is shown in Figure \ref{planetengulfment}. For Jupiter-like composition, a mass of $\mathrm{M_{planet}}=10.1\,\mathrm{M_{Jupiter}}=1.92\times10^{31}\,\mathrm{g}$ is needed, and if the engulfed object had an Earth-like composition, it would require a mass of $\mathrm{M_{planet}}=120\,\mathrm{M_{Earth}}=7.17\times10^{29}\,\mathrm{g}$. Although the amount of Earth masses needed is large, the mass of the Jupiter-like planet required is in range of masses of exoplanets known that can orbit close to their parent star. Monitoring the radial velocity of this star would be interesting to understand if its enhancement comes from planet engulfment, or if a binary companion is responsible for its high Li abundance.

\section{Summary} \label{sec:summary}
We calculated stellar parameters and measured Li and Na abundances of LRGB stars of 5 Galactic globular clusters, covering a wide range of metallicities from [Fe/H]=$-0.72$ to [Fe/H]=$-2.34$ dex.
We find a LRGB plateau in all of the clusters at different levels, all between A(Li)=$0.84-1.03$ dex, consistent with what has been found for other globular clusters previously. 

Using theoretical models, we calculate the initial, primordial Li abundance in these clusters. The abundances found are similar to the Spite plateau value of halo stars, with A(Li)=$2.14-2.28$. However, we note that the exact predicted value could change by using either a different temperature scale or different model. Thus, we use these predictions to conclude about the overall trends in Li abundances and not the exact value of the cosmological Li. As a caveat, we have to consider the possibility that the additional mixing operating during the main sequence which affects the efficiency of diffusion, might cause the transport of some extra Li in the burning regions \citep[e.g.][]{Richard2005}. This would result in an underestimate of the initial A(Li) from measurements of LRGB using standard stellar models.

Considering the uncertainties in Li abundances, our main conclusion is that all of the clusters are consistent with models that have evolved from the same initial Li abundance. This agrees with the idea of a constant Li abundance of stars at this metallicity range, confirming the large discrepancy between Big Bang Nucleosynthesis predictions and observations of main sequence field stars.

We find no correlation between the Li value in the LRGB plateau and metallicity. To further study a possible correlation, we also use literature data available for other clusters, finding no relation between A(Li) and [Fe/H].

The measured sodium abundance is used to distinguish between first and second populations in each cluster. We find no clear difference in Li abundance between Na-rich and Na-poor stars in any of the clusters. If this is confirmed, it could point towards a class of polluter stars that are able to produce Li, such as AGB stars, or the mixing of the processed Li-poor medium with additional unprocessed matter.

We summarize the main results for each of the studied clusters:
\begin{itemize}
    \item {\bf NGC4590}: The median Li of LRGB stars in this metal-poor cluster is A(Li)=$0.90\pm0.08$ dex. Considering its metallicity of [Fe/H]$=-2.34\pm0.10$ dex, we calculate an initial Li abundance $\mathrm{A(Li)_0}=2.16$. There is no clear correlation between Na and Li in this cluster when we only consider LRGB stars.
    \item {\bf NGC6809}: The distribution of LRGB Li abundances in the cluster presents a peak at A(Li)=$1.03\pm0.08$ dex. The primordial value predicted, considering a RGB mass of $0.8\,\mathrm{M_\odot}$ is $\mathrm{A(Li)_0}=2.28$. This is the cluster with the highest Li abundances in the LRGB plateau in our sample. It does not present a Li-Na correlation either.
    \item {\bf NGC6656}: In this cluster, the LRGB plateau is located at A(Li)=$0.88\pm0.09$ dex, with a predicted initial value of $\mathrm{A(Li)_0}=2.14$. 
    \item {\bf NGC3201}: The RGB Li plateau is harder to identify in the cluster, as there appears to always be a small decrease in the abundance at decreasing temperatures. Considering this, we can define the LRGB stars between $\sim 4900$ K and $\sim 5050$ K and find a median abundance of A(Li)$=0.97\pm0.10$ dex. Calibrating models to that value we find an initial $\mathrm{A(Li)_0}=2.21$. There is a Li-rich giant in this cluster with A(Li)=$1.63\pm0.18$ dex, located before the luminosity function bump. Its evolutionary state indicated that its high Li abundance might be the product of external pollution and possibly, an accreted planet.
    \item {\bf NGC6838}: Although the number of stars in this cluster is small, we are able to find a Li plateau value of A(Li)$=0.84\pm0.10$ dex. This implies a primordial $\mathrm{A(Li)_0}=2.17$. Being the most metal-rich cluster in our sample, with [Fe/H]$=-0.72\pm0.07$ dex, we can compare its abundances with 47 Tuc. The similar LRGB Li abundance of NGC6838 with other clusters of lower metallicities suggests that, if all higher metallicity globular clusters experience main sequence depletion similar to 47 Tuc, the effect is mostly erased when they evolve to the RGB phase. It is also possible that 47 Tuc is a peculiar case of main sequence depletion.
    More abundance measurements of clusters at this high metallicity are needed to understand if either NGC6838 or 47 Tuc are unusual when compared to similar clusters.   
\end{itemize}

\begin{acknowledgements}
We would like to thank the anonymous referee for their careful reading of the manuscript and helpful comments and suggestions.
C.A.G.  acknowledges  support  from  the  National  Agency  for  Research  and  Development  (ANID) FONDECYT   Postdoctoral   Fellowship 2018 Project number 3180668. This research was supported in part by the National Science Foundation under Grant No. PHY-1430152 (JINA Center for the Evolution of the Elements).
\end{acknowledgements}

\bibliographystyle{aa.bst}
\bibliography{reference}


\begin{appendix}
\clearpage
\section{Parameters and abundances} \label{ap:param}
\longtab[1]{\begin{longtable}{lp{1.0cm}p{0.8cm}p{1.2cm}p{1.0cm}p{0.6cm}p{1.3cm}llll} 
\caption{Measured atmospheric parameters and abundances for stars in the 5 globular clusters. Stetson IDs, V, and I magnitudes are from \citet{Stetson2019}. \label{table:params}} \\
\hline
Gaia Source ID & Stetson ID & $\mathrm{T_{eff}} $ (K) & $\log g$ (dex) & $v_t$ $\,\mathrm{(km/s)}$ & Flag $\mathrm{A(Li)}$ & $\mathrm{A(Li)_{NLTE}}$ (dex) & $\mathrm{[Na/Fe]_{NLTE}}$ & $\mathrm{V mag}$ & $\mathrm{I mag}$ & SNR\\
\hline \hline
\endfirsthead
\caption{Continued.} \\
\hline
Gaia Source ID & Stetson ID & $\mathrm{T_{eff}} $ (K) & $\log g$ (dex) & $v_t$ $\,\mathrm{(km/s)}$ & Flag $\mathrm{A(Li)}$ & $\mathrm{A(Li)_{NLTE}}$ (dex) & $\mathrm{[Na/Fe]_{NLTE}}$ & $\mathrm{V mag}$ & $\mathrm{I mag}$ & SNR\\
\hline
\endhead \\
\multicolumn{9}{c}{NGC4590 $\mathrm{\langle [Fe/H]\rangle=-2.34\pm0.10}$}\\
\hline
3496356167836934016 & 67010 & 4780 & 1.68 & 1.67 & < & 0.53 & -0.28 & 14.832 & 13.775 & 260.4 \\
3496368228105102720 & 43508 & 4809 & 1.72 & 1.66 &   & 0.41 & 0.46 & 14.851 & 13.807 & 267.7 \\
3496369254600313728 & 26784 & 4493 & 1.02 & 1.92 & < & 0.05 & -0.02 & 13.7 & 12.501 & 439.5 \\
3496369877372570752 & 41758 & 4866 & 1.87 & 1.62 &   & 0.79 & -0.35 & 15.158 & 14.139 & 233.6 \\
3496371183042641152 & 59548 & 4896 & 1.93 & 1.59 &   & 0.91 & 0.4 & 15.28 & 14.274 & 209.4 \\
3496372931092316288 & 40799 & 4140 & 0.52 & 2.19 & < & -0.65 & - & 12.673 & 11.252 & 760.1 \\
3496366338319493504 & 24032 & 4827 & 1.6 & 1.7 & < & 0.5 & -0.3 & 14.409 & 13.373 & 306.4 \\
3496399422451210624 & 75588 & 4873 & 1.86 & 1.62 &   & 0.78 & 0.49 & 15.122 & 14.106 & 215.6 \\
3496371079963753088 & 77334 & 4957 & 2.15 & 1.52 &   & 0.78 & 0.34 & 15.763 & 14.782 & 161.2 \\
3496394616384561664 & 79098 & 4712 & 1.51 & 1.73 & < & 0.43 & 0.26 & 14.511 & 13.423 & 290.8 \\
3496375684168708096 & 40678 & 4139 & 0.52 & 2.19 & < & -0.67 & - & 12.717 & 11.295 & 783.7 \\
3496375203132394112 & 69220 & 4942 & 2.1 & 1.54 &   & 0.87 & 0.15 & 15.66 & 14.673 & 184.2 \\
3496371973316933888 & 70630 & 4892 & 1.95 & 1.59 &   & 0.96 & 0.1 & 15.344 & 14.336 & 197.9 \\
3496372106461287936 & 73969 & 4845 & 1.71 & 1.66 & < & 0.57 & -0.61 & 14.716 & 13.688 & 275.6 \\
3496371664079202944 & 73337 & 4952 & 2.1 & 1.54 &   & 0.99 & 0.33 & 15.621 & 14.638 & 171.3 \\
3496373244626964096 & 31721 & 4871 & 1.7 & 1.66 & < & 0.56 & 0.41 & 14.615 & 13.598 & 297.1 \\
3496374687736171392 & 36851 & 4857 & 1.86 & 1.62 &   & 0.84 & 0.48 & 15.151 & 14.128 & 222.0 \\
3496373210267235968 & 34954 & 4400 & 0.88 & 2.0 &   & -0.52 & - & 13.348 & 12.097 & 567.1 \\
3496374786518184960 & 36489 & 4954 & 2.07 & 1.55 &   & 1.0 & 0.03 & 15.527 & 14.545 & 185.2 \\
3496372935389301888 & 37517 & 4769 & 1.61 & 1.69 &   & 0.3 & -0.21 & 14.672 & 13.61 & 285.7 \\
3496374000541223040 & 28713 & 4687 & 1.45 & 1.75 & < & 0.35 & 0.33 & 14.405 & 13.305 & 326.2 \\
3496373760023015936 & 22693 & 4745 & 1.59 & 1.71 &   & 0.41 & -0.45 & 14.646 & 13.573 & 202.2 \\
3496374619014696704 & 49051 & 4957 & 2.05 & 1.56 &   & 0.93 & 0.3 & 15.47 & 14.489 & 205.0 \\
3496374649080244096 & 50919 & 5252 & 2.98 & 1.21 & < & 1.4 & 0.43 & 17.54 & 16.671 & 82.0 \\
3496374649080247808 & 52127 & 5289 & 3.06 & 1.19 & < & 1.45 & -0.59 & 17.736 & 16.88 & 79.6 \\
3496374653376525952 & 53030 & 5026 & 2.27 & 1.49 &   & 0.98 & 0.11 & 15.921 & 14.968 & 170.0 \\
3496374653376535424 & 55788 & 4989 & 2.18 & 1.52 &   & 0.86 & 0.16 & 15.755 & 14.787 & 179.0 \\
3496374962612519808 & 58277 & 5171 & 2.78 & 1.27 & < & 1.23 & -0.22 & 17.139 & 16.241 & 103.2 \\
3496374962612865152 & 59742 & 5252 & 3.01 & 1.18 & < & 1.35 & 0.15 & 17.688 & 16.819 & 91.9 \\
3496374722095925888 & 39365 & 5144 & 2.6 & 1.39 & < & 1.15 & 0.2 & 16.62 & 15.712 & 119.9 \\
3496374511641271552 & 40876 & 5240 & 2.93 & 1.23 & < & 1.35 & -0.11 & 17.43 & 16.557 & 89.5 \\
3496374619014475520 & 46780 & 5057 & 2.19 & 1.52 &   & 1.0 & 0.37 & 15.654 & 14.713 & 177.1 \\
3496374619014599296 & 49038 & 5067 & 2.4 & 1.46 &   & 0.94 & 0.2 & 16.2 & 15.263 & 149.2 \\
3496374580360787840 & 58425 & 5260 & 2.92 & 1.25 & < & 1.35 & - & 17.313 & 16.447 & 94.7 \\
3496374515937475584 & 42407 & 4868 & 1.88 & 1.61 &   & 0.94 & 0.3 & 15.19 & 14.172 & 218.0 \\
3496374408562059136 & 43074 & 4996 & 2.16 & 1.52 &   & 0.93 & 0.38 & 15.705 & 14.74 & 176.3 \\
3496374515935103104 & 45404 & 4908 & 1.95 & 1.59 &   & 0.83 & -0.03 & 15.315 & 14.314 & 178.3 \\
3496374447215621632 & 48174 & 4991 & 2.06 & 1.56 &   & 1.05 & 0.16 & 15.404 & 14.437 & 191.4 \\
3496374546001036288 & 53410 & 5036 & 2.26 & 1.5 &   & 0.98 & 0.5 & 15.838 & 14.889 & 163.3 \\
3496374546001036672 & 53535 & 5065 & 2.31 & 1.48 &   & 0.96 & 0.61 & 15.932 & 14.994 & 153.4 \\
3496371595357372416 & 56263 & 4786 & 1.63 & 1.69 & < & 0.48 & 0.43 & 14.673 & 13.619 & 295.9 \\
3496371561000005248 & 64967 & 5312 & 3.08 & 1.19 & < & 1.48 & - & 17.736 & 16.888 & 78.0 \\
3496371492280322560 & 56063 & 4918 & 2.0 & 1.58 &   & 0.93 & -0.23 & 15.409 & 14.412 & 203.4 \\
3496371457920560512 & 48774 & 5204 & 2.79 & 1.29 & < & 1.21 & -0.16 & 17.064 & 16.178 & 104.1 \\
3496371487984315008 & 51213 & 5210 & 2.85 & 1.25 & < & 1.28 & 0.63 & 17.261 & 16.377 & 97.4 \\
\hline \hline
\multicolumn{9}{c}{NGC6809 $\mathrm{\langle [Fe/H]\rangle=-1.79\pm0.10}$}\\
\hline
6751344034158411264 & 326335 & 5174 & 2.74 & 1.34 &   & 1.01 & 0.12 & 15.81 & 14.82 & 122.0 \\
6751342453610402048 & 324080 & 5235 & 2.94 & 1.27 & < & 1.08 & -0.5 & 16.246 & 15.278 & 107.5 \\
6751343995499656320 & 324051 & 5179 & 2.74 & 1.34 & < & 0.97 & - & 15.811 & 14.823 & 129.8 \\
6751342556689633792 & 320469 & 5324 & 3.17 & 1.22 & < & 1.26 & 0.49 & 16.775 & 15.838 & 80.0 \\
6751343484402601088 & 319643 & 5272 & 3.03 & 1.23 & < & 1.22 & 0.14 & 16.509 & 15.554 & 79.0 \\
6751391686812938496 & 315902 & 5450 & 3.3 & 1.2 & < & 1.36 & 0.32 & 17.051 & 16.156 & 77.2 \\
6751341109281472768 & 313156 & 5359 & 3.19 & 1.22 &   & 1.13 & -0.12 & 16.801 & 15.876 & 81.5 \\
6751390797762448640 & 311803 & 5309 & 2.98 & 1.29 &   & 1.24 & -0.13 & 16.279 & 15.337 & 105.2 \\
6751343724920865152 & 309106 & 5238 & 2.87 & 1.29 & < & 1.1 & 0.23 & 16.144 & 15.177 & 106.2 \\
6751341830836349440 & 308529 & 5083 & 2.42 & 1.44 &   & 0.99 & 0.09 & 15.182 & 14.158 & 155.1 \\
6751392305288190592 & 307788 & 5332 & 3.12 & 1.23 & < & 1.2 & 0.26 & 16.64 & 15.706 & 94.8 \\
6751390763402728064 & 307656 & 5347 & 3.18 & 1.22 &   & 1.2 & 0.39 & 16.795 & 15.866 & 75.9 \\
6751342140074767232 & 306943 & 5456 & 3.23 & 1.23 &   & 1.46 & 0.38 & 16.878 & 15.985 & 72.0 \\
6751390656025341056 & 306141 & 5297 & 2.98 & 1.28 & < & 1.16 & -0.38 & 16.286 & 15.34 & 96.1 \\
6751390660317035008 & 306056 & 5318 & 3.17 & 1.21 & < & 1.23 & -0.05 & 16.795 & 15.856 & 83.7 \\
6751342140074761600 & 305242 & 5067 & 2.28 & 1.47 &   & 1.13 & -0.46 & 14.851 & 13.821 & 206.9 \\
6751340941781941376 & 305193 & 5438 & 3.05 & 1.28 & < & 1.28 & 0.06 & 16.402 & 15.503 & 92.2 \\
6751391038280721664 & 303831 & 5029 & 2.18 & 1.51 &   & 1.12 & 0.03 & 14.624 & 13.579 & 219.8 \\
6751342110013112960 & 302916 & 5215 & 2.8 & 1.32 &   & 1.2 & 0.36 & 15.965 & 14.99 & 103.0 \\
6751341903854668800 & 302540 & 4981 & 2.1 & 1.53 &   & 1.09 & 0.12 & 14.509 & 13.445 & 228.4 \\
6751390694683279744 & 301860 & 5026 & 2.04 & 1.55 &   & 1.13 & 0.07 & 14.382 & 13.336 & 259.2 \\
6751342208794224512 & 301300 & 4804 & 1.61 & 1.68 &   & 0.24 & -0.16 & 13.529 & 12.39 & 403.8 \\
6751341693397940864 & 300879 & 5286 & 3.01 & 1.25 & < & 1.2 & 0.03 & 16.419 & 15.469 & 88.6 \\
6751341319739079168 & 298844 & 5318 & 3.15 & 1.22 & < & 1.23 & -0.68 & 16.706 & 15.767 & 83.5 \\
6751389144196825856 & 298790 & 5266 & 2.99 & 1.26 & < & 1.16 & -0.07 & 16.392 & 15.435 & 94.0 \\
6751342174434476032 & 298598 & 5289 & 2.99 & 1.27 & < & 1.17 & -0.34 & 16.347 & 15.398 & 96.5 \\
6751389144196814464 & 296015 & 5330 & 3.2 & 1.19 & < & 1.26 & 0.52 & 16.907 & 15.972 & 81.7 \\
6751341285379341056 & 293901 & 5243 & 3.0 & 1.24 &   & 1.19 & 0.08 & 16.434 & 15.469 & 97.4 \\
6751389114135277952 & 293635 & 5229 & 2.59 & 1.41 &   & 1.21 & 0.24 & 15.411 & 14.441 & 140.7 \\
6751341968275842432 & 293220 & 5179 & 2.63 & 1.38 & < & 0.97 & 0.41 & 15.608 & 14.62 & 134.2 \\
6751390900841793920 & 292670 & 5088 & 2.37 & 1.46 &   & 1.05 & -0.21 & 14.986 & 13.964 & 207.9 \\
6751341349799595904 & 292563 & 5356 & 3.21 & 1.2 & < & 1.27 & 0.19 & 16.922 & 15.996 & 81.6 \\
6751391003921076608 & 292256 & 4999 & 2.2 & 1.51 &   & 1.03 & -0.63 & 14.687 & 13.63 & 239.3 \\
6751341182300980992 & 290518 & 5109 & 2.53 & 1.4 &   & 0.99 & 0.15 & 15.368 & 14.354 & 141.9 \\
6751389419074690176 & 289029 & 5389 & 3.2 & 1.22 & < & 1.28 & - & 16.825 & 15.91 & 88.9 \\
6751342036995315584 & 288835 & 5168 & 2.64 & 1.37 &   & 1.07 & - & 15.652 & 14.66 & 108.2 \\
6751389182848799360 & 288516 & 5321 & 2.72 & 1.38 & < & 1.04 & -0.5 & 15.638 & 14.7 & 143.9 \\
6751388972398082816 & 287053 & 5306 & 3.15 & 1.21 & < & 1.26 & 0.42 & 16.74 & 15.797 & 75.4 \\
6751341659038190208 & 286192 & 5295 & 3.0 & 1.26 & < & 1.21 & - & 16.381 & 15.434 & 85.9 \\
6751389389007466752 & 286170 & 5257 & 2.83 & 1.31 & < & 1.04 & 0.06 & 16.033 & 15.073 & 128.9 \\
6751389384714938112 & 286012 & 5463 & 2.92 & 1.35 &   & 1.22 & -0.17 & 15.989 & 15.098 & 129.2 \\
6751389079769304320 & 285750 & 5277 & 2.91 & 1.29 &   & 1.18 & 0.31 & 16.171 & 15.218 & 93.9 \\
6751389389007492864 & 283117 & 5052 & 2.16 & 1.52 &   & 1.08 & 0.06 & 14.557 & 13.521 & 242.0 \\
6751389041117539328 & 282936 & 5303 & 2.82 & 1.33 &   & 1.02 & -0.43 & 15.938 & 14.994 & 119.2 \\
6751389389013324416 & 282347 & 5274 & 2.83 & 1.32 &   & 1.21 & -0.09 & 16.006 & 15.052 & 118.8 \\
6751391721177129344 & 282318 & 5179 & 2.68 & 1.35 &   & 0.97 & 0.01 & 15.718 & 14.73 & 145.9 \\
6751389487794135040 & 282263 & 5198 & 2.78 & 1.33 &   & 1.03 & -0.57 & 15.911 & 14.93 & 119.7 \\
6751341560257307904 & 282240 & 5120 & 2.46 & 1.42 &   & 1.04 & -0.26 & 15.254 & 14.244 & 138.9 \\
6751389075477272448 & 281826 & 5383 & 2.65 & 1.4 &   & 1.14 & - & 15.439 & 14.522 & 129.6 \\
6751390248006850048 & 281308 & 5383 & 3.19 & 1.23 &   & 1.3 & 0.11 & 16.777 & 15.86 & 84.1 \\
6751389045415866752 & 280881 & 5235 & 2.55 & 1.42 &   & 1.11 & 0.22 & 15.282 & 14.314 & 142.0 \\
6751389281635697024 & 280653 & 5212 & 2.73 & 1.35 &   & 0.98 & 0.31 & 15.758 & 14.782 & 136.0 \\
6751389453434385024 & 279936 & 5212 & 2.76 & 1.35 &   & 1.02 & 0.01 & 15.803 & 14.827 & 136.3 \\
6751389453434382464 & 279541 & 5315 & 2.81 & 1.34 &   & 1.16 & -0.39 & 15.891 & 14.951 & 103.9 \\
6751389281635686528 & 278571 & 5218 & 2.76 & 1.35 &   & 1.12 & -0.39 & 15.81 & 14.836 & 127.7 \\
6751389251574759552 & 277507 & 5289 & 3.04 & 1.23 &   & 1.18 & 0.14 & 16.521 & 15.572 & 95.2 \\
6751389453434371456 & 277070 & 5152 & 2.51 & 1.41 &   & 1.05 & 0.11 & 15.301 & 14.303 & 167.8 \\
6751389251568413056 & 276681 & 5410 & 2.9 & 1.33 &   & 1.27 & -0.24 & 16.022 & 15.114 & 110.5 \\
6751388564373105920 & 275964 & 5263 & 2.81 & 1.33 &   & 1.14 & 0.26 & 15.946 & 14.988 & 106.8 \\
6751389354653600384 & 275405 & 5315 & 2.97 & 1.29 & < & 1.16 & 0.05 & 16.251 & 15.311 & 101.6 \\
6751388667455550848 & 274415 & 5324 & 3.21 & 1.18 & < & 1.31 & - & 16.957 & 16.02 & 73.0 \\
6751388564380202368 & 273315 & 5131 & 2.56 & 1.4 & < & 0.93 & 0.37 & 15.421 & 14.415 & 135.7 \\
6751389247275926272 & 273262 & 5422 & 3.18 & 1.25 & < & 1.33 & 0.46 & 16.722 & 15.818 & 79.9 \\
6751388560081158656 & 273227 & 5144 & 2.58 & 1.4 &   & 1.08 & 0.06 & 15.454 & 14.453 & 107.8 \\
6751388628800630272 & 271839 & 5327 & 3.19 & 1.2 & < & 1.3 & - & 16.875 & 15.939 & 72.7 \\
6751388663160365312 & 270899 & 5330 & 3.02 & 1.26 & < & 1.21 & -0.33 & 16.419 & 15.484 & 93.8 \\
6751389320294242944 & 270742 & 5185 & 2.59 & 1.4 & < & 0.92 & 0.29 & 15.446 & 14.46 & 152.8 \\
6751388873617246464 & 269381 & 5356 & 3.08 & 1.25 &   & 1.08 & - & 16.546 & 15.62 & 84.8 \\
6751390518588716160 & 268772 & 5263 & 3.0 & 1.25 & < & 1.13 & -0.38 & 16.43 & 15.472 & 99.2 \\
6751395225866017536 & 267230 & 5353 & 3.19 & 1.21 & < & 1.26 & 0.06 & 16.819 & 15.892 & 86.5 \\
6751388457001828864 & 266435 & 5318 & 3.15 & 1.22 & < & 1.27 & 0.39 & 16.713 & 15.774 & 77.2 \\
6751388938038247040 & 264847 & 5120 & 2.37 & 1.46 &   & 0.95 & -0.12 & 14.971 & 13.961 & 173.3 \\
6751390591604327168 & 264022 & 5096 & 2.43 & 1.43 &   & 1.06 & -0.05 & 15.204 & 14.185 & 151.4 \\
6751294586192199424 & 260457 & 5246 & 2.85 & 1.29 & < & 1.12 & -0.24 & 16.108 & 15.144 & 98.7 \\
6751388731879732864 & 260403 & 5226 & 2.86 & 1.28 &   & 1.16 & -0.15 & 16.146 & 15.175 & 98.2 \\
6751389698245827456 & 260043 & 5286 & 3.14 & 1.21 &   & 1.28 & 0.1 & 16.741 & 15.791 & 131.9 \\
6751388907977025280 & 260039 & 5196 & 2.76 & 1.34 & < & 1.04 & -0.33 & 15.822 & 14.84 & 112.4 \\
6751389663892790912 & 259514 & 5235 & 2.66 & 1.37 &   & 1.31 & -0.34 & 15.634 & 14.666 & 113.6 \\
6751390140629070208 & 259067 & 4873 & 1.8 & 1.63 &   & 0.48 & 0.15 & 13.891 & 12.782 & 335.6 \\
6751389968830367872 & 255421 & 5232 & 3.03 & 1.21 &   & 1.19 & 0.01 & 16.575 & 15.606 & 94.3 \\
6751389728312198144 & 255015 & 5274 & 2.97 & 1.27 &   & 1.12 & -0.34 & 16.306 & 15.352 & 104.1 \\
6751390415505772672 & 249222 & 5212 & 2.86 & 1.28 &   & 1.02 & 0.2 & 16.151 & 15.175 & 114.6 \\
6751389934470118272 & 247467 & 5210 & 2.93 & 1.26 & < & 1.12 & -0.04 & 16.266 & 15.289 & 93.6 \\
6751393370443742080 & 247265 & 5204 & 2.96 & 1.25 & < & 1.09 & 0.11 & 16.364 & 15.385 & 98.6 \\
6751389831390224256 & 245323 & 5218 & 2.99 & 1.23 & < & 1.11 & 0.09 & 16.447 & 15.473 & 95.8 \\
6751294895432000512 & 245081 & 5141 & 2.64 & 1.37 &   & 0.88 & 0.32 & 15.685 & 14.683 & 132.1 \\
6751392958126845568 & 245026 & 5254 & 3.07 & 1.21 & < & 1.23 & 0.11 & 16.659 & 15.698 & 74.7 \\
6751392958126849024 & 240566 & 5254 & 3.05 & 1.21 &   & 1.23 & -0.13 & 16.62 & 15.659 & 93.0 \\
6751389865750374784 & 239906 & 5182 & 2.95 & 1.24 &   & 0.84 & 0.05 & 16.378 & 15.391 & 111.3 \\
6751295999240906752 & 236052 & 5049 & 2.44 & 1.43 &   & 0.93 & -0.17 & 15.286 & 14.249 & 180.2 \\
\hline \hline
\multicolumn{9}{c}{NGC6656 $\mathrm{\langle [Fe/H]\rangle=-1.77\pm0.12}$}\\
\hline
4077588663251815680 & 368904 & 4860 & 1.81 & 1.62 &   & 0.94 & -0.27 & 14.443 & 13.044 & 252.5 \\
4077588658876838272 & 377616 & 4981 & 1.98 & 1.58 &   & 0.84 & - & 14.629 & 13.287 & 224.9 \\
4077588761956054656 & 379517 & 5207 & 2.25 & 1.5 &   & 1.09 & -0.13 & 15.147 & 13.89 & 172.4 \\
4077588560086108800 & 410829 & 4823 & 1.74 & 1.64 &   & 1.0 & -0.35 & 14.193 & 12.822 & 283.2 \\
4077588590157385856 & 412896 & 5005 & 2.07 & 1.55 &   & 0.95 & 0.04 & 14.811 & 13.515 & 182.3 \\
4077588590157388160 & 416194 & 5335 & 2.37 & 1.47 & < & 1.14 & 0.65 & 15.251 & 14.071 & 141.4 \\
4077588628892104960 & 350337 & 4881 & 1.8 & 1.63 &   & 0.88 & 0.18 & 14.231 & 12.871 & 255.5 \\
4077588457007213184 & 373829 & 5230 & 2.04 & 1.55 &   & 1.09 & 0.45 & 14.577 & 13.341 & 213.8 \\
4077588383998945792 & 398364 & 5200 & 2.04 & 1.55 &   & 1.08 & 0.83 & 14.646 & 13.394 & 218.4 \\
4077588491374970880 & 408580 & 4933 & 1.87 & 1.6 &   & 0.86 & 0.37 & 14.376 & 13.043 & 242.4 \\
4077588487078079488 & 414857 & 4964 & 1.83 & 1.61 &   & 0.89 & 0.66 & 14.091 & 12.785 & 250.9 \\
4077588521437920896 & 430379 & 5315 & 2.11 & 1.55 &   & 1.15 & - & 14.493 & 13.318 & 228.9 \\
4077588418358572160 & 357743 & 5174 & 2.1 & 1.54 &   & 1.02 & - & 14.812 & 13.556 & 211.9 \\
4077588418358573824 & 364260 & 5021 & 2.13 & 1.53 &   & 0.89 & 0.33 & 15.061 & 13.744 & 180.3 \\
4077588422647379584 & 370752 & 5057 & 2.28 & 1.48 &   & 0.91 & - & 15.301 & 13.993 & 171.0 \\
4077588353927668608 & 378794 & 4990 & 2.17 & 1.51 &   & 0.74 & - & 15.241 & 13.895 & 181.2 \\
4077588383998846592 & 389508 & 4848 & 1.79 & 1.63 &   & 0.78 & 0.18 & 14.409 & 13.001 & 257.5 \\
4077588383998852736 & 397395 & 4887 & 1.82 & 1.62 &   & 0.98 & -0.41 & 14.384 & 13.003 & 227.7 \\
4077588388287066112 & 399172 & 5150 & 2.01 & 1.57 &   & 0.91 & 0.76 & 14.488 & 13.219 & 234.4 \\
4077588383998856448 & 403369 & 5114 & 2.08 & 1.54 &   & 0.82 & 0.54 & 14.804 & 13.527 & 191.2 \\
4077587632372496384 & 416210 & 4942 & 2.21 & 1.5 &   & 0.9 & 0.2 & 15.347 & 14.001 & 161.2 \\
4077587731163850624 & 432474 & 5096 & 2.1 & 1.54 &   & 1.01 & 0.42 & 14.817 & 13.555 & 197.2 \\
4077494513194645248 & 370100 & 4947 & 1.81 & 1.62 &   & 1.01 & 0.01 & 14.123 & 12.775 & 294.7 \\
4077494513194647936 & 376994 & 5391 & 2.67 & 1.4 &   & 1.1 & 0.58 & 15.748 & 14.553 & 139.2 \\
4077588349639105664 & 384662 & 4992 & 1.85 & 1.6 &   & 0.94 & 0.57 & 14.266 & 12.913 & 271.6 \\
4077588354014015616 & 390011 & 4939 & 2.0 & 1.57 &   & 0.8 & -0.37 & 14.825 & 13.457 & 218.0 \\
4077587628084617984 & 411350 & 5079 & 2.0 & 1.57 &   & 1.1 & -0.27 & 14.582 & 13.281 & 209.2 \\
4077587529292978560 & 414393 & 5332 & 2.87 & 1.33 & < & 1.2 & 0.55 & 16.246 & 15.043 & 113.7 \\
4077587563652581248 & 432135 & 5055 & 2.46 & 1.43 & < & 0.93 & 0.2 & 15.731 & 14.43 & 132.8 \\
4077587662444374016 & 432601 & 5179 & 2.23 & 1.5 &   & 1.04 & 0.47 & 15.11 & 13.857 & 169.0 \\
4077494483219998336 & 360399 & 5286 & 2.46 & 1.45 &   & 1.15 & -0.03 & 15.421 & 14.206 & 160.9 \\
4077587593724864512 & 389612 & 5210 & 2.1 & 1.54 &   & 1.22 & -0.23 & 14.815 & 13.564 & 203.2 \\
4077497644316147840 & 276208 & 5168 & 2.73 & 1.32 & < & 1.08 & -0.43 & 16.242 & 14.994 & 113.7 \\
4077591996146712320 & 283673 & 5076 & 2.59 & 1.38 &   & 1.0 & 0.51 & 16.139 & 14.833 & 114.4 \\
4077497747395400448 & 276100 & 5131 & 2.54 & 1.41 & < & 0.99 & -0.68 & 15.757 & 14.49 & 136.2 \\
4077497708642558464 & 260869 & 5106 & 2.09 & 1.54 &   & 1.04 & 0.51 & 14.819 & 13.54 & 215.3 \\
4077592511542754816 & 358787 & 5241 & 2.88 & 1.29 &   & 1.08 & 0.52 & 16.507 & 15.284 & 96.6 \\
4077591789902826112 & 356349 & 5171 & 2.81 & 1.28 & < & 1.15 & 0.44 & 16.471 & 15.236 & 95.0 \\
4076836352399393408 & 481388 & 5202 & 2.81 & 1.3 & < & 1.2 & - & 16.368 & 15.144 & 94.9 \\
4077591819965737600 & 372658 & 5122 & 2.75 & 1.29 &   & 1.08 & 0.4 & 16.371 & 15.132 & 88.0 \\
4077493967753050752 & 314736 & 5054 & 2.66 & 1.28 &   & 0.93 & 0.08 & 16.333 & 15.09 & 107.1 \\
4077497777360246016 & 300161 & 5202 & 2.79 & 1.31 & < & 1.16 & 0.6 & 16.273 & 15.053 & 102.8 \\
4077592202305011584 & 375932 & 5164 & 2.74 & 1.32 &   & 1.11 & 0.33 & 16.233 & 15.003 & 109.9 \\
4077494654948461824 & 313278 & 5188 & 2.71 & 1.35 &   & 1.05 & 0.19 & 16.106 & 14.868 & 110.2 \\
4077587662444381952 & 444075 & 5187 & 2.7 & 1.35 &   & 1.1 & 0.24 & 16.098 & 14.853 & 98.8 \\
4077588113495723008 & 481069 & 5191 & 2.7 & 1.36 &   & 1.28 & 0.43 & 16.052 & 14.819 & 107.4 \\
4077590617373368064 & 475444 & 5099 & 2.59 & 1.38 &   & 0.75 & 0.45 & 16.073 & 14.776 & 113.8 \\
4077589006771268864 & 447304 & 5132 & 2.63 & 1.37 &   & 0.81 & -0.54 & 16.019 & 14.767 & 121.3 \\
4077494478834898176 & 352147 & 5141 & 2.61 & 1.38 &   & 1.13 & 0.39 & 15.948 & 14.689 & 121.3 \\
4077588968106307712 & 432629 & 5177 & 2.68 & 1.36 & < & 1.1 & 0.6 & 15.988 & 14.768 & 110.7 \\
4077591648174052224 & 324068 & 5109 & 2.57 & 1.39 &   & 0.8 & 0.63 & 15.931 & 14.66 & 120.8 \\
4077493727234677760 & 368201 & 5134 & 2.57 & 1.4 &   & 1.11 & 0.33 & 15.855 & 14.592 & 121.6 \\
4077589316000968192 & 445345 & 5272 & 2.73 & 1.36 &   & 1.17 & 0.01 & 15.947 & 14.739 & 105.7 \\
4077592202219665280 & 380144 & 5090 & 2.59 & 1.38 & < & 1.0 & 0.4 & 15.966 & 14.711 & 115.8 \\
4077494380069993856 & 339684 & 5083 & 2.56 & 1.39 &   & 0.8 & 0.12 & 15.928 & 14.665 & 126.9 \\
4077591648165458304 & 325389 & 5144 & 2.6 & 1.39 & < & 1.01 & -0.17 & 15.898 & 14.64 & 129.2 \\
4077588761956051968 & 375635 & 5130 & 2.55 & 1.4 &   & 0.99 & 0.21 & 15.847 & 14.568 & 130.6 \\
4077493280556921600 & 409713 & 5049 & 2.52 & 1.4 &   & 0.9 & 0.43 & 15.847 & 14.601 & 126.9 \\
4077592129201872512 & 314072 & 5083 & 2.53 & 1.4 &   & 0.82 & 0.64 & 15.855 & 14.579 & 126.5 \\
4077588079048728192 & 470314 & 5084 & 2.53 & 1.4 &   & 0.82 & 0.47 & 15.83 & 14.565 & 117.7 \\
4077588938043174912 & 447899 & 5049 & 2.49 & 1.41 &   & 0.97 & 0.33 & 15.81 & 14.53 & 131.3 \\
4077494822424525952 & 303010 & 5087 & 2.52 & 1.41 & < & 1.09 & 0.45 & 15.764 & 14.504 & 97.5 \\
4077493727305709184 & 362412 & 5290 & 2.66 & 1.39 &   & 1.28 & 0.25 & 15.689 & 14.491 & 129.7 \\
4077592271024514176 & 370654 & 5044 & 2.48 & 1.41 &   & 0.88 & 0.11 & 15.802 & 14.515 & 135.1 \\
4077494861110024448 & 322055 & 5147 & 2.55 & 1.4 &   & 1.03 & 0.25 & 15.724 & 14.475 & 137.5 \\
4077589105544866432 & 447306 & 5156 & 2.56 & 1.4 &   & 0.99 & 0.29 & 15.739 & 14.497 & 116.3 \\
4077587421917657984 & 472852 & 5076 & 2.47 & 1.43 &   & 0.97 & -0.23 & 15.683 & 14.394 & 106.2 \\
4077591824347848064 & 370494 & 5223 & 2.61 & 1.4 &   & 1.01 & 0.2 & 15.669 & 14.458 & 133.0 \\
4077587563652352384 & 427901 & 5222 & 2.54 & 1.42 &   & 0.95 & 0.48 & 15.619 & 14.375 & 133.6 \\
4077591648174053376 & 332770 & 5031 & 2.41 & 1.44 &   & 0.78 & 0.03 & 15.632 & 14.336 & 137.3 \\
4077591824262504192 & 369580 & 5211 & 2.56 & 1.41 &   & 1.03 & -0.55 & 15.581 & 14.372 & 127.9 \\
4077591923043403648 & 305397 & 5317 & 2.5 & 1.43 &   & 1.17 & 0.46 & 15.511 & 14.299 & 120.3 \\
4077587357581444480 & 442522 & 5047 & 2.38 & 1.45 & < & 0.93 & 0.12 & 15.538 & 14.226 & 140.0 \\
4077494822432267904 & 323639 & 5319 & 2.52 & 1.43 &   & 1.05 & 0.67 & 15.489 & 14.301 & 147.3 \\
4077493413674983552 & 420886 & 5096 & 2.47 & 1.43 &   & 0.73 & - & 15.534 & 14.289 & 150.1 \\
4077493555506972416 & 365795 & 5082 & 2.45 & 1.43 &   & 1.03 & -0.43 & 15.511 & 14.266 & 144.6 \\
4076742584671349248 & 429983 & 5158 & 2.49 & 1.43 &   & 1.06 & -0.04 & 15.464 & 14.245 & 156.3 \\
4077587662436171136 & 447193 & 5045 & 2.36 & 1.46 &   & 0.86 & 0.04 & 15.473 & 14.172 & 137.1 \\
4077588143480814720 & 456131 & 5093 & 2.43 & 1.44 &   & 0.88 & 0.52 & 15.462 & 14.208 & 139.6 \\
4077588834965047424 & 360069 & 5125 & 2.44 & 1.44 &   & 0.92 & 0.61 & 15.462 & 14.202 & 145.4 \\
4077587391853159552 & 469946 & 5071 & 2.38 & 1.46 &   & 0.93 & 0.59 & 15.454 & 14.163 & 136.2 \\
4077588143472508544 & 452973 & 5071 & 2.39 & 1.45 & < & 0.98 & - & 15.432 & 14.166 & 125.1 \\
4077591682533796864 & 345027 & 4985 & 2.27 & 1.48 &   & 0.93 & 0.11 & 15.327 & 14.01 & 158.9 \\
4077587524997281152 & 411988 & 4987 & 2.16 & 1.51 & < & 0.8 & - & 15.16 & 13.84 & 171.5 \\
4077494753712791424 & 324783 & 5088 & 2.22 & 1.5 &   & 0.7 & 0.7 & 15.122 & 13.852 & 187.7 \\
4077494856782482304 & 322939 & 5344 & 2.23 & 1.5 &   & 1.19 & 0.58 & 15.05 & 13.868 & 169.6 \\
4077588525812688896 & 439821 & 5196 & 2.11 & 1.53 &   & 0.91 & 0.83 & 14.764 & 13.547 & 184.2 \\
4077493967823900544 & 324422 & 5046 & 2.07 & 1.55 &   & 1.01 & 0.22 & 14.722 & 13.467 & 205.9 \\
4077494444467310592 & 330527 & 4931 & 1.98 & 1.58 &   & 0.64 & - & 14.679 & 13.344 & 240.4 \\
4077591888683618816 & 350137 & 4901 & 1.93 & 1.58 &   & 0.85 & -0.41 & 14.618 & 13.273 & 213.7 \\
4077591751253266816 & 317142 & 4930 & 1.91 & 1.59 &   & 1.0 & -0.61 & 14.525 & 13.174 & 233.2 \\
4077589144288088448 & 461833 & 4905 & 1.88 & 1.6 &   & 0.65 & 0.69 & 14.479 & 13.129 & 212.4 \\
4077494375780165120 & 332065 & 4925 & 1.89 & 1.59 &   & 0.89 & 0.52 & 14.385 & 13.071 & 255.3 \\
4077587288861927808 & 445577 & 4918 & 1.84 & 1.61 &   & 0.94 & -0.03 & 14.345 & 12.989 & 222.0 \\
4077588800612974848 & 410770 & 4894 & 1.85 & 1.61 &   & 0.87 & 0.52 & 14.355 & 13.025 & 250.5 \\
4076837112684613504 & 513824 & 5124 & 2.69 & 1.33 & < & 1.06 & 0.44 & 16.291 & 15.011 & 106.5 \\
4077588212192057856 & 498572 & 5046 & 2.3 & 1.48 &   & 0.71 & 0.55 & 15.291 & 14.001 & 156.2 \\
4076836803447012864 & 500055 & 5002 & 1.99 & 1.58 &   & 0.77 & 0.44 & 14.587 & 13.273 & 224.7 \\
\hline \hline
\multicolumn{9}{c}{NGC3201 $\mathrm{\langle [Fe/H]\rangle=-1.58\pm0.06}$}\\
\hline
5413528757309438976 & 117898 & 4966 & 2.18 & 1.5 &   & 0.86 & - & 15.338 & 14.081 & 138.2 \\
5413574764994407296 & 60587 & 5037 & 2.49 & 1.39 & < & 1.07 & -0.19 & 15.892 & 14.727 & 112.4 \\
5413576620427026688 & 62920 & 5215 & 2.8 & 1.31 &   & 1.24 & 0.29 & 16.651 & 15.462 & 78.6 \\
5413574936799800960 & 64905 & 4910 & 2.05 & 1.53 &   & 0.76 & - & 15.184 & 13.9 & 119.9 \\
5413576757865689472 & 65575 & 5028 & 2.44 & 1.41 &   & 1.04 & -0.09 & 15.946 & 14.71 & 105.8 \\
5413576757860772352 & 67621 & 5046 & 2.35 & 1.45 &   & 1.19 & -0.17 & 15.647 & 14.417 & 105.8 \\
5413573528050401920 & 67864 & 5099 & 2.62 & 1.37 & < & 1.17 & 0.41 & 16.139 & 15.0 & 100.3 \\
5413577165872587904 & 68916 & 4982 & 2.26 & 1.47 &   & 0.86 & - & 15.546 & 14.294 & 125.8 \\
5413573562410144896 & 69996 & 4782 & 1.75 & 1.64 &   & 0.54 & -0.33 & 14.371 & 13.096 & 232.6 \\
5413573940367652736 & 70894 & 5193 & 2.99 & 1.24 & < & 1.36 & - & 16.94 & 15.838 & 69.3 \\
5413576547403214336 & 74008 & 5071 & 2.57 & 1.38 &   & 1.12 & 0.17 & 16.117 & 14.913 & 96.3 \\
5413573974727042944 & 74399 & 5047 & 2.39 & 1.44 & < & 1.03 & 0.48 & 15.616 & 14.457 & 125.4 \\
5413576586067387904 & 74864 & 5007 & 2.24 & 1.47 &   & 1.1 & -0.09 & 15.452 & 14.234 & 128.6 \\
5413573145784283520 & 74998 & 5123 & 2.72 & 1.33 &   & 1.26 & 0.24 & 16.364 & 15.228 & 87.1 \\
5413573493690665728 & 75717 & 5092 & 2.86 & 1.24 & < & 1.28 & -0.12 & 16.824 & 15.676 & 72.0 \\
5413573867339062144 & 76722 & 5174 & 2.87 & 1.27 & < & 1.34 & 0.42 & 16.711 & 15.596 & 77.3 \\
5413574073501937408 & 77070 & 5041 & 2.38 & 1.44 &   & 1.03 & 0.19 & 15.646 & 14.461 & 125.4 \\
5413574039142213120 & 81284 & 5163 & 2.84 & 1.28 & < & 1.33 & 0.3 & 16.75 & 15.59 & 76.4 \\
5413576998384278784 & 81845 & 4993 & 2.16 & 1.51 &   & 0.82 & - & 15.176 & 13.979 & 150.2 \\
5413573661180542848 & 81952 & 5173 & 2.9 & 1.26 & < & 1.36 & -0.04 & 16.76 & 15.645 & 74.1 \\
5413573665489370624 & 82941 & 5193 & 2.83 & 1.29 &   & 1.29 & -0.01 & 16.604 & 15.496 & 77.7 \\
5413576891000614272 & 83375 & 4921 & 2.09 & 1.52 &   & 1.04 & - & 15.125 & 13.899 & 145.8 \\
5413576517347896192 & 83561 & 5054 & 2.47 & 1.4 &   & 1.16 & -0.31 & 15.911 & 14.717 & 86.5 \\
5413576895305030144 & 84320 & 5069 & 2.48 & 1.4 & < & 1.14 & 0.31 & 15.87 & 14.698 & 102.3 \\
5413576891000617088 & 85224 & 5028 & 2.52 & 1.39 &   & 0.94 & -0.26 & 15.979 & 14.799 & 83.3 \\
5413576929664797056 & 85420 & 5114 & 2.69 & 1.33 &   & 1.23 & -0.14 & 16.332 & 15.184 & 83.8 \\
5413576139390780160 & 88006 & 5022 & 2.41 & 1.43 & < & 1.06 & 0.33 & 15.817 & 14.604 & 113.7 \\
5413576139385749248 & 88986 & 5182 & 2.76 & 1.33 &   & 1.23 & 0.23 & 16.39 & 15.258 & 83.1 \\
5413577067103768448 & 90932 & 4989 & 2.19 & 1.5 &   & 1.08 & - & 15.23 & 14.032 & 149.3 \\
5413573837288105984 & 91115 & 5099 & 2.51 & 1.4 &   & 1.03 & -0.2 & 15.912 & 14.74 & 114.4 \\
5413573734208868096 & 91828 & 5058 & 2.43 & 1.42 & < & 1.09 & 0.39 & 15.783 & 14.604 & 113.9 \\
5413575314756912768 & 93635 & 4970 & 2.21 & 1.49 &   & 0.77 & -0.0 & 15.426 & 14.173 & 139.1 \\
5413573768568647552 & 94417 & 5076 & 2.43 & 1.42 &   & 1.04 & 0.35 & 15.826 & 14.627 & 92.5 \\
5413526657058305152 & 94562 & 5185 & 2.82 & 1.29 & < & 1.36 & -0.11 & 16.683 & 15.524 & 75.0 \\
5413576070671310720 & 96397 & 5001 & 2.3 & 1.46 &   & 0.94 & -0.44 & 15.552 & 14.329 & 121.8 \\
5413575447891536256 & 96886 & 4895 & 2.12 & 1.51 &   & 0.95 & -0.09 & 15.311 & 14.017 & 151.8 \\
5413575447891536896 & 97114 & 5027 & 2.42 & 1.42 & < & 1.09 & -0.08 & 15.905 & 14.658 & 106.3 \\
5413582186704791168 & 97812 & 4937 & 2.03 & 1.54 &   & 1.63 & - & 15.029 & 13.81 & 163.3 \\
5413575241728645120 & 97963 & 4833 & 1.89 & 1.58 &   & 0.98 & 0.34 & 14.846 & 13.54 & 189.6 \\
5413575447891539072 & 98029 & 5028 & 2.29 & 1.46 &   & 1.07 & -0.15 & 15.565 & 14.328 & 126.9 \\
5413575280397190016 & 99159 & 4985 & 2.33 & 1.45 &   & 0.89 & - & 15.644 & 14.402 & 127.3 \\
5413575894568230656 & 101600 & 5071 & 2.84 & 1.25 & < & 1.14 & -0.14 & 16.949 & 15.72 & 62.4 \\
5413575383476473088 & 101793 & 4999 & 2.31 & 1.45 &   & 0.84 & - & 15.638 & 14.387 & 114.7 \\
5413575276092858368 & 101809 & 5048 & 2.47 & 1.4 &   & 1.0 & -0.1 & 15.979 & 14.759 & 105.5 \\
5413575379172073856 & 102080 & 5126 & 2.44 & 1.41 & < & 1.15 & -0.13 & 15.834 & 14.639 & 111.6 \\
5413576001951864064 & 104083 & 4989 & 2.29 & 1.46 &   & 1.09 & -0.02 & 15.636 & 14.369 & 124.9 \\
5413581907522447488 & 104928 & 4987 & 2.15 & 1.51 &   & 1.03 & -0.4 & 15.244 & 14.01 & 142.9 \\
5413575933232381696 & 106829 & 5077 & 2.56 & 1.39 & < & 1.18 & 0.09 & 16.151 & 14.924 & 93.1 \\
5413528482431516672 & 107985 & 5041 & 2.46 & 1.4 &   & 0.99 & -0.19 & 15.972 & 14.745 & 108.3 \\
5413528787362489856 & 110428 & 5053 & 2.46 & 1.4 &   & 1.09 & -0.2 & 15.987 & 14.755 & 100.2 \\
5413528791669165440 & 110970 & 5004 & 2.38 & 1.44 &   & 0.91 & 0.35 & 15.771 & 14.523 & 131.9 \\
5413529032187348608 & 111064 & 5122 & 2.79 & 1.3 & < & 1.28 & 0.09 & 16.677 & 15.47 & 79.9 \\
5413528826028906880 & 113875 & 4949 & 2.05 & 1.53 &   & 0.95 & - & 15.158 & 13.895 & 159.1 \\
5413577548139593984 & 43732 & 5177 & 2.88 & 1.27 & < & 1.44 & -0.26 & 16.692 & 15.593 & 63.5 \\
5413577685578562944 & 45453 & 5249 & 2.92 & 1.28 & < & 1.43 & - & 16.69 & 15.61 & 73.8 \\
5413577616859080960 & 47619 & 5221 & 2.9 & 1.27 & < & 1.44 & 0.15 & 16.701 & 15.61 & 67.0 \\
5413574421403594240 & 51296 & 5233 & 3.0 & 1.25 & < & 1.44 & 0.04 & 16.903 & 15.822 & 69.8 \\
5413574558842563328 & 54279 & 5133 & 2.77 & 1.32 &   & 1.28 & -0.06 & 16.399 & 15.291 & 84.7 \\
5413574971159438592 & 56935 & 5198 & 2.84 & 1.28 & < & 1.35 & 0.23 & 16.607 & 15.509 & 79.1 \\
5413579678443436288 & 57259 & 4971 & 2.31 & 1.45 &   & 1.07 & -0.33 & 15.644 & 14.388 & 123.1 \\
5413576482988137600 & 80609 & 4993 & 2.3 & 1.45 &   & 0.93 & - & 15.567 & 14.338 & 126.1 \\
5413576379903287680 & 74438 & 5122 & 2.79 & 1.3 &   & 1.21 & -0.32 & 16.632 & 15.445 & 80.5 \\
5413575761428289664 & 83738 & 5068 & 2.25 & 1.47 &   & 1.01 & -0.04 & 15.436 & 14.233 & 137.6 \\
5413575722769510528 & 83906 & 4955 & 2.01 & 1.55 &   & 0.91 & - & 15.009 & 13.763 & 160.7 \\
5413575692714165888 & 90831 & 4971 & 2.34 & 1.45 &   & 1.11 & -0.22 & 15.666 & 14.427 & 123.2 \\
5413576341244664448 & 75098 & 5107 & 2.31 & 1.46 &   & 0.97 & - & 15.527 & 14.328 & 128.7 \\
5413576375604411136 & 78059 & 4957 & 2.05 & 1.53 &   & 0.98 & -0.25 & 15.155 & 13.895 & 151.2 \\
5413575619690179072 & 81662 & 5064 & 2.39 & 1.44 & < & 1.07 & 0.04 & 15.725 & 14.518 & 121.5 \\
5413575722769408896 & 86611 & 5076 & 2.44 & 1.41 & < & 1.11 & 0.47 & 15.859 & 14.657 & 112.7 \\
5413575658348856192 & 89130 & 5087 & 2.6 & 1.38 & < & 1.19 & - & 16.211 & 15.017 & 95.7 \\
5413575692714152192 & 90541 & 4943 & 2.0 & 1.55 &   & 0.74 & - & 15.023 & 13.76 & 173.0 \\
5413575589634849408 & 77604 & 5028 & 2.35 & 1.45 & < & 1.01 & 0.41 & 15.612 & 14.399 & 129.5 \\
5413575623988713984 & 78651 & 4965 & 2.05 & 1.53 &   & 0.72 & - & 15.125 & 13.879 & 163.9 \\
5413575585330436608 & 80314 & 5020 & 2.31 & 1.45 & < & 0.99 & -0.12 & 15.538 & 14.332 & 132.0 \\
5413575550970716928 & 86502 & 5030 & 2.22 & 1.48 &   & 1.17 & -0.39 & 15.385 & 14.171 & 137.8 \\
5413575654049935360 & 87245 & 5118 & 2.45 & 1.41 &   & 1.1 & 0.13 & 15.871 & 14.673 & 111.0 \\
5413575447891531008 & 95041 & 5063 & 2.51 & 1.39 &   & 0.93 & -0.21 & 16.055 & 14.826 & 108.1 \\
5413574863775897216 & 69121 & 5044 & 2.43 & 1.41 &   & 0.78 & 0.29 & 15.837 & 14.644 & 118.3 \\
5413574863775908736 & 73445 & 5032 & 2.36 & 1.45 &   & 1.11 & -0.23 & 15.576 & 14.383 & 133.0 \\
5413574863775910784 & 73955 & 5076 & 2.74 & 1.32 & < & 1.22 & 0.12 & 16.489 & 15.314 & 83.3 \\
5413575555275124352 & 90802 & 4952 & 2.21 & 1.49 &   & 0.98 & - & 15.504 & 14.221 & 118.9 \\
5413574043446583552 & 82836 & 5023 & 2.4 & 1.44 &   & 0.9 & -0.18 & 15.813 & 14.574 & 121.4 \\
5413575349116671744 & 91356 & 5061 & 2.61 & 1.37 &   & 1.2 & -0.2 & 16.321 & 15.09 & 92.7 \\
\hline \hline
\multicolumn{9}{c}{NGC6838 $\mathrm{\langle [Fe/H]\rangle=-0.72\pm0.07}$} \\
\hline
1821606719608983552 & 159767 & 5007 & 2.3 & 1.44 &   & 0.87 & - & 15.181 & 13.938 & 198.1 \\
1821607028846759936 & 139910 & 5053 & 2.61 & 1.32 &   & 0.76 & - & 16.059 & 14.834 & 131.8 \\
1821606921430534144 & 132783 & 4872 & 2.21 & 1.44 & < & 0.83 & - & 15.307 & 14.008 & 176.3 \\
1821618642439491072 & 123258 & 4890 & 2.23 & 1.44 &   & 0.74 & - & 15.295 & 14.004 & 183.4 \\
1821608849912747904 & 189653 & 5082 & 2.39 & 1.39 &   & 0.82 & - & 15.65 & 14.436 & 148.5 \\
1821608781193211392 & 195102 & 5095 & 2.4 & 1.39 &   & 0.88 & - & 15.681 & 14.472 & 144.9 \\
1821609124790724736 & 184627 & 4997 & 2.29 & 1.44 &   & 0.85 & - & 15.219 & 13.972 & 172.9 \\
1821608918632106368 & 212643 & 5103 & 2.46 & 1.36 & < & 1.04 & - & 15.905 & 14.699 & 135.0 \\
1821614862867880192 & 206187 & 5159 & 2.41 & 1.37 &   & 1.19 & - & 15.823 & 14.638 & 138.3 \\
1821620978901638144 & 177436 & 4995 & 2.29 & 1.44 &   & 0.91 & - & 15.235 & 13.987 & 188.3 \\
1821620978901633152 & 178029 & 4950 & 2.28 & 1.43 &   & 0.92 & - & 15.337 & 14.071 & 178.1 \\
1821608334516883968 & 182638 & 5236 & 2.43 & 1.32 &   & 1.07 & - & 16.053 & 14.896 & 128.9 \\
1821607990919181696 & 196866 & 4921 & 2.36 & 1.4 &   & 0.73 & - & 15.607 & 14.329 & 158.3 \\
1821607200645158400 & 190514 & 5079 & 2.91 & 1.23 & < & 1.09 & - & 16.423 & 15.208 & 79.2 \\
1821605894975152128 & 175885 & 5025 & 2.41 & 1.39 &   & 0.82 & - & 15.701 & 14.465 & 158.1 \\
1821627232374952704 & 170015 & 4881 & 2.24 & 1.44 &   & 0.98 & - & 15.353 & 14.058 & 186.1 \\
1821620326067079680 & 151882 & 5124 & 2.38 & 1.42 & < & 0.98 & - & 15.425 & 14.227 & 167.9 \\
1821620772743302912 & 139388 & 4924 & 2.59 & 1.34 &   & 0.82 & - & 16.012 & 14.735 & 127.0 \\
1821609120458083456 & 185301 & 5030 & 2.38 & 1.41 &   & 0.67 & - & 15.541 & 14.307 & 157.4 \\
1821620841462681984 & 167131 & 4948 & 2.31 & 1.42 &   & 0.88 & - & 15.43 & 14.163 & 162.6 \\
1821609051738460288 & 187503 & 5159 & 2.67 & 1.27 &   & 1.09 & - & 16.192 & 15.007 & 117.7 \\
1821620424811791744 & 155286 & 5095 & 2.86 & 1.25 & < & 1.1 & - & 16.309 & 15.1 & 109.6 \\
1821608712473894784 & 169861 & 5124 & 2.5 & 1.34 &   & 0.95 & - & 15.979 & 14.781 & 136.4 \\
1821608643720219776 & 174250 & 4973 & 2.29 & 1.43 &   & 1.0 & - & 15.273 & 14.016 & 180.7 \\
1821608433263154176 & 180780 & 4973 & 2.31 & 1.43 &   & 0.75 & - & 15.375 & 14.118 & 173.3 \\
1821608815518575872 & 184337 & 5103 & 2.54 & 1.34 &   & 0.94 & - & 15.986 & 14.78 & 139.1 \\
1821608811220291200 & 187296 & 5197 & 3.08 & 1.21 & < & 1.22 & - & 16.594 & 15.423 & 99.7 \\
1821608605061812224 & 166246 & 5079 & 2.38 & 1.42 &   & 0.74 & - & 15.448 & 14.233 & 176.5 \\
1821608433263141248 & 174895 & 5151 & 2.4 & 1.39 &   & 1.12 & - & 15.634 & 14.446 & 150.7 \\
1821608437595913472 & 177054 & 5020 & 2.37 & 1.41 &   & 0.9 & - & 15.534 & 14.296 & 154.8 \\
1821608437561457536 & 181230 & 5124 & 2.42 & 1.36 &   & 0.97 & - & 15.874 & 14.676 & 136.7 \\
1821608403236176000 & 173214 & 5095 & 2.36 & 1.44 &   & 1.16 & - & 15.086 & 13.877 & 195.7 \\
\end{longtable}
}
\end{appendix}

\end{document}